\documentclass[aps,
               prd,
               twocolumn,
               showpacs,
               superscriptaddress,
               nofootinbib,
               longbibliography,
               floatfix]{revtex4-1}

\usepackage{braket}
\usepackage{amsfonts}
\usepackage{amsmath}
\usepackage{amssymb}
\usepackage{amsthm}
\usepackage{bm}
\usepackage{dcolumn}
\usepackage{epsfig}
\usepackage{graphicx}
\usepackage{graphics}
\usepackage[latin1]{inputenc}
\usepackage{latexsym}
\usepackage{rotating}
\usepackage[dvipsnames]{xcolor}
\usepackage{mathrsfs}
\usepackage{microtype}
\usepackage{verbatim}
\usepackage{url}

\usepackage[caption=false]{subfig}
\usepackage[normalem]{ulem}

\usepackage[breaklinks=true]{hyperref}
\hypersetup{
    colorlinks=true,
    linkcolor=NavyBlue,
    filecolor=Magenta,
    urlcolor=NavyBlue,
    citecolor=NavyBlue
}

\usepackage{yfonts}
\usepackage{float}
\usepackage{xspace} 
\usepackage{mathrsfs}
\usepackage[toc,page]{appendix}
\usepackage{siunitx}
\usepackage{array}
\usepackage[normalem]{ulem}

\newcommand{\be}{\begin{equation}}
\newcommand{\ee}{\end{equation}}
\newcommand{\EH}{{\mbox{\tiny EH}}}
\newcommand{\R}{{\mbox{\tiny R}}}
\newcommand{\RW}{{\mbox{\tiny RW}}}

\newcommand{\ZM}{{\mbox{\tiny ZM}}}
\newcommand{\CS}{{\mbox{\tiny CS}}}
\newcommand{\GR}{{\mbox{\tiny GR}}}

\newcommand{\p}{\prime}
\newcommand{\pp}{\prime\prime}

\newcommand{\K}{{\mbox{\tiny K}}}
\newcommand{\sr}{{\mbox{\tiny SR}}}
\newcommand{\h}{{\mbox{\tiny H}}}
\newcommand{\I}{{\mbox{\tiny I}}}

\newcommand{\ba}{\begin{align}}
\newcommand{\ea}{\end{align}}

\AtBeginDocument{%
    \newwrite\bibnotes
    \def\bibnotesext{Notes.bib}
    \immediate\openout\bibnotes=\jobname\bibnotesext
    \immediate\write\bibnotes{@CONTROL{REVTEX41Control}}
    \immediate\write\bibnotes{@CONTROL{%
    apsrev41Control,author="08",editor="1",pages="1",title="0",year="1"}}
    \if@filesw
    \immediate\write\@auxout{\string\citation{apsrev41Control}}%
    \fi
}

\newcommand{\UIUC}{Illinois  Center  for  Advanced  Studies  of  the  Universe \&
Department of Physics, University of Illinois at Urbana-Champaign, Urbana, Illinois 61801, USA}
\newcommand{\AEI}{Max Planck Institute for Gravitational Physics (Albert Einstein Institute), Am M\"uhlenberg 1, 14476 Potsdam, Germany}

\begin{document}
\title{Quasinormal modes of slowly-rotating black holes in dynamical Chern-Simons gravity}

\author{Pratik Wagle}
\affiliation{\UIUC}
\email{wagle2@illinois.edu}

\author{Nicol\'as Yunes}
\affiliation{\UIUC}

\author{Hector O. Silva}
\affiliation{\AEI}
\affiliation{\UIUC}

\date{\today}

\begin{abstract}
The detection of gravitational waves from compact binary mergers by the
LIGO/Virgo collaboration has, for the first time, allowed us to test relativistic
gravity in its strong, dynamical and nonlinear regime, thus opening a new
arena to confront general relativity (and modifications thereof) against
observations.
We consider a theory which modifies general relativity by introducing a scalar
field coupled to a parity-violating curvature term known as dynamical Chern-Simons gravity.
In this theory, spinning black holes are different from their general
relativistic counterparts and can thus serve as probes to this theory.
We study linear gravito-scalar perturbations of black holes
in dynamical Chern-Simons gravity at leading-order in spin and 
(i) obtain the perturbed field equations describing the evolution of the perturbed gravitational and scalar fields,
(ii) numerically solve these equations by direct integration to calculate the quasinormal mode frequencies for the dominant and higher multipoles and tabulate them,
(iii) find strong evidence that these rotating black holes are linearly stable, and
(iv) present general fitting functions for different multipoles for gravitational and scalar quasinormal mode frequencies in terms of spin and Chern-Simons coupling parameter.
Our results can be used to validate the ringdown of small-spin remnants of numerical relativity simulations of black hole binaries in dynamical Chern-Simons gravity and
pave the way towards future tests of this theory with gravitational wave ringdown observations.
\end{abstract}

\maketitle

\section{Introduction}
\label{sec:intro}

General Relativity (GR) has passed a plethora of experimental tests both in the Solar
System~\cite{Will2014} and in binary pulsars systems~\cite{Stairs2003,Wex:2020ald} making it one of the most successful physical theories.
These tests probe situations in which gravitational fields
are either weak, as in the Solar System, or systems where the field is strong but the system
is slowly-varying, as in binary pulsars.
However, the observation of gravitational waves (GW) by the LIGO/Virgo collaboration offers a new arena,
where the spacetime is highly dynamical and strongly curved, in which the predictions of Einstein's theory have 
being once more shown to agree with observations~\cite{TheLIGOScientific:2016src}.
Complementary, GW observations also allow one to constrain modifications to GR~\cite{Yunes:2016jcc,Yagi:2016jml,Berti:2018cxi} and
with more ground-based and space-based detectors in the future, these constraints will become more 
stringent (see e.g.,~Refs.~\cite{Gnocchi:2019jzp,Toubiana:2020vtf,Perkins:2020tra,Datta:2020vcj,Gupta:2020lxa}).

But why should one study modifications to GR? There are  
observational and theoretical anomalies that GR in its simplest form (i.e.,~without additional ``dark'' components or a UV completion) fails to answer. 
These include the late-time acceleration of the Universe~\cite{Perlmutter:1998np,Riess:1998cb}, 
the anomalous galaxy rotation curves~\cite{Sofue:2000jx,Bertone:2016nfn} and
the matter-antimatter asymmetry of the Universe~\cite{Canetti:2012zc}
and the incompatibility of quantum mechanics with GR.
A resolution to these anomalies may reside in a modification to Einstein's
theory that remains consistent with all current observational tests, 
yet yields deviations in other extreme
regimes where the gravitational interaction is simultaneously strong,
non-linear and highly dynamical.
On the theoretical side, the incompatibility of GR with quantum
mechanics has prompted efforts in a variety of unified theories, including string
theory and loop quantum gravity.

These issues have
served as motivation to study various extensions to
GR~\cite{Clifton:2011jh,Berti:2015itd}, such as $f(R)$ gravity and scalar-tensor theories~\cite{Sotiriou:2006hs,Kobayashi:2019hrl}, tensor-vector-scalar theories~\cite{Skordis:2009bf}, massive
gravity~\cite{deRham:2014zqa} and bi-gravity~\cite{Schmidt-May:2015vnx}.
Whether these attempts at modifying GR have any physical implications, requires one to first derive the predictions of such theories (in
a given scenario) which should be followed by a comparison of these predictions against observations.

Although the correct completion of GR is yet unknown, GWs from compact binary coalescence observations can help in constraining and excluding entire arrays of modified theories of gravity.
For instance, the GWs emitted in the inspiral of black hole binaries can tell us about the presence of extra radiative degrees of freedom, which provide an extra energy sink to which orbital energy and angular momentum can be extracted from the binary and hence affecting the system's orbital evolution (see e.g.,~Ref.~\cite{Berti:2018cxi}).
Here we concentrate on GWs emitted when the newly formed black hole relaxes towards 
its final equilibrium state, the so-called ringdown.
The GWs emitted during the ringdown can be described by a set of quasinormal modes (QNMs) -- complex-valued frequencies whose
imaginary part dictate how fast the mode decays in time. 
In GR, the observation of two or more QNMs in the ringdown signal allows one to uniquely infer the properties of the remnant Kerr black hole, similarly to how the observation of emission lines allows one to identify chemical elements~\cite{Detweiler:1980gk}.
This  ``black hole spectroscopy'' thus allows one to test the ``Kerr hypothesis''~\cite{Dreyer:2003bv,Berti:2005ys,Berti:2007zu} i.e., that the BHs found in Nature are described by the Kerr metric.
In general, modified theories of gravity do not admit the Kerr metric as a solution (see e.g. Ref.~\cite{Berti:2015itd}) and even when 
they do so~\cite{Sotiriou:2011dz,Herdeiro:2015waa,Motohashi:2018wdq}, the presence of the modifications to GR can be probed by perturbations 
to the Kerr metric~\cite{Barausse:2008xv}.
This makes BH spectroscopy a powerful probe into beyond-GR physics.

Here we concentrate on modifications to GR which introduce a scalar field non-minimally coupled to squared
curvature scalars, known as quadratic gravity theories~\cite{Yunes:2011we,Yagi:2015oca}.
One subset of these theories, known as dynamical Chern-Simons (dCS) gravity~\cite{Jackiw:2003pm}, 
was proposed as an explanation to the matter-antimatter asymmetry of the
universe by introducing additional parity-violating gravitational
interactions, challenging a fundamental pillar of GR~\cite{Alexander:2004xd,Alexander:2009tp}.
The theory is poorly constrained by Solar System experiments (see~\cite{Nakamura:2018yaw} for an overview), and remains unconstrained 
by both binary pulsars~\cite{Yagi:2013mbt} and GWs~\cite{Nair:2019iur} observations. 
Nonetheless, first constraints on dCS were obtained through multi-messenger neutron star observations in Ref.~\cite{Silva:2020acr}.

In dCS, nonrotating BHs are identical to their GR counterparts, but when spun a nontrivial scalar field configuration arises and whose presence affects the spacetime metric. 
Perturbations of spherically symmetric BHs in dCS were first studied in Ref.~\cite{Yunes:2007ss} who found the system of equation to be coupled and complicated. 
Later work decoupled these equations and studied them extensively~\cite{Cardoso:2009pk,Molina:2010fb,Kimura:2018nxk}. 
Here we extend all of these results to axisymmetric, slowly-rotating BHs in dCS gravity and 
study their QNM spectra and stability.


\subsection*{Executive summary}

We study the QNM spectra of slowly-rotating BHs in dCS gravity,
generalizing Refs.~\cite{Cardoso:2009pk,Molina:2010fb} which focused on the
non-rotating case.
To do so, we consider as a background the BH solution found
in Refs.~\cite{Yunes:2009hc,Konno:2009kg} that captures the leading-order corrections due to the dimensionless spin ($a/M$) and  Chern-Simons (CS) dimensionless coupling strength ($\alpha / M^2$) to the Schwarzschild spacetime. Here $a=J/M$ with $J$ and $M$ the Arnowitt-Deser-Misner angular momentum and mass respectively, while $\alpha$ is the coupling parameter (with dimensions of length squared in geometric units) between of the CS (pseudo-)scalar field and the Pontryagin density in the theory's action.

We study the most general linear perturbations to this solution, taking into consideration
both gravitational and scalar field perturbations. 
The outcome of this calculation is a pair of coupled, inhomogenous ordinary differential equations (ODEs) for the axial gravitational and scalar perturbations and a single homogeneous
equation for the polar gravitational perturbations. 
All these equations are found to have CS modifications when compared with their counterparts for a slowly-rotating Kerr BH in GR~\cite{Pani:2013pma}.

With these equations in hand, we numerically calculate the QNM frequencies
$\omega$, exploring their dependence on spin and coupling strength.
We find that the QNM spectra can be split into two branches:
(i) the scalar-led modes, whose frequencies in the limit $\alpha / M^2 \to 0$
reduce to that of a test scalar field on a slowly-rotating Kerr BH background~\cite{Witek:2018dmd,Pani:2013pma,Okounkova:2019dfo} and
(ii) the gravitational-led modes, whose frequencies in the limit $\alpha / M^2 \to 0$
reduce to that of the gravitational modes of a slowly-rotating Kerr background~\cite{Pani:2013pma}.

Our results show that the isospectrality existent in GR between axial and polar-parity gravitational
modes~\cite{Maggiore:2018sht} (i.e., the equivalence between the QNM spectra of each parity) is broken due to the scalar field.
This was first observed in~\cite{Molina:2010fb} in the non-rotating limit and is
shown here to persist when rotation is added.
The leading order corrections to QNM frequencies introduced by the CS coupling are found to enter at the quadratic order in the CS coupling for both the gravitational and the scalar modes.

\begin{table}[t]
	\begin{center}
		\begin{tabular}{c @{\hskip 0.15in} c @{\hskip 0.1in} c }
			\hline 
			\hline
			& $a/M$ & $\alpha/M^2$  \\ \hline
			$\textrm{Re}(\omega_g^\textrm{axial})$	& $ \uparrow$ & $ \uparrow $  \\[0.2cm]
			$\textrm{Im}(\omega_g^\textrm{axial})$	& $ \downarrow $ & $ \sim $  \\ [0.2cm]
			$\textrm{Re}(\omega_s)$	& $ \downarrow $ & $ \downarrow $  \\ [0.2cm]
			$\textrm{Im}(\omega_s)$	& $ \uparrow $ & $ \sim $  \\ [0.2cm]
			$\textrm{Re}(\omega_g^\textrm{polar})$	& $ \uparrow $ & $ \sim $  \\ [0.2cm]
			$\textrm{Im}(\omega_g^\textrm{polar})$	& $ \downarrow $ & $ \sim $  \\
			\hline
            \hline
            \\[-0.2cm]
		\end{tabular}
	\end{center}
	\caption{Summary of the behavior of fundamental dominant QNM frequency as we increase either the dimensionless spin $a/M$ or the dimensionless CS coupling $\alpha/M^2$ while keeping the other constant. $\uparrow$, $\downarrow$ and $\sim$ indicate an increase, decrease or almost constant behavior of the QNM frequency, respectively. Here, the subscript $g$ represents the gravitational-led and the subscript $s$, the scalar-led mode.
	}
	\label{table:summary}
\end{table}

We also found (at fixed spin $a/M$) that the axial gravitational modes decay slower and oscillate faster in dCS than in GR, with the latter being more sensitive to the CS coupling.
We find a positive correlation between $a/M$ and $\alpha/M^2$ on how they affect the real part of the QNMs: the oscillation frequency increases in the same way by increasing either of the two parameters while keeping the other constant.
This correlation is broken when considering decay rates, because changing the CS coupling has a negligible affect on the decay rate whereas a change in the spin parameter leads to longer lived modes. 
This result is important because it breaks the degeneracy between spin and CS coupling effects present in 
the real part of the QNM, allowing one to, in principle, constraint dCS gravity with the ringdown part of sufficiently high signal-to-noise-ratio GW events.
At fixed spin, these corrections scale with $(\alpha / M^2)^2$ due to dependence of the effective potential on the CS coupling and the coupling between the scalar and axial modes.
In Table~\ref{table:summary}, we summarize how each of the three set of modes behave as we increase either spin or CS coupling  while keeping the other constant.

We calculated a large set of QNM frequencies (see Appendix~\ref{appendix:qnm})
which we used to obtain fitting formulas for their real and imaginary parts as function of dimensionless spin and CS coupling [cf.~Eqs.~\eqref{eq:fullfitgrav} and \eqref{eq:fullfitsca}].
Our exploration of the parameter space indicates that the QNMs decay for all values of spin and CS coupling within the limits of the slow rotation and small coupling approximation we use.
This provides evidence that slowly rotating BH solutions in dCS gravity are linearly stable against gravito-scalar perturbations.

In the rest of this paper we show how these results were obtained.
In Sec.~\ref{sec:dcs} we give a short overview of dCS gravity and present 
the slowly-rotating BH solutions in this theory, whose QNM frequencies we are 
interested in computing.
In Sec.~\ref{sec:perturbation_eqs} we review some general aspects of BH
perturbation theory and derive the master equations that govern linear
perturbations of our background BH spacetime.
In Sec.~\ref{sec:int} we explain how these equations can be integrated numerically
and in
Sec.~\ref{sec:results} we present our numerical results.
Finally, in Sec.~\ref{sec:discussions} we summarize our main findings and discuss some avenues for future work.

%
We adopt the following conventions unless
stated otherwise: we work in 4-dimensions with metric signature $(-,+,+,+)$
as in~\cite{Misner:1974qy}.
Greek indices ($\alpha, \beta ....$)
represent spacetime indices, round brackets around indices represent
symmetrization, $\partial_\mu$ partial derivatives, 
$\nabla_\mu$ covariant derivatives
and $\Box = \nabla_\mu \nabla^\mu$ 
the d'Alembertian operator.
The Einstein summation convention is used throughout and we work in geometrical
units in which $G=1=c$.

\section{\label{sec:dcs} Dynamical Chern-Simons Gravity}

\subsection{Basics}

Let us start with a brief review of dCS gravity and establish some
notation~\cite{Alexander:2009tp}. In vacuum, the theory is described by the action
\be \label{eq:Action1}
S = S_{\EH} + S_{\vartheta} + S_{\CS}\,,
\ee
where the Einstein-Hilbert term is
\be \label{eq:EH}
S_{\EH} = \kappa \int d^4 x \sqrt{-g} \; R\,,
\ee
where $\kappa = (16 \pi )^{-1}$, $R$ is the Ricci scalar and $g$ is the determinant of
the metric $g_{\alpha \beta}$.
The action for the scalar field is
\be \label{eq:sf1}
S_{\vartheta} = - \frac{1}{2} \int d^4 x  \sqrt{-g} \left[ g^{\mu \nu} (\nabla_\mu \vartheta) (\nabla_\nu \vartheta) + 2 V(\vartheta) \right]\,,
\ee
where $\nabla_{\mu}$ is the covariant derivative operator compatible with the
metric, while
$V(\vartheta)$ is a potential for the scalar that we set to zero.
The scalar field is nonminimally coupled to the Pontryagin density $^*\!R R$
as
\be
S_{\CS} = \frac{\alpha}{4} \int d^4 x \sqrt{-g} \vartheta  ~^*\!R R\,,
\,\,
^*\!R R = ~^*\! R^{\mu}{}_{\nu}{}^{\kappa \delta} R^\nu{}_{\mu \kappa \delta}\,,
\ee
where $\alpha$ is the CS coupling constant with units of [Length]$^2$ and
$^*\!R^{\mu}{}_{\nu}{}^{\kappa \delta}$ is the dual Riemann tensor
\be
^*\! R^{\mu}{}_{\nu}{}^{\kappa \delta} = \frac{1}{2} \epsilon^\mu{}_{\nu \alpha \beta} ~ R^{\alpha \beta}{}^{\kappa \delta}  \,,
\ee
and $\epsilon^{\mu \nu \alpha \beta}$ is the Levi-Civita tensor.

The field equations are obtained by varying the action in Eq.~\eqref{eq:Action1} with
respect to the (inverse) metric $g^{\mu\nu}$ and scalar field $\vartheta$.
Variation with respect to $g^{\mu\nu}$ gives
\begin{equation}
G_{\mu \nu} + \frac{\alpha}{\kappa} C_{\mu \nu} = \frac{1}{2 \kappa} T^{\vartheta}_{\mu \nu}  \,,
\label{eq:field-eq}
\end{equation}
where
$G_{\mu \nu}$ is the Einstein tensor,
$C_{\mu\nu}$ is the (trace-free) C-tensor
\begin{equation}
C_{\mu\nu} = (\nabla_\sigma \vartheta) \epsilon^{\sigma \delta \alpha} {}_{(\mu} \nabla_\alpha R_{\nu) \delta}
+
(\nabla_\sigma \nabla_\delta \vartheta) ~ {}^*R^{\delta} {}_{(\mu \nu)} {}^{\sigma},
\label{eq:CTensor}
\end{equation}
which contains derivatives of the scalar
field, and
$T^{\vartheta}_{\mu\nu}$ is the canonical scalar field stress-energy tensor
\be \label{eq:SETfield}
T^{\vartheta}_{\mu \nu} =  \left[ (\nabla_\mu \vartheta)(\nabla_\nu \vartheta) - \frac{1}{2} g_{\mu \nu} (\nabla^\sigma \vartheta) (\nabla_\sigma \vartheta) \right]\,.
\ee
Variation with respect to $\vartheta$ gives the inhomogenous wave equation
\begin{equation}
\label{eq:CSfield}	 \Box \vartheta = - \frac{\alpha}{4} ~^*\! R R \,.
\end{equation}
One can show that $^*\! R R$ vanishes for static, spherically symmetric
spacetimes, resulting in $\vartheta = \textrm{const.}$ as the only regular solutions
of dCS in BH spacetimes with these symmetries~\cite{Jackiw:2003pm,Yunes:2007ss}.
This is no longer the case when these symmetries are lifted as we will see
next.

Throughout this paper, we treat dCS gravity as a low energy effective field theory. What this means is that we will be working perturbatively in the coupling of the theory, considering only small deformations away from GR. Since this topic has been covered in literature in thorough detail, we redirect a reader unfamiliar with this topic to these references~\cite{Alexander:2021ssr,Delsate:2014hba,Motohashi:2011pw,Motohashi:2011ds}.


\subsection{Slowly-rotating black holes} \label{sec:kerrmet}

Slowly-rotating BH solutions in dCS are known both
analytically~\cite{Konno:2009kg,Yunes:2009hc,Yagi:2012ya,Maselli:2017kic} and
numerically~\cite{Delsate:2018ome}.
Here we will consider the solution found in Refs.~\cite{Yunes:2009hc,Konno:2009kg},
which was obtained by solving the field equations~\eqref{eq:field-eq}
and~\eqref{eq:SETfield} perturbatively to linear order in spin $a$ and to
quadratic order in the coupling strength $\alpha$.
Following the notation of~\cite{Yunes:2009hc}, the line element of this solution is
\begin{align} \label{eq:ds2}
{d\bar{s}}^2 = ds^2_\sr + \frac{5}{4} \frac{\alpha^2}{ \kappa} \frac{a}{r^4} \left(1 + \frac{12}{7} \frac{M}{r} + \frac{27}{10} \frac{M^2}{r^2}\right) \sin^2 \theta dt d\phi \,,
\nonumber \\
\end{align}
where $d\bar{s}^2$ is the background line element for the slowly rotating BH in dCS gravity to leading order in the dimensionless spin parameter and CS coupling constant and
$ds^2_\sr$ is the line element for a slowly-rotating Kerr BH
\begin{align} \label{eq:skerrgr}
ds^2_\sr &= - f(r)  dt^2 -\frac{4 M a \sin^2 \theta}{r} dt d\phi + f(r)^{-1} dr^2
\nonumber \\
&\quad + r^2 d\theta^2  + r^2 \sin^2  \theta d\phi^2 \,,
\end{align}
and $ f(r) = 1 - 2 M/r$ is the Schwarzschild factor.
The solution also contains a nontrivial scalar field,
\be \label{eq:sfdcs}
\bar{\vartheta} = \frac{5}{8} \frac{a \alpha}{M}\frac{\cos \theta}{r^2} \left(1 + \frac{2M}{r} + \frac{18 M^2}{5 r^2}\right)\,.
\ee
A far-field analysis shows that the BH has a scalar dipole charge of value $-(5/8)(a \alpha/M)$.

We see from Eq.~\eqref{eq:ds2} that the metric for slowly-rotating BHs in dCS
gravity has an additional modification in the ($t\phi$)-component when compared
to the metric in GR given in Eq.~\eqref{eq:skerrgr}.
This modifies the horizon angular frequency $\Omega_{\h}$ as observed by a zero angular momentum observer 
at the horizon location $r_\h$ at first order in spin and second order in $\alpha$~\cite{Yunes:2009hc} as
\begin{equation}
\Omega_\h = \frac{a}{4 M^2}-\frac{709 \alpha ^2 a}{28672   \kappa  M^6} = \frac{a}{4 M^2} \left(1 - \frac{709}{7168} \zeta\right)\,,
\label{eq:ZAMO}
\end{equation}
where we defined
\begin{equation}
\zeta = \alpha^2/(M^4  \kappa).
\end{equation}
As we will see later, this $\alpha^2$-term affects the QNMs.

\section{Black Hole Perturbation Theory}
\label{sec:perturbation_eqs}

\subsection{Decomposition of the fundamental fields} 
\label{sec:bhp}

We consider linear perturbations
\begin{equation} \label{eq:perts}
    g_{\mu\nu} = \bar{g}_{\mu\nu} + \epsilon \, \delta_{\mu\nu},
    \quad
    \vartheta = \bar{\vartheta} + \epsilon \, \delta \vartheta\,,
\end{equation}
to the background BH spacetime [cf.~Eqs.~\eqref{eq:ds2} and~\eqref{eq:sfdcs}],
where $\epsilon$ is a bookkeeping parameter and both $\delta g_{\mu\nu}$ and
$\delta\vartheta$ are functions of the coordinates of the metric. 

The angular dependence of these perturbations can be described by
scalar, vector and tensor spherical harmonics.
The metric decomposition comes from the transformation properties of the ten
components of the perturbation tensor $\delta g_{\mu\nu}$ under a rotation of
the frame of origin~\cite{Regge:PhysRev.108.1063}.
These quantities transform as three SO(2) scalars $\delta g_{MN}$, two
SO(2) vectors $\delta g_{mN}$ and one SO(2) second rank tensor $\delta
g_{ab}$ which can be expanded into a complete basis formed by spherical
harmonics of different corresponding ranks.
Under a parity transformation (i.e., the simultaneous shifts
$\theta \to \pi - \theta$ and $\phi \to \phi + \pi$) the aforementioned metric quantities can be separated into odd (or ``axial'') and even (or ``polar'') sectors respectively, depending on whether they 
pick up a factor of $(-1)^{\ell+1}$ and $(-1)^\ell$.
This allows us to decompose $\delta g_{\mu\nu}$ as
\be
\delta g_{\mu\nu} (t,r,\theta,\phi) = \delta g^{\rm odd}_{\mu\nu} (t,r,\theta,\phi) + \delta g^{\rm even}_{\mu\nu} (t,r,\theta,\phi) \,,
\ee
where
\begin{widetext}
\be \label{eq:odd}
	 \delta g^{\rm odd}_{\mu\nu} =
\begin{pmatrix}
 0 & 0 & h^{\ell m}_0(t,r) S^{\ell m}_\theta(\theta, \phi) &  h^{\ell m}_0(t,r) S^{\ell m}_\phi(\theta, \phi) \\
	 							* & 0 & h^{\ell m}_1(t,r) S^{\ell m}_\theta(\theta ,\phi) &  h^{\ell m}_1(t,r) S^{\ell m}_\phi(\theta, \phi) \\
	 							* & * & 0 & 0  \\
	 							* & * & * & 0
\end{pmatrix}
\,,
\ee
and
\be \label{eq:even}
\delta g^{\rm even}_{\mu\nu} =
\begin{pmatrix}
	H_0^{\ell m}(t,r) Y_{\ell m}(\theta ,\phi) & H_1^{\ell m}(t,r) Y_{\ell m}(\theta, \phi) & 0 &  0 \\
	* & H_2^{\ell m}(t,r) Y_{\ell m}(\theta, \phi) & 0 &  0 \\
	* & * & r^2 K^{\ell m}(t,r) Y_{\ell m}(\theta, \phi) & 0  \\
	* & * & *  & r^2 \sin^2 \theta K^{\ell m}(t,r) Y_{\ell m}(\theta, \phi)
\end{pmatrix}
\,,
\ee
\end{widetext}
where the asterisk denotes symmetric components, $Y_{\ell m}(\theta, \phi)$ are
the scalar spherical harmonics, while
\begin{align}
S_{\theta}^{\ell m} (\theta, \phi) &= - \frac{1}{\sin\theta} \partial_{\phi} Y_{\ell m}(\theta, \phi)\,, \\
S_{\phi}^{\ell m}(\theta,\phi) &= \sin \theta \partial_\theta Y_{\ell m}(\theta, \phi) \,,
\end{align}
and a sum over $\ell$ and $m$ in the usual sense ($\ell \geqslant 0$ and $|m| \leqslant \ell$)
is implicit.
Equations~\eqref{eq:odd} and~\eqref{eq:even} hold under the Regge-Wheeler
gauge~\cite{Regge:PhysRev.108.1063}, which can be applied in theories with a massless graviton that support the usual two polarizations.
In certain modified theories, a graviton can
propagate with up to six polarizations, thereby leaving no residual gauge freedom.
However, for the case of dCS gravity, GWs continue to propagate with only two polarizations 
(as measured at future null infinity)~\cite{Sopuerta:2009iy,Wagle:2019mdq}, and thus, one retains enough gauge freedom to impose the 
Regge-Wheeler gauge.
Such a decomposition separates the axial and polar perturbations with different
harmonic index $\ell$, i.e., for a given $\ell$, we have two
systems of evolution equations, one for the axial sector and one for the polar sector.
These separate sets of equations completely characterize the linear response of
the system.

Additional fields in the system, such as 
vectors and scalars, can be decomposed into spherical harmonics of the corresponding
type.
For scalar fields, we use scalar spherical harmonics and
the perturbed scalar field reads
\be \label{eq: sfpert}
\delta \vartheta(t,r,\theta,\phi) = \frac{R_{\ell m}(r,t)}{r} Y_{\ell m}(\theta,\phi)\,.
\ee

\subsection{Evolution equations for perturbations of a slowly rotating black hole}
\label{sec:slowrotpert}

Having established the background spacetime and explained how the linear
perturbations can be decomposed into scalar and tensor harmonics, we 
can now derive the perturbed form of the field equations in dCS gravity.
The procedure is as follows:
\begin{itemize}
\item Substitute the linear perturbations~\eqref{eq:perts} into the field 
equations~\eqref{eq:field-eq} and~\eqref{eq:CSfield} and expand to 
linear order in~$\epsilon$.
\item Expand the perturbed field equations to linear order in the spin parameter $a$ (slow-rotation approximation) and quadratic order in the coupling parameter $\alpha$ (small-coupling approximation). 
\item Use the orthogonality properties of the spherical harmonics presented in Appendix~\ref{appendix:sphharm} to eliminate the angular dependence 
of the functions $A^{a} \in \lbrace h_0 , h_1, H_0 , H_1, H_2 , K , R \rbrace $, making them functions of $t$ and $r$ only. Moreover, assume an harmonic time-dependence in time, i.e.,
\be \label{eq:dec}
A^{a}_{\ell m}(t,r) = e^{-i \omega t} A_{\ell m}^{a} (r)\,.
\ee
\item The previous steps yields a system of 11 equations; 10 from the metric field equation~\eqref{eq:field-eq} and one from the scalar field equation~\eqref{eq:CSfield}. The latter gives the evolution equation for the scalar field perturbations, whereas 3 of the metric equations reduce to the axial gravitational perturbation equation and the remaining 7 give an expression for the polar gravitational perturbation equation.
\item 
This system of equations can then be expressed in general by an equation of the form
\be \label{eq:pereqrep}
\mathcal{D} \Psi_j + V_j \Psi_j = S_j[\Psi_k,\partial_r \Psi_k] \,,
\ee
where $j \in \{ \R, \RW, \ZM \}$, $k \in \{\{ \R,\RW,\ZM \} - j \}$ with $\{ \R,\RW,\ZM \}$ denoting the scalar, axial and polar gravitational perturbations respectively, $\mathcal{D}$ is a second-order radial differential operator, which in tortoise coordinates $(r_{*})$ reduces to $d^2/dr_{*}^{2}$, and $V_j$ is the effective potential. 
The source term $S_j$ is found to be a linear combination of the $\Psi_k$ master functions and its first radial derivatives, e.g.~when $j = \R$, the master function $\Psi_\R = R_{\ell m}$, and the source term $S_\R$ is a function of $\Psi_{\RW}$ and $\Psi_{\ZM}$ and their first radial derivatives.

\end{itemize}

In the next subsections, we provide the final expressions for the perturbation equations for the scalar and gravitational sectors. 


\subsubsection{Scalar sector}
\label{sec:scalar}

The full equation describing the scalar field perturbation $R_{\ell m}$ is given by
\begin{align} \label{eq:sfpeq}
&f(r)^2 \partial_{rr} R_{\ell m} + \frac{2 M}{r^2}f(r) \partial_{r} R_{\ell m}  + \left[\omega^2 - V^{S}_{\rm eff}(r,a,\alpha^2) \right] R_{\ell m}
\nonumber \\
&= \alpha  f(r) \left\lbrace  \left[g(r) + a m h(r) \right]   \Psi^{\RW}_{\ell m} +  a m j(r) \partial_{r}\Psi^{\RW}_{\ell m}  \right\rbrace
\nonumber \\
&\quad + \alpha a \left\lbrace q_{\ell m}  \left[ k_{1}(r) \Psi^{\ZM}_{\ell -1,m} + k_{2}(r) \partial_{r} \Psi^{\ZM}_{\ell -1,m} \right] \right.
\nonumber \\
&\quad + \left. q_{\ell +1,m} \left[ k_{3}(r) \Psi^{\ZM}_{\ell +1,m} + k_{4}(r) \partial_{r} \Psi^{\ZM}_{\ell +1,m} \right] \right\rbrace \,,
\end{align}
where $f(r) = 1-2 M/r$,
\be \label{eq:qlm}
q_{\ell m} = \sqrt{\frac{\ell ^2 - m^2}{4\ell ^2 -1}}\,,
\ee
and $\partial_r$ denotes radial derivatives. In Eq.~\eqref{eq:sfpeq}, the functions $g$, $h$, $j$ and $k_{i}$
($i=1,\dots,4$) also depend on $\ell$ and the mass of the black hole $M$
in addition to the radial coordinate.
Their explicit forms are shown in Appendix~\ref{appendix:coefficientsofPE} and in a Mathematica notebook~\cite{PratikRingdown}.

We also followed~\cite{Pani:2013pma} and defined a CS modified Regge-Wheeler function $\Psi^{\RW}_{\ell m}$, which to leading order in $a$ is given by
\be
\Psi^{\RW}_{\ell m} = \frac{f(r)}{r} \left( 1 + \frac{2mMa}{r^3 \omega} - \alpha^2 a \delta\Psi^{\RW}(r) \right) h_{1}^{\ell m}\,,
\ee
where $h_{1}^{\ell m}$ comes from Eq.~\eqref{eq:odd}; the quantiy $\delta \Psi^{RW}(r)$ is a function of the radial coordinate included in the definition to allow us to maintain a form of the left hand side of Eq.~\eqref{eq:sfpeq} and other perturbed field equations similar to the form held by the Regge-Wheeler and Zerilli-Moncrief equation in GR.
Similarly, we introduced the Zerilli-Moncrief function $\Psi^{\ZM}_{\ell m}$, which to leading order in $a$ is given by
\be \label{eq:zm}
\Psi^{\ZM}_{\ell m} = ({z_1}/{z_2}) - a ({z_3}/{z_4}) + \alpha^2 a \delta\Psi^{\ZM}(r) \,,
\ee
where
\begin{subequations}
\begin{align}
z_1 &= - 2i H_1^{\ell m} (r-2M) + 2 K^{\ell m} r^2 \omega \,, 
\\
z_2 &= \left( 6M + \lambda_{\ell} r \right) \omega \,, 
\\
z_3 &= -4imM \omega \left[H_1^{\ell m} (r-2M) + i K^{\ell m} r^2 \omega \right]
\nonumber \\
&\quad \times \left\lbrace 48 M^3 - 24 M^2 r + \lambda_{\ell} r^3 (\lambda_{\ell} + 2r^2 \omega^2) \right.
\nonumber \\
&\quad + \left.  2Mr^2 \left( \lambda_{\ell}^2 + 6r^2 \omega^2 \right) \right\rbrace \,,
\\
z_4 &= \ell(\ell+1)r^4 z_2^3  \,,
\end{align}
\end{subequations}
with $\lambda_{\ell}= (\ell+2)(\ell-1)$, and $H_1^{\ell m}$ and $K^{\ell m}$ coming from Eq.~\eqref{eq:even}. The quantity $\delta\Psi^{\ZM}(r)$ serves a similar purpose to $\delta \Psi^{RW}(r)$. The forms of both $\delta \Psi^{RW}(r)$ and $\delta\Psi^{\ZM}(r)$ are provided in a separate Mathematica notebook~\cite{PratikRingdown}.

\subsubsection{Metric sector}
\label{sec:metric}

The full equation describing the axial gravitational perturbation $\lambda_{\ell m}$ is given by
\begin{align} \label{eq:axpeq}
f(r)^2& \partial_{rr} \Psi^{\RW}_{\ell m} + \frac{2 M}{r^2}f(r) \partial_{r}\Psi^{\RW}_{\ell m}  + \left[\omega^2 -  V^A_{\rm eff}(r,a,\alpha^2)\right]\Psi^{\RW}_{\ell m}
\nonumber \\
= \alpha & f(r) \left\lbrace \left[v(r) + a m n(r)\right] R_{\ell m} + a m p(r) \partial_{r} R_{\ell m}^{\p}  \right\rbrace
\nonumber \\
+ & a \left\lbrace q_{\ell m}  \left[ p_{1}(r) \Psi^{\ZM}_{\ell -1,m} + p_{2}(r) \partial_{r} \Psi^{\ZM}_{\ell -1,m} \right] \right.
\nonumber \\
+ & \left. q_{\ell +1,m} \left[ p_{3}(r) \Psi^{\ZM}_{\ell +1,m} + p_{4}(r) \partial_{r}\Psi^{\ZM}_{\ell +1,m} \right] \right\rbrace \,,
\end{align}
where the functions $v$, $n$, $p$ and $p_{i}$ ($i=1, \dots, 4$) also depend on
$\ell$ and $M$ in addition to the radial coordinate. 
Their explicit forms are given in Appendix~\ref{appendix:coefficientsofPE} and 
in a Mathematica notebook available upon request.

Finally, the polar gravitational perturbation $Z_{\ell m}$ satisfies the equation
\begin{align}
f&(r)^2 \partial_{rr}\Psi^{\ZM}_{\ell m} + \frac{2 M}{r^2}f(r) \partial_{r}\Psi^{\ZM}_{\ell m}  + \left[\omega^2 - V^P_{\rm eff}(r,a,\alpha^2)\right]\Psi^{\ZM}_{\ell m}
\nonumber \\
&= \alpha a  f(r) \left\lbrace q_{\ell m} \left( s_1(r) R_{\ell -1,m} + s_2(r) \partial_{r}R_{\ell -1,m}\right) \right.
\nonumber \\
&\quad  + q_{\ell +1,m} \left. \left( s_3(r) R_{\ell +1,m} +s_4(r) \partial_{r}R_{\ell +1,m} \right) \right\rbrace 
\nonumber \\
&\quad +  a  \left\lbrace q_{\ell m} \left( r_1(r) \Psi^{\RW}_{\ell -1,m} + r_2(r) \partial_{r}\Psi^{\RW}_{\ell -1,m} \right) \right.
\nonumber \\
&\quad + \left. q_{\ell +1,m} \left( r_3(r) \Psi^{\RW}_{\ell +1,m} + r_4(r) \partial_{r}\Psi^{\RW}_{\ell +1,m} \right) \right \rbrace \,,
\label{eq:popeq}
\end{align}
where the functions $s_{i}$ and $r_{i}$ ($i = 1, \dots, 4$) depend on $r$, $\ell$ and $M$.
Their explicit forms are shown in Appendix~\ref{appendix:coefficientsofPE} and in a Mathematica notebook~\cite{PratikRingdown}.

\subsubsection{Selection rule and propensity rule}
\label{sec:analysisperteq}

The perturbation equations~\eqref{eq:sfpeq},~\eqref{eq:axpeq} and \eqref{eq:popeq} show explicitly that $\ell$ modes couple to $\ell$ modes \textit{and} $\ell \pm 1$ modes.
However, as for slowly-rotating BHs in GR, these equations possess a selection rule~\cite{1991RSPSA.433..423C,Pani:2012bp}. 
In GR and at linear order in spin, the selection rule is that the $\ell$-th axial (polar) mode couples to the $\ell \pm 1$ polar (axial) mode; at second order in spin, this simple selection rule needs to be modified~\cite{Pani:2012bp}. 
Similarly, in dCS gravity and at linear order in spin, the same selection rule applies to the perturbation equations. 
The only modification is that in dCS gravity we have two fields with axial parity: the scalar field (encoded in $R_{\ell m}$) and the Regge-Wheeler function (encoded in $\Psi_{\ell m}^{\RW}$). 
Thus, in dCS gravity, $R_{\ell m}$ ($\Psi_{\ell m}^{\RW}$) couples to $\Psi_{\ell m}^{\RW}$ ($R_{\ell m}$) and to $\Psi_{\ell \pm 1, m}^{\ZM}$, while $\Psi_{\ell m}^{\ZM}$ couples to both $R_{\ell \pm 1, m}$  and $\Psi_{\ell \pm 1, m}^{\RW}$.

In addition to this selection rule, the perturbation equations also suggest a propensity rule. More specifically, we see that when $\ell=|m|$, which dominates the linear response of the system, Eq.~\eqref{eq:qlm} yields $q_{\ell m} = 0$, and thus the coupling of $\ell$ modes with $\ell-1$ modes is suppressed. This is similar in nature to the propensity rule of atomic physics, which states that transitions involving $\ell \to \ell+1$ are favored over those involving $\ell \to \ell-1$~\cite{1991RSPSA.433..423C}. Thus, we expect that the dominant modes are coupled only to the $\ell$ or the $\ell + 1$ modes, after imposing the selection rule. 

\subsection{Simplification of the perturbation equations} \label{sec:omegacontribute}

Mode coupling between perturbation of different $\ell$-modes leads to a rich spectrum of solutions, but this paper is concerned with the QNM frequencies, which, as it turns out, are not affected by mode coupling to leading order in spin. Indeed, in GR this has been known since the 1990s, thanks to the work of Kojima and others~\cite{Kojima:1993,Ferrari:2007rc,Pani:2012vp,Pani:2012bp}. Let us then establish the same result in dCS gravity to leading order in spin and second order in coupling parameter. Without loss of generality, let us rewrite the perturbation equations as
\begin{align}
\label{eq:aeq}	A_{\ell m} &+ ma \tilde{A}_{\ell m} + a (q_\ell \hat{P}_{\ell-1,m} + q_{\ell+1} \hat{P}_{\ell+1, m}) = 0 \,,  \\
\label{eq:peq}	P_{\ell m} &+ ma \tilde{P}_{\ell m} \,+  a (q_\ell \hat{A}_{\ell-1,m} + q_{\ell+1} \hat{A}_{\ell+1, m}) = 0 	\,.
\end{align}
In these equations, $A_{\ell m}$, $\tilde{A}_{\ell m}$ and $\hat{A}_{\ell \pm 1, m}$ are linear
combinations of odd (axial) perturbations, which include the Regge-Wheeler
function $\Psi_{\ell m}^{\RW}$ and its derivatives, and also the scalar field
perturbation $R_{\ell m}$ and its derivatives; we remind the reader that
the scalar field perturbations are of odd (axial) parity as seen from Eq.~\eqref{eq:CSfield}.
The prefactors of $\alpha^0$ and $\alpha^1$ have also been suppressed in the above functions for simplicity of notation.
Similarly, $P_{\ell m}$, $\tilde{P}_{\ell m}$ and $\hat{P}_{\ell \pm 1, m}$ are linear
combinations of polar perturbations encoded in the Zerilli-Moncrief function $\Psi_{\ell m}^{\ZM}$ and
its derivatives.

In GR, Kojima~\cite{Kojima:1993} showed using symmetry arguments for the $m=0$ mode that the terms $\hat{P}_{\ell \pm 1,m}$ and $\hat{A}_{\ell \pm 1,m}$ in
Eqs.~\eqref{eq:aeq} and~\eqref{eq:peq} make no contribution to the
QNMs. 
This argument was later extended to other values of $m$ in Ref.~\cite{Pani:2012bp}
for massive vector field perturbations of the slowly-rotating Kerr metric
in GR.
Following Ref.~\cite{Pani:2012bp}, we now extend this argument to slowly rotating BHs in dCS gravity. 

Consider a simultaneous transformation,
\begin{align} \label{eq:trans}
	x_{\ell ,m} \to& \mp x_{\ell ,-m} \,, & y_{\ell ,m} \to& \pm y_{\ell ,-m} \,, \nonumber \\
	m \to& -m \,, &a \to& -a \,,
\end{align}
where $x_{\ell m}$ and $y_{\ell m}$ represent the axial and polar perturbation
variables respectively, with indices $(\ell,m)$ given in Eqs.~\eqref{eq:aeq}
and~\eqref{eq:peq} which remain invariant under such a transformation.
The boundary conditions for QNMs of slowly rotating BHs in dCS (see~Sec.~\ref{sec:bdd}) are also invariant under such transformation. 
Then, in the slow rotation limit, the QNM frequencies can be expanded as
\be \label{eq:genw}
\omega = \omega_0 + m a \omega_1 + a \omega_2 + \mathcal{O}(a^2) \,,
\ee
where $\omega_0$ is the eigenfrequency of the non-rotating BH in dCS gravity, which in our case is just a Schwarzschild BH.
The effective potential presented in Sec.~\ref{sec:slowrotpert} is proportional to $a^0$ and $ma$, but not to $a$ alone (see Appendix~\ref{appendix:coefficientsofPE}).
Hence, $\omega_2=0$, because the above potential would not source such a term\footnote{Another (more physical) way to see this is by considering the $a \to -a$ transformation of Eq.~\eqref{eq:trans}. Such a transformation would physically correspond to inverting the direction of the BH's spin angular momentum. However, the QNM frequencies should not change due to the spin orientation. Hence, in general, $\omega_2=0$ making Eq.~\eqref{eq:genw} invariant under the symmetry in Eq.~\eqref{eq:trans}.}.

The only terms that could source $\omega_2$, at least in principle, are $(\hat{P}_{\ell \pm 1, m},\hat{A}_{\ell \pm 1, m})$. This is because the second terms in Eqs.~\eqref{eq:aeq} and~\eqref{eq:peq} are explicitly proportional to $m$, and because $(\tilde{A}_{\ell m},\tilde{P}_{\ell m})$ cannot be inversely proportional to $m$. Whether $(\hat{P}_{\ell \pm 1, m},\hat{A}_{\ell \pm 1, m})$ source $\omega_2$ depends on if they are independent of $m$ or not. However, since the second and third terms in  Eqs.~\eqref{eq:aeq} and~\eqref{eq:peq} are linear in $a$, both the functions $(\tilde{A}_{\ell m},\tilde{P}_{\ell m})$ and $(\hat{P}_{\ell \pm 1, m},\hat{A}_{\ell \pm 1, m})$ must be kept only to ${\cal{O}}(a^0)$. Therefore, they correspond to perturbations of a non-rotating BH, which is necessarily spherically symmetric, implying in particular that $(\hat{P}_{\ell \pm 1, m},\hat{A}_{\ell \pm 1, m})$ are actually independent of $m$. But since $\omega_2$ vanishes by the arguments presented above, it follows that $(\hat{P}_{\ell \pm 1, m},\hat{A}_{\ell \pm 1, m})$ need not be included when computing the QNM spectrum.   

Since the mode coupling terms can be neglected, our perturbation equations reduce to
\begin{subequations}
\label{eqs:perturbed_final}
\begin{align}
\label{eq:SF}	f(r)^2& \partial_{rr}R_{\ell m} + \frac{2 M}{r^2}f(r) \partial_{r}R_{\ell m}  + \left[\omega^2 - V^S_{\rm eff}(r,a,\alpha^2)\right] R_{\ell m}
\nonumber \\ &= \alpha f(r) \left \lbrace \left[g(r) + a m h(r) \right] \Psi^{\RW}_{\ell m} +  a m j(r) \partial_{r}\Psi^{\RW}_{\ell m}  \right \rbrace  \,, \nonumber \\ \\
\label{eq:RW}	f(r)^2 & \partial_{rr}\Psi^{\RW}_{\ell m} + \frac{2 M}{r^2}f(r) \partial_{r}\Psi^{\RW}_{\ell m}  + \left[\omega^2 - V^A_{\rm eff}(r,a,\alpha^2)\right]\Psi^{\RW}_{\ell m}
\nonumber \\ &= \alpha f(r) \left\lbrace \left[v(r) + a m n(r)\right] R_{\ell m} +  a m p(r) \partial_{r} R_{\ell m}  \right \rbrace \,, \nonumber \\ \\
\label{eq:ZM}	f(r)^2& \partial_{rr}\Psi^{\ZM}_{\ell m} + \frac{2 M}{r^2}f(r) \partial_{r}\Psi^{\ZM}_{\ell m} 
+ \left[ \omega^2 - V^P_{\rm eff}(r,a,\alpha^2) \right]\Psi^{\ZM}_{\ell m} \nonumber \\ & =0 \,,
\end{align}
\end{subequations}
where the values of all these functions are the same as before 
and are given in Appendix~\ref{appendix:coefficientsofPE}. 

Equations~\eqref{eqs:perturbed_final} can be recast in a Schr\"odinger-like form by introducing 
tortoise coordinates $r_{\ast}$, 
\be
r_* = r + 2 M \log \left(\frac{r}{2 M}-1\right)\,.
\ee
However, when solving these equations, we will stick to the form given above to avoid confusion. Note in passing that the tortoise coordinate is typically used to map Schwarzschild coordinates to a horizon-penetrating (typically ingoing Eddington-Finkelstein) coordinate system, which is well adapted to imposing boundary conditions at the BH horizon. The standard transformation known in GR, however, may need to be modified by terms of ${\cal{O}}(\alpha^2 a^2)$ to transform a dCS BH to horizon-penetrating coordinates. Since we are here working to a lower order in perturbation theory, we do not need to worry about such details. 

\subsection{dCS coupling dependence of $\omega$ through a Fermi estimate}
\label{sec:Fermi}
As seen from the previous section, for the calculation of the QNM frequencies, the perturbed equations take the form of Eqs.~\eqref{eqs:perturbed_final}. In this subsection, we discuss the dependence of the QNM frequency $\omega$ on the dCS coupling parameter $\alpha$.

From Eq.~\eqref{eq:ZM}, we see that the polar gravitational sector satisfies a homogeneous equation that lacks a direct coupling to the scalar field,  although the effective potential $V_{\rm eff}^P$ is dependent on the CS coupling parameter $\alpha$. Thus, the QNM frequencies for the polar sector are proportional to second order in the CS coupling parameter.
On the other hand, Eqs.~\eqref{eq:SF} and \eqref{eq:RW} have a non-vanishing linear-in-$\alpha$ source term in addition to the effective potential having quadratic dependence on the CS coupling parameter.

What does this imply on the dCS corrections to the QNM frequencies $\omega$? We can answer this question with a Fermi estimate in which we replace any radial derivative by a characteristic radius $\partial_r \to 1/{\cal{R}}$ and evaluate the equation at this characteristic radius. Doing this in the perturbation equations \eqref{eq:SF} and \eqref{eq:RW} gives
\begin{subequations}\label{eq:fermieq}
\begin{align} \label{eq:fermieqR}
&\left[G({\cal{R}}) + \omega^2 - \left( V^S_{1}({\cal{R}}) + \alpha^2  V^S_{2}({\cal{R}}) \right) \right] R_{lm} =  \alpha H({\cal{R}}) \Psi^{lm}_{\RW}\,,
 \\
\label{eq:fermieqRW}
&\left[G({\cal{R}}) + \omega^2 - \left( V^A_{1}({\cal{R}}) + \alpha^2  V^A_{2}({\cal{R}}) \right) \right] \Psi^{lm}_{\RW} = 
\alpha I({\cal{R}}) R^{lm}\,,
\end{align}
\end{subequations}
where the effective potentials have been written as a linear combination of terms at zeroth and second order in CS coupling parameter and $G({\cal{R}}), H({\cal{R}}), I({\cal{R}})$ are polynomials in the radial coordinate $r$ evaluated at $r = {\cal{R}}$, namely 
\begin{align}
    G({\cal{R}}) &= \frac{f({\cal{R}})^2}{r^2} + \frac{2M f({\cal{R}})}{{\cal{R}}^3} \,, \\
    H({\cal{R}}) &= f({\cal{R}}) \left[ g({\cal{R}}) + a m \left( h({\cal{R}}) + \frac{j({\cal{R}})}{{\cal{R}}} \right) \right] \,, \\
    I({\cal{R}}) &= f({\cal{R}}) \left[ v({\cal{R}}) + a m \left( n({\cal{R}}) + \frac{p({\cal{R}})}{{\cal{R}}} \right) \right] \,.
    \label{eq:fermi-poly}
\end{align}
Equations~\eqref{eq:fermieqR} can be solved for $R_{\ell m}$ and then inserted into Eq.~\eqref{eq:fermieqRW}, which leads to an equation for $\omega$ only, since $\Psi^{\RW}_{\ell m}$ cancels and the $\omega$ dependence in $H({\cal{R}})$ and $I({\cal{R}})$ cancels when they are multiplied together. Doing so, one finds
\begin{align}
\left[G + \omega^2 - \left( V^S_{1} + \alpha^2  V^S_{2} \right)\right]& \left[G + \omega^2 - \left( V^A_{1} + \alpha^2  V^A_{2} \right) \right] \nonumber \\ &=  
\alpha^2 I \, H\,,
\end{align}
where we have suppressed the argument of the functions. This is a quadratic equation for $\omega^2$, which we can solve perturbatively in $\alpha \ll 1$ to find
\begin{equation} \label{eq:fermimain}
\omega = \omega_{\GR}  + \zeta \, \delta \omega + \mathcal{O}(\alpha^{3}) \,,
\end{equation}
where, recall, $\zeta = \alpha^2/(M^4 \kappa)$
and
\begin{align}
    \omega_{\GR} &= \left( V^A_{\rm eff} - G\right)^{1/2}\,,
    \\
    \label{eq:delta-w-Fermi}
    \delta \omega &= \pm \frac{I \, H + (V^A_{1}-V^S_{1})V^A_{2}}{2 \left(V^A_{1} - G\right)^{1/2} (V^A_{1} - V^S_{1})}\,.
\end{align}
One may worry that the above Fermi estimate for $\delta \omega$ may diverge because the denominator of Eq.~\eqref{eq:delta-w-Fermi} may vanish, but this does not occur anywhere outside the horizon.
If we evaluate these expressions at ${\cal{R}} = 3M$, which is close to where the effective potentials $V^A_{\rm eff}$ and $V^S_{\rm eff}$ are extremized, we find for the $\ell=2$ mode at $a=0$ that $M \omega_{\GR} = 1/3$~($\approx 0.333$) and $M \delta \omega = -2/81$~($\approx -0.025$), both of which are close to the real part of the correct numerical answers we will find later. The precise numerical factors, however, do not matter here. What matters is that the above Fermi estimate shows explicitly that the dCS corrections to the QNM frequencies will be of ${\cal{O}}(\alpha^2)$. These corrections for the axial gravitational sector come from the second order correction to the effective potential as well as linear order coupling with scalar field. 

Equation~\eqref{eq:fermimain} also suggests that the QNM frequencies are even in the CS coupling. We have verified this with our numerical calculations shown in Sec.~\ref{sec:results} by taking $ \alpha \to -\alpha$ and found no change in the QNM frequency thereby supporting the results from the Fermi estimate.

\section{Calculation of the Quasinormal Modes} 
\label{sec:int}

The late-time GW signal from a perturbed BH is generally
dominated by a sum of exponentially damped sinusoids known as the QNMs. These
correspond to the characteristic vibrational modes of the
spacetime~\cite{Press:1971wr} and are complex valued.
The real part represents the temporal oscillations whereas the imaginary
part represents an exponentially decaying temporal part of the oscillations.
Using the slowly-rotating approximation for finding QNMs of BHs allows us to use well-established 
numerical methods for their calculation.
In this section, we show the boundary conditions and the numerical integration technique that will be used to calculate the QNMs. 

\subsection{Boundary conditions} \label{sec:bdd}

The QNMs are solutions of the inhomogeneous wave equations [Eqs.~\eqref{eq:SF} to
\eqref{eq:ZM}] with appropriate boundary conditions.
For the case of slowly-rotating BHs, we have two boundaries: one at spatial
infinity and the other at the horizon $r_\h$.
The horizon of the BH described by the metric
of Eq.~\eqref{eq:ds2} coincides with the Schwarzschild horizon (they are both located at $r=2M$ in Schwarzschild-like coordinates) at
leading order in spin.

A QNM represents waves which are of purely ingoing at the
horizon $r_\h$ and purely outgoing at spatial infinity, which are 
characterized by complex frequency $\omega$ with ${\rm Re}(\omega)\geqslant 0$.
These boundary conditions are,
\begin{align} \label{eq:bdd1}
\Psi_j \propto \bigg\{ ~
\begin{matrix}
e^{-i \omega_\h r_*}\,, &  r \to r_\h \,, \\
e^{i \omega r_*} \,, & r \to \infty \,, \\
\end{matrix}
\end{align}
where recall that $\Psi_j=\{R_{\ell m},\Psi^{\RW}_{\ell m},\Psi^{\ZM}_{\ell m} \}$, $r_*$ is the tortoise coordinate and
$\omega_\h = \omega - m \Omega_\h$, with $\Omega_\h$, given by Eq.~\eqref{eq:ZAMO}, is the horizon angular frequency for the BH under consideration.

\subsection{Evaluation of the QNMs: direct integration} 
\label{sec:intmethod}
To compute the QNMs, we use the direct integration method.
In this method, we integrate the equations twice, once from finite distance outside the horizon towards spatial infinity, and once from a finite distance far from the horizon but inwards towards the event horizon. 
The integrations are started using the boundary conditions presented above, with a given choice of $\omega$. We then compare the two numerical solutions at a matching point $r_m$ that is somewhere between the horizon and spatial infinity, to check whether the master functions and their radial derivatives are continuous at $r_m$. Typically, this is not the case, so we then iterate this process over various values of $\omega$ until one finds a choice of the complex frequency that leads to  continuous and differentiable solutions. In practice, this can be done by finding the value of $\omega$ for which the Wronskian $W$ of the two solutions vanishes at $r_m$~\cite{Maggiore:2018sht}.

However, as simple as this approach appears, it is numerically difficult to implement due to inherent numerical instabilities. If one chooses the wrong value for $\omega$ then the trial integrations will tend to diverge as one approaches spatial infinity (for the outward solution) or as one approaches the BH horizon (for the inward solution).  Also, since numerically we cannot start the integrations exactly at the horizon or at spatial infinity, the boundary conditions are not sufficiently accurate in general. 

To improve numerical stability, one can improve the boundary conditions by finding analytic asymptotic solutions to the perturbation equations about spatial infinity and the horizon. More specifically, 
we can write~\cite{Chandra:10.2307/78902},
\begin{equation} \label{eq:bdd2}
\Psi_j \propto \Bigg\{ ~
\begin{matrix}
 e^{-i \omega_\h r_*}  \sum_{k=0}^{\infty} (\gamma_k)_j (r-r_\h)^{k}  \,, &  r \to r_\h \,, \\ \\
 e^{i \omega r_*}  \sum_{k=0}^{\infty} (\eta_k)_j r^{-k}  \,, & r \to \infty \,, \\
\end{matrix}
\end{equation}
where the coefficients $\gamma_k$ and $\eta_k$ can be determined
order by order in a series expansion of the perturbation
equations around spatial infinity or the horizon. 
These coefficients can all be written entirely in terms of $(\gamma_0)_j$ and $(\eta_0)_j$, but they are long and not illuminating, so we will not present them in the paper, and instead they will be made available through a Mathematica notebook upon request. We will use these boundary conditions for our numerical integrations. 

We will here compute the QNMs adapting the procedure in~\cite{Blazquez-Salcedo:2016enn} to our case.
We begin by constructing two square
matrices $W_{\rm o}$ and $W_{\rm e}$, which is four dimensional for the axial case due to the coupling
between $\Psi^{\RW}_{\ell m}$ and $R_{\ell m}$, and two dimensional for the polar case due to the lack of coupling.
The columns of $W_{\rm o,e}$ are independent solutions of the perturbation equations, so 
\begin{align} \label{eq:W}
	W_{\rm o} &=
\begin{pmatrix}
{}_{\h}\Psi^{\RW\, (1)}_{\ell m} &  {}_{\I}\Psi^{\RW\, (1)}_{\ell m} &  {}_{\h}\Psi^{\RW\, (2)}_{\ell m} &  {}_{\I}\Psi^{\RW\,(2)}_{\ell m}  \\[0.5em]
 \partial_r {}_{\h}\Psi^{\RW\, (1)}_{\ell m} & \partial_r {}_{\I}\Psi^{\RW\, (1)}_{\ell m} & \partial_r {}_{\h}\Psi^{\RW\, (2)}_{\ell m} & \partial_r {}_{\I}\Psi^{\RW\, (2)}_{\ell m} \\[0.5em]
 {}_{\h}R_{\ell m}^{(1)} & {}_{\I}R_{\ell m}^{(1)} & {}_{\h}R_{\ell m}^{(2)} & {}_{\I}R_{\ell m}^{(2)} \\[0.5em]
 \partial_r {}_{\h}R_{\ell m}^{(1)} & \partial_r {}_{\I}R_{\ell m}^{(1)} & \partial_r {}_{\h}R_{\ell m}^{(2)} & \partial_r {}_{\I}R_{\ell m}^{(2)} \\[0.5em]
\end{pmatrix}
, \nonumber \\ \\
	W_{\rm e} &=
\begin{pmatrix}
 {}_{\h}\Psi^{\ZM}_{\ell m} &  {}_{\I}\Psi^{\ZM}_{\ell m} \\[0.5em]
 \partial_r {}_{\h}\Psi^{\ZM}_{\ell m} & \partial_r {}_{\I}\Psi^{\ZM}_{\ell m}  \\
\end{pmatrix}\,,
\end{align}
where, the pre-subscript to the perturbation function ${}_{\h,\I}\Psi$ denote whether the solutions is obtained by integration from the horizon to $r_m$, or from spatial infinity to $r_m$, while the superscripts $\Psi^{(1),(2)}$ denote two solutions evaluated with different initial conditions at the boundaries.

In principle, any set of independent solutions will do for the calculation of these $W_{\rm o,e}$ matrices, but in this paper we make the following choices. 
For the even sector, we choose the solution to be that obtained by integrating the perturbation equations with the boundary conditions in Eq.~\eqref{eq:bdd2} and $[(\gamma_0)_{\ZM},(\eta_0)_{\ZM}] = (1,1)$.
For the odd sector, we choose the two solutions to be those obtained by integrating with the boundary conditions in Eq.~\eqref{eq:bdd2} and with $[(\gamma_0)_{\RW},(\eta_0)_{\RW},(\gamma_0)_{\R},(\eta_0)_{\R}] = (1,1,0,0)$ or with
$[(\gamma_0)_{\RW},(\eta_0)_{\RW},(\gamma_0)_{\R},(\eta_0)_{\R}] = (0,0,1,1)$.
In general, these solutions are linearly independent, unless $\omega$
is the correct QNM frequency, in which case 
\be \label{eq:wron}
{\rm det}(W)|_{r=r_m} = 0 \,.
\ee
We can use a root-finding algorithm to find the $\omega$ such that the Wronskian vanishes at the matching point. 

Using this method, we have calculated the QNMs of a slowly-rotating BH in dCS
gravity.
In practice, all numerical integrations that start at the horizon are initiated at $r_{\rm initial,r_+} = (2 + 10^{-4}) M$, while those that start at spatial infinity are initiated at $r_{\rm initial,i_0} = 60 M$, with the matching always performed at $r_m = 20 M$. 
We have checked the numerical stability of the QNM frequencies against changes
in the values of $r_{\rm initial,r_+}$, $r_{\rm initial,i_0}$ and $r_m$. All numerical integrations are done with the NDSolve package of Mathematica, with  accuracy and precision set to 10 digits. 

\begin{figure*}[t]
\centering
\includegraphics[width=\linewidth]{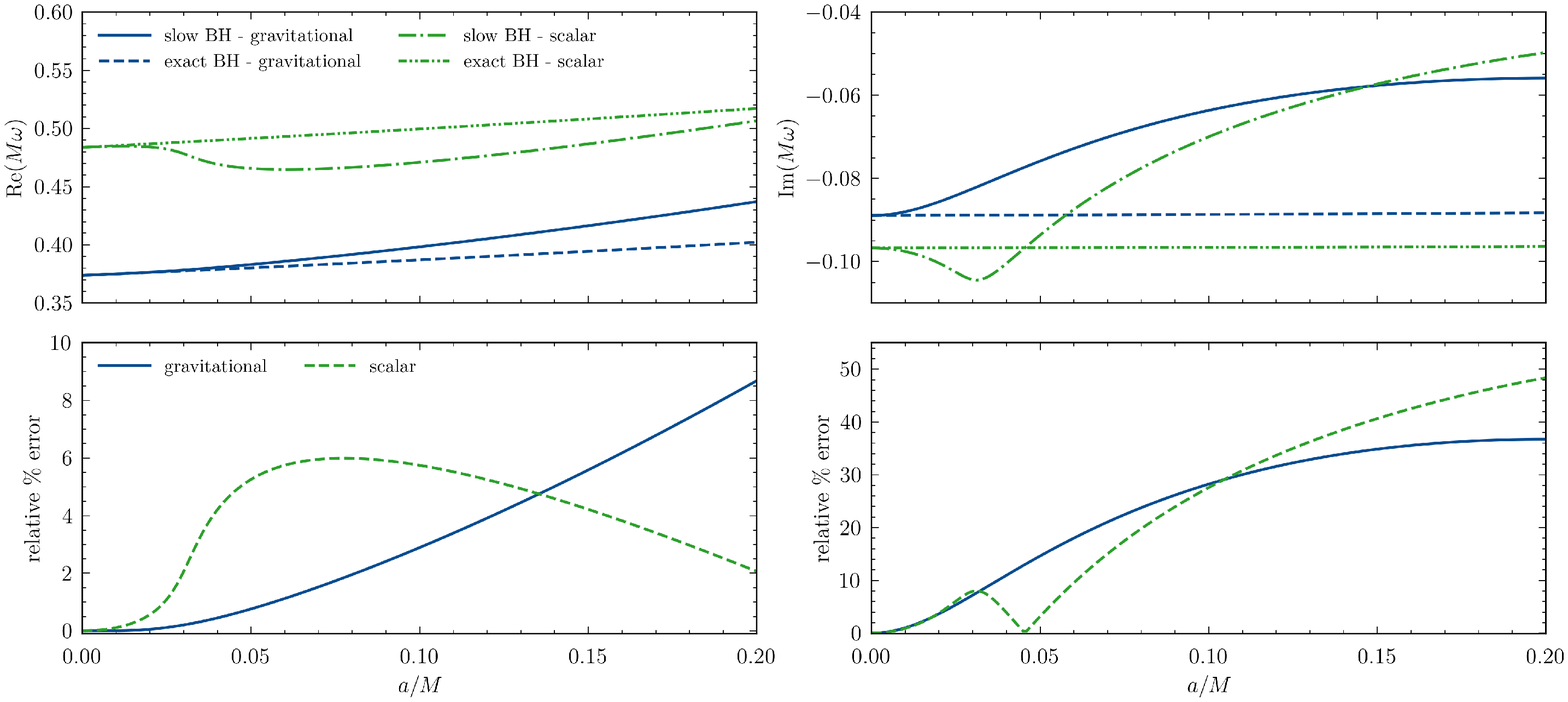}
\caption{
Comparison between the fundamental dominant $\ell=m=2$ gravitational and scalar QNMs calculated using the exact
Kerr metric and its expansion to leading-order in $a$. The dependence
of the real (imaginary) part of the frequency on the spin $a$ is shown in the left (right) top panel.
The left (right) bottom panel shows the relative error defined by Eq.~\eqref{eq:relerror}.
The colors distinguish the scalar (green) and gravitational (blue) QNMs.
Unsurprisingly, we see that the relative percent error $\delta \omega$
increases with spin $a/M$, doing so faster for the imaginary part of the
frequency, although the absolute error is $\mathcal{O}(a^2)$.
The legends shown in the left panels apply to the corresponding right panels as well.
}
\label{fig:error}
\end{figure*}

\section{Numerical Results}
\label{sec:results}

In this section, we present our numerical results for the QNM frequencies.
For clarity, we have divided this section into three parts.
First, we take the GR limit (i.e., $\alpha / M^2 = 0$) of our equation and
discuss the associated QNMs of a slowly-rotating Kerr BH with the metric expanded
to leading-order in spin.
We compare these results with the exact Kerr QNMs obtained using Leaver's
continued fraction method~\cite{Leaver:1985ax}.
This preliminary step will give us an estimate of where the slow-rotation
approximation breaks down and how the numerical errors due to this approximation
compare against the modifications to the QNM frequencies due to the CS
coupling.
Second, we consider a nonzero CS coupling $\alpha / M^2$ and study in detail how
the QNMs behave as functions of this coupling and of the BH spin.
Third, we construct fitting formulas for the real and imaginary parts of the QNMs,
valid within the errors associated to the slow-rotation approximation.

In the main body of this paper, we will show numerical results for the fundamental $n=0$, $\ell = m = 2$ frequencies since these are dominant for
tensorial perturbations. 
We leave our results for the QNM frequencies for the fundamental mode with $\ell=2$, $\ell=3$ and $\ell=4$ modes, and all $m$ modes to Appendix~\ref{appendix:qnm}. 
To aid in the presentation of these numerical results, we will work with dimensionless parameters by rescaling $a$, $\alpha$ and $\omega$ as
\be
a \to a/M, \quad \alpha \to \alpha/M^2, \quad \textrm{and} \quad \omega \to \omega M\,.
\ee
In our numerical calculations, we work in code units, in which $M=1$, thus making 
the code quantities $a$, $\alpha$ and $\omega$ dimensionless.

\subsection{Slow rotation: GR} \label{sec:slowgr}

The calculation of the QNMs of a slow-rotating Kerr BH in GR were presented in~\cite{Pani:2013pma}, but to our knowledge a comparison
between these results and those obtained using the exact Kerr metric has not appeared in the literature. 
See~\cite{Maselli:2019mjd,Hatsuda:2020egs} for complementary studies.
This comparison is important for our purposes for two reasons.
First, it will tell us up to what values of $a$ we can trust our
slow-rotation approximation.
Second, it will tell us whether the errors due to this approximation are
degenerate with modifications to the GR QNMs introduced by the CS coupling. 
We refer the interested reader to~\cite{Ayzenberg:2016ynm} for a similar analysis, but in a different context.

In the GR limit ($\alpha / M^2 = 0$) the perturbation
equations [cf.~Eqs.~\eqref{eq:sfpeq},~\eqref{eq:axpeq} and~\eqref{eq:popeq}] decouple and
each of them reduces to equations of the form
\be
\label{eq:pert_gr_limit}
f(r)^2 y^{\pp}_{\ell m} + (2 / r^2) f(r) y^{\p}_{\ell m}
+ [\omega^2 + v_{\rm eff}(r,a)] y_{\ell m} = 0\,,
\ee
for the field variables and effective potential pairs
$\{ y_{\ell m}, \, v_{\rm eff} \} = \{ R_{\ell m}, \, V^S_{\rm eff} \}$,
$\{ \Psi^{\RW}_{\ell m}, \, V^A_{\rm eff}\}$
and
$\{ \Psi^{\ZM}_{\ell m}, \, V^P_{\rm eff}\}$.
For the gravitational perturbations these equations agree with those 
in~\cite{Pani:2013pma}, whereas for the scalar perturbation they agree with the
small-$a$ limit of that in~\cite{Brill:1972xj}. Moreover, in this limit,
the axial and the polar gravitational equation become isospectral.

As a benchmark for our numerical code, we calculated the gravitational and
scalar QNMs of a Schwarzschild BH by taking the nonrotating limit ($a = 0$).
We find excellent agreement with the well-known result summarized e.g.~in~\cite{Berti:2009kk,BertiRingdown}.
Next, we computed the QNMs associated with Eq.~\eqref{eq:pert_gr_limit} and
compared them against the QNMs obtained using the exact Kerr metric (i.e.,
without performing any small-$a$ expansion) by means of the continuous fraction
method~\cite{Leaver:1985ax} and tabulated in~\cite{Berti:2009kk,BertiRingdown}

Figure~\ref{fig:error} presents the results of this exercise, focusing on the fundamental mode with $\ell = m = 2$.
The top panels show a comparison between the behavior
of the real (left) and the imaginary (right) parts of the QNM frequencies for
the slowly-rotating Kerr metric (solid) and the exact Kerr metric (dashed) as a
function of the spin parameter $a$ for both the scalar (green curves) and the
gravitational modes (blue curves).
The bottom panels show the relative percent error due to the slow rotation
approximation, which we define via
\begin{subequations}\label{eq:relerror}
\begin{align}
\delta (\rm{Im} \, \omega) &= \left| 1 - (\rm{Im}\,\omega_{\sr}) / (\rm{Im}\,\omega_{\K}) \right| \times 100 \,, \\
\delta (\rm{Re} \,\omega) &= \left| 1 - (\rm{Re}\,\omega_{\sr}) / (\rm{Re}\,\omega_{\K}) \right| \times 100 \,, 
\end{align}
\end{subequations}
where, $\omega_{\K}$ are the QNM frequencies calculated from the exact Kerr
metric~\cite{Leaver:1985ax,Berti:2009kk}, whereas $\omega_{\sr}$ are the QNM
frequencies we calculated for the Kerr metric expanded to linear order in
$a/M$.

\begin{figure*}[htb]
	\includegraphics[width=\linewidth]{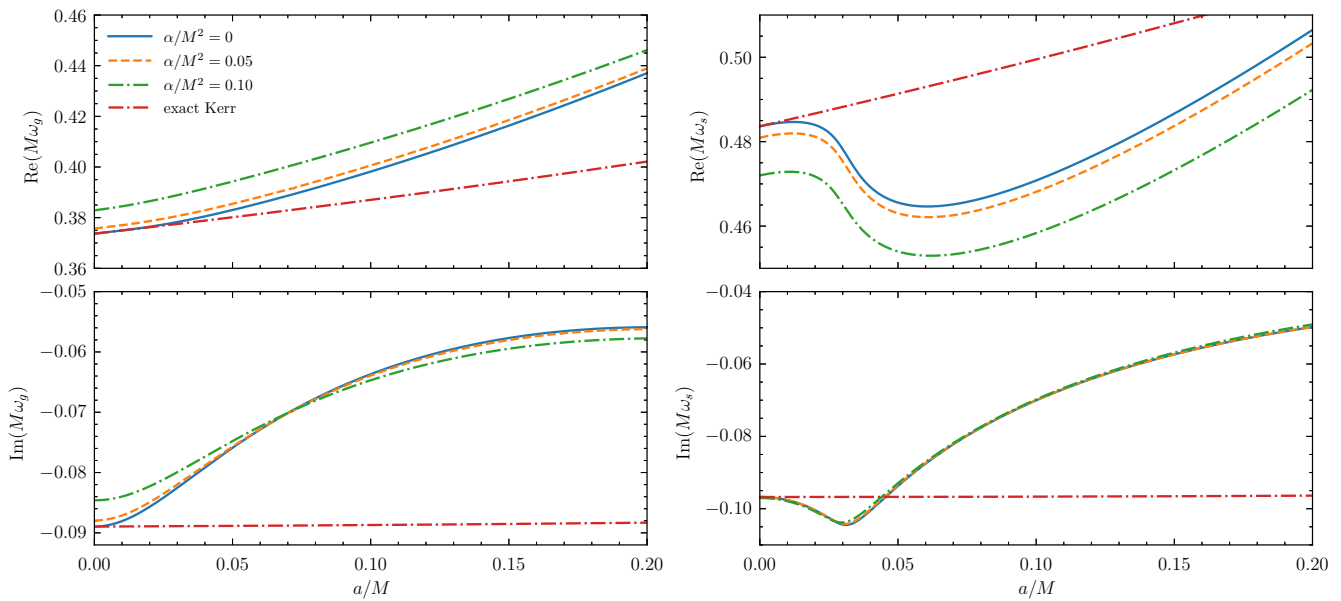}
    \caption{
    Real (Imaginary) parts of the QNM frequencies for the $n=0$, $\ell=m=2$ mode as a function of the spin parameter for slowly rotating BHs in dCS gravity with different CS couplings $\alpha/M^2$ are shown in the top (bottom) panels for the axial gravitation-led (left) and scalar-led (right) modes. Additionally, we have also shown the exact Kerr QNMs for BHs in GR for quick comparison calculated in~\cite{Berti:2009kk}.
    The individual QNM frequencies increase for the gravitational-led modes
    with spin parameter $a/M$ as well as with the CS coupling $\alpha/M^2$, whereas the scalar-led modes increase with spin parameter, but decrease with the CS coupling.
    The imaginary part of $M \omega_{g}^{\rm axial}$ decreases slightly in magnitude with increasing $\alpha/M^2$ for spins $a/M \lesssim 0.06$, while it increases slightly in magnitude for spins $a/M \gtrsim 0.06$.
    Since our slow-rotation approximation is valid up to spin values of approximately at most $0.04$, overall these modes become less damped in dCS gravity. 
    The legends apply to all panels.
    }
	\label{fig:balpha}
\end{figure*}

We can extract several conclusions from this figure. First, the relative error introduced by the slow-rotation
approximation is larger for the imaginary part of the frequency than for the real part.
For instance, when $a/M = 0.1$, the relative error is approximately eight times larger
on the imaginary part than on the real part for both gravitational and
scalar-led modes.
In addition, we also find that the absolute error $|\omega_{\sr} - \omega_{\K}|$
is of $\mathcal{O}(a^2)$ for both the real and imaginary parts of the QNM frequencies.
This naturally follows from the fact that the error should indeed be of $\mathcal{O}(a^2)$, since we have
evaluated $\omega$ to leading order in the spin parameter.
The results show that the approximation introduces errors smaller than $10\%$ when $a/M \lesssim 0.2$ for the real part and $a/M \lesssim 0.04$ for the imaginary part.
Here, we have not included the errors due to our numerical integration scheme since the total error is dominated by the error introduced in the slow rotation approximation.

\subsection{Slow rotation: dCS} \label{sec:slowdcs}

In this subsection, we show how the QNM frequencies discussed in the previous subsection are modified by the presence of a nonzero CS coupling $\alpha / M^2$ for the dominant $n=0$, $\ell=m=2$ mode. Tabulated values for the QNM frequencies for the fundamental mode with $\ell =2,~ \ell=3,~$ and $\ell=4$ and all $m$ can be found in Appendix~\ref{appendix:qnm}.  

As we discussed in Sec.~\ref{sec:analysisperteq}, our calculations are valid to linear order
in spin $a/M$, but second-order in the CS coupling $\alpha / M^2$.
Recall that the polar QNM frequencies [governed by Eq.~\eqref{eq:ZM}] are independent of the coupling with the scalar field or with the axial parity sector
due to the dependence of the effective potential on the CS coupling \footnote{We thank Pablo~A.~Cano and Thomas~Hertog for bringing this to our attention.}~\cite{Cano:2020cao}.
We therefore focus on the coupled system of Eqs.~\eqref{eq:RW} and \eqref{eq:SF} and the Eq.~\eqref{eq:ZM} separately and use the numerical procedure described in Sec.~\ref{sec:int} to calculate
the QNMs in different parity sectors. 

\subsubsection{Axial gravitational and scalar sector}

\begin{figure*}[htb]
	\includegraphics[width=\linewidth]{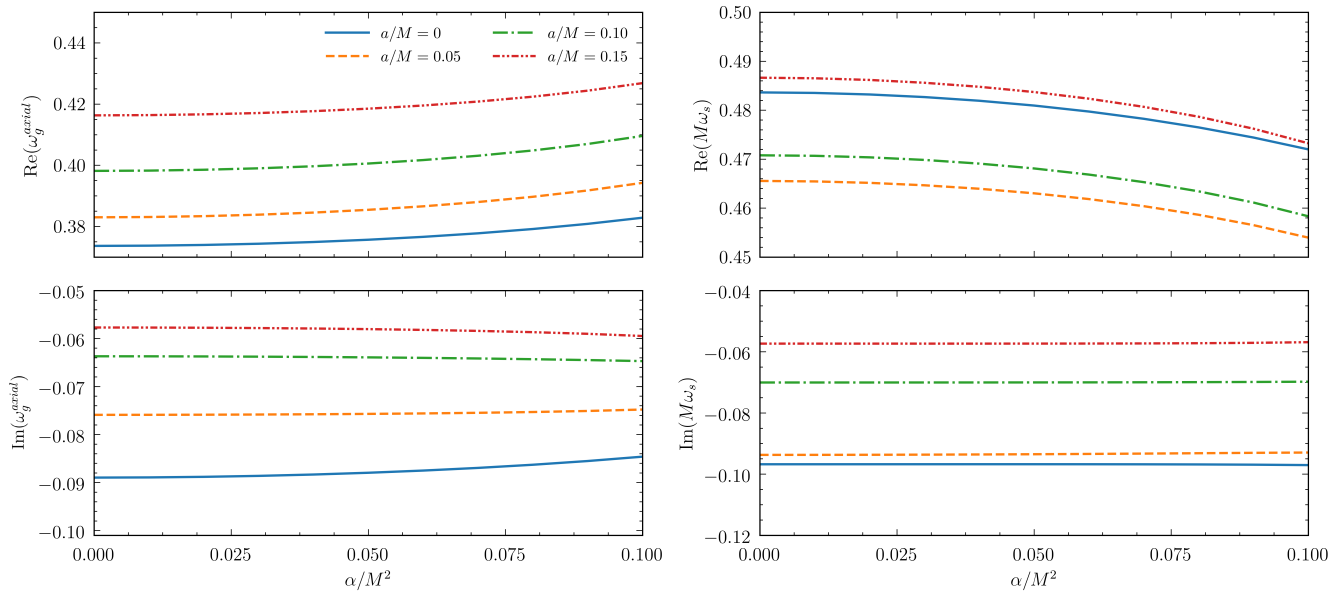}
    \caption{
    Real (Imaginary) parts of 
    of the QNM frequencies as a function of the CS coupling for
    slowly rotating BH in dCS gravity for different values of the spin parameter
    $a$ are shown in the top (bottom) panels 
    for the gravitational-led (left) and scalar-led (right) modes. 
    We see that increasing the spin $a/M$ shifts the curves 
    upward in all panels.
    In addition, we see that the real part of $M \omega_{g}^{\rm axial}$ increases as we increase the CS coupling, whereas it decreases for the real part of $M \omega_{s}$.
    This is behavior happens for all constant-spin curves.
    For higher spins, we see that the changes to the decay rates are negligible as the CS coupling is increased for both sectors. 
    This behavior may be a consequence of the slow rotation approximation we use.
    Finally, for a fixed value of spin, the curves exhibit a quadratic order dependence on the CS coupling supporting our estimate in Sec.~\ref{sec:Fermi}.
    %
    %
    The legends apply to all panels.
    }
	\label{fig:bspin}
\end{figure*}

Our results for the fundamental modes are summarized in Figs.~\ref{fig:balpha} and~\ref{fig:bspin}.
In Fig.~\ref{fig:balpha} we show the dependence of the real and imaginary values
of gravitational- and scalar-led QNMs as function of spin parameter $a/M$ for
$\alpha / M^2 = \{0, 0.05, 0.10\}$. 
In Fig.~\ref{fig:bspin} we complement this analysis, by showing the dependence of the QNMs
on $\alpha/M^2$ for $a / M = \{0, 0.05, 0.10, 0.15\}$.
These two figures can be thought of as showing the slow-rotation corrections to
the non-rotating BH QNM frequencies in dCS and the dCS modifications to the GR
QNM frequencies respectively.

As seen from Figs.~\ref{fig:balpha} and~\ref{fig:bspin}, the behaviors of the real and the imaginary part
for gravitational-led and the scalar-led sectors are distinct. 
Thus, we present individual analysis for each.
\begin{itemize}
\item \textbf{Real gravitational-led QNM} (top left panel of Figs.~\ref{fig:balpha} and \ref{fig:bspin}).
The $\textrm{Re}(\omega_{g}^{\rm axial})$ increases with BH spin just as in GR (see Sec.~\ref{sec:slowgr}) for constant CS coupling, and it increases as the CS coupling increases for constant spin. 
\item \textbf{Imaginary gravitational-led QNM} (bottom left panel of Figs.~\ref{fig:balpha} and \ref{fig:bspin}).
The $\textrm{Im}(\omega_{g}^{\rm axial})$ decreases with BH spin for a constant CS coupling just like in GR, whereas it remains constant with increasing CS coupling for a constant spin.
\item \textbf{Real scalar-led QNM} (top right panel of Figs.~\ref{fig:balpha} and \ref{fig:bspin}).
The $\textrm{Re}(\omega_{s})$ initially shows a sinusoidal behavior as we increase spin and hold the CS coupling constant up to $a=0.05$, and then it increases with BH spin.
On the other hand, the $\textrm{Re}(\omega_{s})$ decreases monotonically with the CS coupling while keeping the BH spin constant. 
\item \textbf{Imaginary scalar-led QNM} (bottom right panel of Figs.~\ref{fig:balpha} and \ref{fig:bspin}).
The $\textrm{Im}(\omega_{s})$ initially increases with BH spin for constant CS coupling, reaches a minimum and then decreases again.
On the other hand, the $\textrm{Im}(\omega_{s})$ remains essentially constant as we change the CS coupling for constant spin. 
\end{itemize}
Figures~\ref{fig:balpha} and~\ref{fig:bspin} shows that the real part of QNM frequencies has a similar functional behavior with spin, for different values of the CS coupling. These frequencies depend quadratically on the CS coupling, as required by the order reduction scheme employed in this paper (see e.g.~Ref.~\cite{Yunes:2009hc}), which is also displayed in the figures.

The modes presented above show a non-monotonic behavior with respect to the spin parameter in contrast with the equations being linear in this parameter. This is mainly because the equations~\eqref{eqs:perturbed_final} are nonlinear in the QNM frequency $\omega$ (this can be seen explicitly from the occurrences of $\omega$ in the various functions these equations). 
Since we are not linearizing $\omega$ with respect to the spin parameter to avoid loss of information from the equations, we obtain a nonlinear behavior for the QNM frequencies of the gravitational-led modes and, more evidently, for the scalar-led modes.

These results have interesting implications for GW observations. GW detectors will be sensitive directly only to the gravitational wave modes, not the scalar modes. Focusing on the real part of $\omega_g^{\rm axial}$, we then see that there is a degeneracy through a positive correlation between $\alpha$ and $a$; increasing $a$ while keeping $\alpha$ constant has the same effect as increasing $\alpha$ while keeping $a$ constant. However, this correlation breaks when considering the imaginary part of $\omega_g^{\rm axial}$; increasing $a$ leads to longer lived modes (smaller imaginary frequencies) for constant $\alpha$, while increasing $\alpha$ does not change much the lifetime for constant $a$. This then suggests that, in principle it may be possible to separate the effects of spin and CS coupling parameter given a ringdown observation that is loud enough.

\subsubsection{Polar gravitational sector}
\begin{figure*}[htb]
	\includegraphics[width=\linewidth]{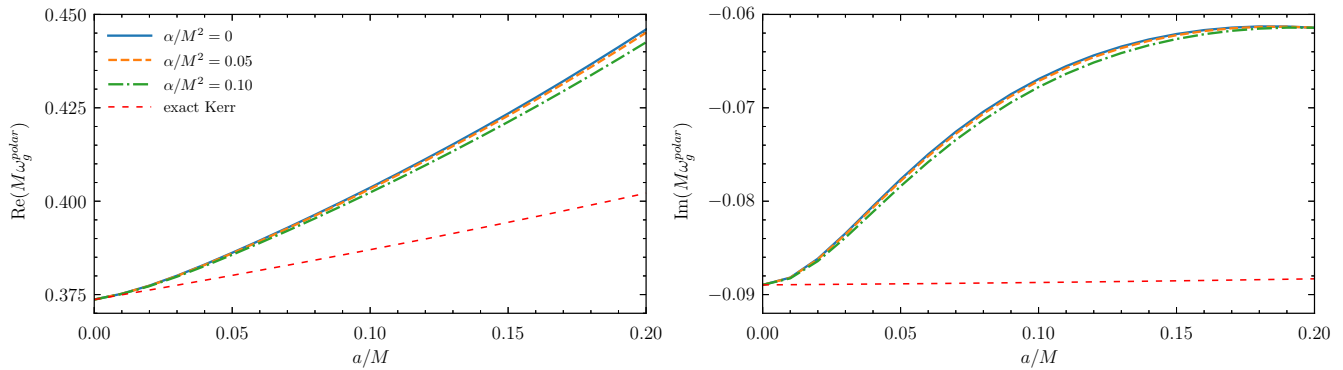}
    \caption{
    Real (left panel) and imaginary parts (right panel) of the QNM frequencies for the $n=0$, $\ell=m=2$ polar gravitational-led mode as functions of the spin parameter for slowly rotating BHs in dCS gravity with different CS couplings $\alpha/M^2$. We have also shown the exact Kerr QNMs for BHs in GR for comparison taken from~\cite{Berti:2009kk}.
    The individual QNM frequencies increase for the polar gravitational-led modes
    with spin parameter $a/M$, but decrease slightly with the CS coupling $\alpha/M^2$.
    The imaginary part is also almost constant in magnitude with increasing $\alpha$, suggesting that damping of modes is almost independent of the CS coupling. 
    The legends apply to all panels.}
	\label{fig:polar}
\end{figure*}

The results for the fundamental modes for the polar gravitational sector is summarized in Fig.~\ref{fig:polar}.
We show the dependence of the real and imaginary values of polar gravitational QNMs as a function of spin parameter $a/M$ for $\alpha / M^2 = \{0, 0.05, 0.10\}$. 
Tabulated values for the QNM frequencies for the fundamental mode with $\ell= 2$ and $\ell= 3$ and $\ell=4$ 
for all $|m| \leqslant \ell$ can be found in Appendix~\ref{appendix:qnm}.

As seen from Fig.~\ref{fig:polar}, the behavior of the real and the imaginary part
for polar gravitational sector is distinct. Therefore, we present individual analysis for each.
\begin{itemize}
\item \textbf{Real gravitational-led QNM} (left panel of Fig.~\ref{fig:polar}).
The $\textrm{Re}(\omega_{g}^{\rm polar})$ increases with BH spin just as in GR (see Sec.~\ref{sec:slowgr}) for constant CS coupling,
while it remains almost constant as the CS coupling increases for constant spin.
\item \textbf{Imaginary gravitational-led QNM} (right panel of Fig.~\ref{fig:polar}).
The $\textrm{Im}(\omega_{g}^{\rm polar})$ decreases with BH spin for a constant CS coupling just like in GR, whereas it remains almost constant with increase in CS coupling for a constant spin.
\end{itemize}

The modes decay for all values of spin parameter and CS coupling within the limits of our approximation, giving strong evidence that slowly-rotating BHs in dCS gravity are linearly stable against gravito-scalar perturbations, extending the results of~\cite{Molina:2010fb,Kimura:2018nxk} to small spins. 

\subsubsection{Regime of validity of numerical results and implications}

In interpreting the figures, especially Fig.~\ref{fig:balpha}, one must be careful to take into account the approximate nature of our results. The numerical calculations we have performed are only valid in the slow-rotation approximation (to leading order in the BH spin and second order in the CS coupling).
This is why we included a discussion in Sec.~\ref{sec:slowgr}, which quantify approximately the error in the slow-rotation approximation. 
When the spin is large enough that this error becomes comparable to the dCS correction, then our numerical results should not be trusted any longer. This occurs roughly at $a/M \gtrsim 0.2$ for $\textrm{Re}(\omega_{g,s})$, $a/M \gtrsim 0.05$ for $\textrm{Im}(\omega_{g})$ and $a/M \gtrsim 0.07$ for $\textrm{Im}(\omega_{s})$, for both axial or polar modes due to their isospectrality, where we have a maximum relative error of around $10\%$ due to the slow rotation approximation as shown in Fig.~\ref{fig:error}. 
Therefore, to be conservative, henceforth we restrict our attention to the regime $a/M \leqslant 0.0375$, where the errors introduced by the slow-rotation approximation are very small and the dCS corrections we have calculated are meaningful. 

Our results allow us to estimate the magnitude of the dCS deviations with respect to the GR 
QNMs. As an example, let us saturate the best constraint to date on the CS coupling parameter,
$\alpha^{1/2} \leqslant 8.5$~km (at 90\% confidence), obtained in~\cite{Silva:2020acr}, and 
consider the smallest remnant BH mass observed so far ($M \approx 18 \, M_{\odot}$), the product of the event GW170608~\cite{Abbott:2017gyy}. 
Combining this value of $M$ with the maximum value of $\alpha$ allowed from~\cite{Silva:2020acr}, we 
find $\alpha/M^2 \approx  0.1$. 
We can now use the data presented in Tables~\ref{tab:grv_l2_all_m}~and~\ref{tab:sca_l2_all_m}
(see also Fig.~\ref{fig:balpha}) to find that the 
maximum deviations from the GR QNMs is about $2\%$ ($2\%$) for the oscillation frequency
and about $9\%$ ($6\%$) for the decay rate for the fundamental dominant $\ell = m = 2$
axial gravitational- (scalar-) led modes and a BH with spin $a/M=0.0375$ 
(i.e., at the upper limit of our slow-rotation approximation).

Let us re-iterate that a GW detector responds only to gravitational degrees of freedom that propagate from the source to the detector (taken to be at spatial infinity), which in dCS gravity was shown to consist of the standard ``plus'' and ``cross''
transverse-traceless polarization modes, just as in GR~\cite{Wagle:2019mdq}.
This means that in practice only deviations to the gravitational-led modes
can be used to test dCS gravity, since these are the modes that affect the detector.
To constrain the modest deviations predicted here one would require a very high 
signal-to-noise ratio event, in addition to detection and characterization of at least two ringdown modes
(to break degeneracies between mass, spin and dCS coupling).
One should bear in mind however, that the small deviations found here may be a 
consequence of the small-spin approximation. 
For BHs in dCS gravity, the larger the spin, the larger the deformations away from the 
Kerr metric become, which in turn may reflect on larger deviations in the QNM frequencies.
This implies that tests of dCS gravity through BH spectroscopy would benefit from
an extension of our work to larger spin values.

\subsection{Fitting formulas for the QNMs}
\label{sec:fits}
We have calculated a large catalog of QNM frequencies from which we can construct ready-to-use fitting formulas for the fundamental mode with $\ell = m =2$, as well as other multipoles. 
This catalog allows for the fast evaluation of QNM frequencies without having to redo the numerical calculations, which are computationally non-trivial, requiring approximately $20$ hours of CPU time per curve in the figures above.
We first focus our attention on the axial gravitational- and scalar-led modes. For a given value of $\alpha$, and a given value of $n$, $\ell$ and $m$, let us use the fitting functions
\begin{subequations}	\label{eq:all_fits}
	\begin{align}
	\label{eq:fitg}	M \omega_{g}(M,a) &= a_g + b_g\left(1-a/M\right)^{c_g} \,, \\
	\label{eq:fits}	M \omega_{s}(M,a) &= a_s + b_s\left(1-a/M\right)^{c_s} \,,
	\end{align}
\end{subequations}
where $(a_g, b_g, c_g)$ and $(a_s, b_s, c_s)$ are fitting coefficients for the axial gravitational-led and the scalar-led sectors of the QNMs, respectively. 
Table~\ref{table:fit1} presents the numerical values of these coefficients for the fundamental dominant $n=0$, $\ell=2$, $m=2$ mode, and selected values of the CS coupling $\alpha/M^2$.

The fitting coefficients show a quadratic dependence on $\alpha$, in agreement with the arguments presented in Sec.~\ref{sec:Fermi}.
Thus, we can fit $(a_g, b_g, c_g)$ and $(a_s, b_s, c_s)$ as functions of 
$\zeta = \alpha^2 / (M^4 \kappa)$, 
recasting the fitting functions in Eq.~\eqref{eq:all_fits} to their final form, given by
\begin{subequations}\label{eq:fullfit}
\begin{align}
	\label{eq:fullfitgrav}	
	 M\omega_{g}^{\rm axial}(M,a,\alpha) &= f_1 + f_2 \kappa\zeta + \left( f_3 + f_4 \kappa \zeta \right)
	 \nonumber \\
	 &\quad \times \left(1- a/M \right)^{f_5 + f_6 \kappa \zeta} \,, \\
	 \label{eq:fullfitsca}
	 M\omega_{s}(M,a,\alpha) &=  g_1 + g_2 \kappa \zeta + \left( g_3 + g_4 \kappa \zeta \right)
	 \nonumber \\
	 &\quad \times \left(1- a/M \right)^{g_5 + g_6 \kappa \zeta} \,,
\end{align} 
\end{subequations}
These fitting coefficients vary for different values of $n$, $\ell$, and $m$. 
For the fundamental dominant mode. i.e., $n=0$ and $\ell = m = 2$, the fitting coefficients 
are presented in Table~\ref{table:fitsl2m2}. 
These functional forms are only chosen for simplicity, and to stay in line with the form for rotating BHs in GR given in~\cite{Berti:2005ys}.

\begin{table*}[htb]
	\begin{center}
		\begin{tabular}{c @{\hskip 0.1in} c @{\hskip 0.1in} c @{\hskip 0.1in} c @{\hskip 0.1in} c @{\hskip 0.1in} c @{\hskip 0.1in} c}
			\hline 
			\hline
			$\alpha/M^2$& $a^r_g$ & $b^r_g$ & $c^r_g$& $a^i_g$ & $b^i_g$ & $c^i_g$ \\ \hline
			$0.00$	& $ 0.7081$ & $-0.3350 $ & $ 0.5502 $ & $ 0.5854 $ & $-0.6756 $ & $ 0.3964 $ \\[0.2cm]
			$0.05$	& $ 0.8836 $ & $-0.5084 $ & $ 0.3787 $ & $ 0.5132 $ & $-0.6023 $ & $ 0.4169 $ \\ [0.2cm]
			$0.10$	& $ 0.8814 $ & $-0.4990 $ & $ 0.4536 $ & $ 0.5042 $ & $-0.5897 $ & $ 0.3399 $ \\
			\hline
            \hline
            \\[-0.2cm]
		\end{tabular}
	\end{center}
	\begin{center}
		\begin{tabular}{c @{\hskip 0.1in} c @{\hskip 0.1in} c @{\hskip 0.1in} c @{\hskip 0.1in} c @{\hskip 0.1in} c @{\hskip 0.1in} c}
			\hline 
			\hline
			$\alpha/M^2$& $a^r_s$ & $b^r_s$ & $c^r_s$& $a^i_s$ & $b^i_s$ & $c^i_s$ \\ \hline
			$0.00$	& $ 0.0309 $ & $ 0.4566 $ & $ 0.8835 $ & $ 0.5062 $ & $ -0.6053 $ & $ 0.0140 $ \\[0.2cm]
			$0.05$	& $ 0.0314 $ & $ 0.4534 $ & $ 0.8836 $ & $ 0.5033 $ & $ -0.6024 $ & $ 0.0240 $ \\ [0.2cm]
			$0.10$	& $ 0.0334 $ & $ 0.4422 $ & $ 0.9184 $ & $ 0.4528 $ & $-0.5523 $ & $ 0.0684 $ \\
			\hline
            \hline
            \\[-0.2cm]
		\end{tabular}
	\end{center}
	\caption{Fitting coefficients for Eqs.~\eqref{eq:fitg} (top) and Eq. \eqref{eq:fits} (bottom) with fixed CS coupling parameter, to model both the real (superscript $r$) and imaginary (superscript $i$) QNM frequencies describing the gravitational (top) and scalar (bottom) sectors.
	}
	\label{table:fit1}
\end{table*}

\begin{table*}[t]
    \begin{center}
    \begin{tabular}{c @{\hskip 0.2in} c @{\hskip 0.1in} c @{\hskip 0.1in} c @{\hskip 0.1in} c @{\hskip 0.1in} c @{\hskip 0.1in} c @{\hskip 0.1in} c}
    \hline \hline
	     & $f_1$ & $f_2$ & $f_3$& $f_4$ & $f_5$ & $f_6$ & \% error\\[0.1cm] \hline
         $\textrm{Re}(\omega_g)$ & $ 0.7814 $ & $ 9.4099 $ & $ -0.4092 $ & $ -8.2154 $ & $ 0.4547 $ & $ 3.5368 $ & $ 0.4 $ \\[0.2cm]
         $\textrm{Im}(\omega_g)$ & $ 0.5454 $ & $ -1.4771 $ & $ -0.6360 $ & $ 2.0641 $ & $ 0.3950 $ & $ -1.2499 $ & $ 0.7 $ \\[0.2cm] 
         \hline 
         \hline
         \\[-0.2cm]
    \end{tabular}
    \end{center}
    \begin{center}
    \begin{tabular}{c @{\hskip 0.2in} c @{\hskip 0.1in} c @{\hskip 0.1in} c @{\hskip 0.1in} c @{\hskip 0.1in} c @{\hskip 0.1in} c @{\hskip 0.1in} c}
    \hline \hline
	     & $g_1$ & $g_2$ & $g_3$& $g_4$ & $g_5$ & $g_6$ & \% error\\[0.1cm] \hline
         $\textrm{Re}(\omega_s)$ & $ 0.0326 $ & $ -1.3089 $ & $ 0.4528 $ & $ -0.1552 $ & $ 0.8750 $ & $ 5.1813 $ & $ 0.4 $ \\[0.2cm]
         $\textrm{Im}(\omega_s)$ & $ 0.4943 $ & $ -0.4121 $ & $ -0.5929 $ & $ 0.2087 $ & $ 0.0348 $ & $ -1.4985 $ & $ 0.7 $ \\[0.2cm] 
         \hline 
         \hline
    \end{tabular}
    \end{center}
    \caption{Fitting coefficients for Eqs.~\eqref{eq:fullfitgrav} (top) and~\eqref{eq:fullfitsca} (bottom) for the $n=0,\,\ell=2,\,m=2$ mode, which allow us to approximate both the real and the imaginary parts of the QNM frequencies as a function of both the spin and the CS coupling.
    }
    \label{table:fitsl2m2}
\end{table*}

The average (maximum) relative error of these fit to our data is $0.4\%$ ($0.6\%$) and $0.7\%$ ($0.9\%$) for the real and imaginary parts of the axial gravitational-led modes 
respectively, and $0.4\%$ ($0.6\%$) and $1.6\%$ ($1.86\%$) for  the $\textrm{Re}(\omega_{s})$ and $\textrm{Im}(\omega_{s})$ respectively in the regime $a/M \leqslant 0.0375$ and $\zeta \leqslant 0.01/(\kappa M^4)$.
We can fit the QNM frequencies of other $\ell=2$ multipoles and other higher multipoles with the same functional form as that presented above. The numerical values of these fitting functions and their average errors are tabulated in Appendix~\ref{appendix:qnm}.

Let us now focus on the polar modes. The functional forms for the fitting function describing the polar gravitational QNMs can be obtained using a similar treatment. We will choose a functional form similar to that in Eq.~\eqref{eq:fullfitgrav} but with $f_2 = 0 = f_4$. The fitting equation for the $\ell=m=2$ will then be
\begin{subequations}
\label{eq:polarfit}
\begin{align} 
     &M \, \textrm{Re}(\omega_{g}^{\rm polar}(M,a,\alpha)) 
     \nonumber \\
     &\quad = 0.907205\, -0.53423 (1-a/M)^{0.483296\, -5.57144 \kappa \zeta} \,, 
	 \\
	 &M \, \textrm{Im}(\omega_{g}^{\rm polar}(M,a,\alpha)) 
     \nonumber \\
     &\quad = 0.511681\, -0.601668 (1-a/M)^{0.382411\, -2.17102 \kappa \zeta} \,,
\end{align}
\end{subequations}
In the limit $a \to 0$, these fitting functions also recover the polar QNM frequencies for non-rotating BH in dCS~\cite{Molina:2010fb,Cardoso:2009pk} with a $0.1\%$ error in both the real and imaginary parts. The average (maximum) relative error of these fits to our data is $0.12\% \, (0.28\%) $ and $0.67\% \, (1.15\%)$ for the real and imaginary parts respectively.

\begin{figure*}[t]
    \centering
     \subfloat[$\textrm{Re}(\omega_g^\textrm{axial})$]{
     \includegraphics[width=0.3\textwidth]{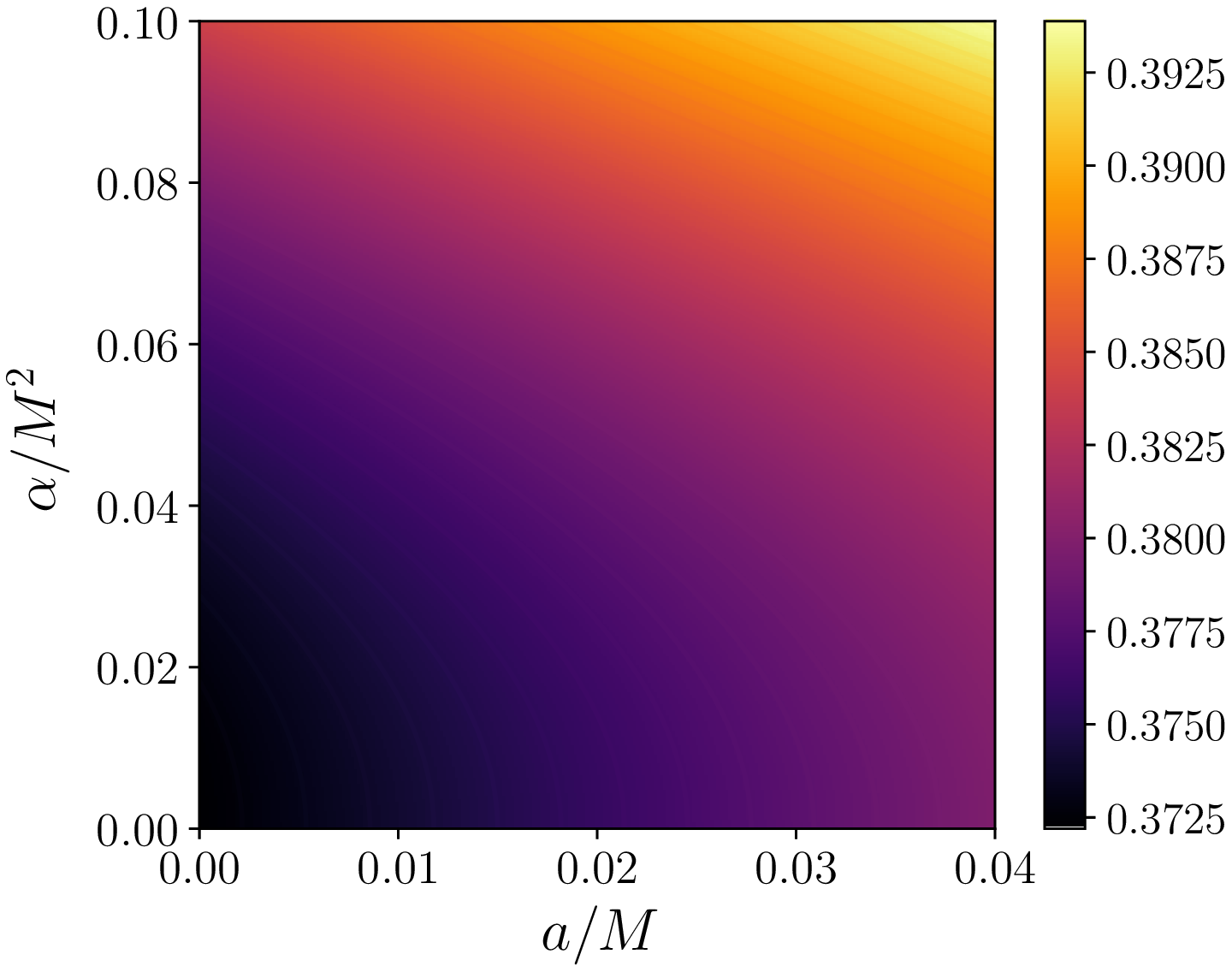}}
     \subfloat[$\textrm{Re}(\omega_s)$]{
     \includegraphics[width=0.3\textwidth]{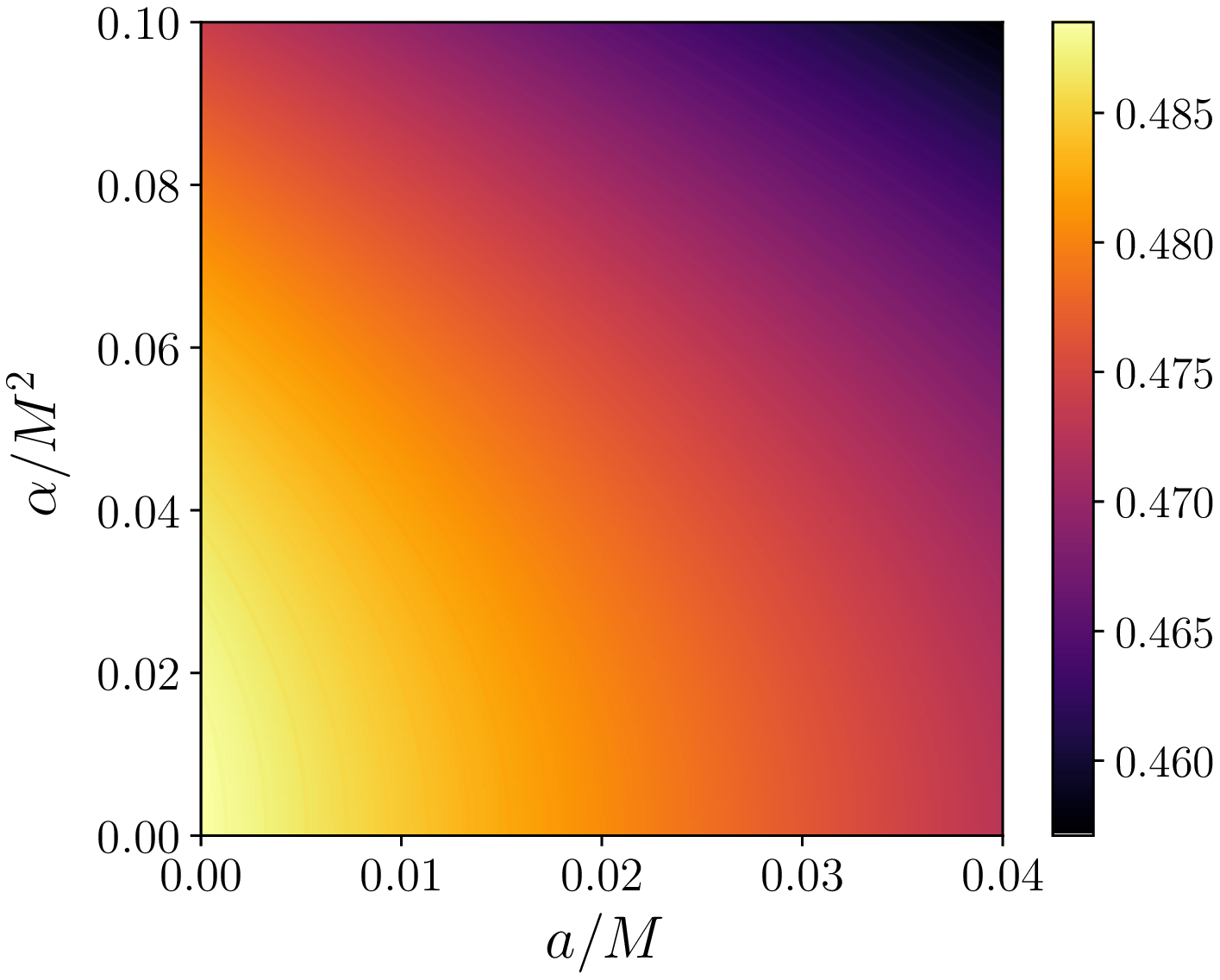}}
     \subfloat[$\textrm{Re}(\omega_g^\textrm{polar})$]{
     \includegraphics[width=0.3\textwidth]{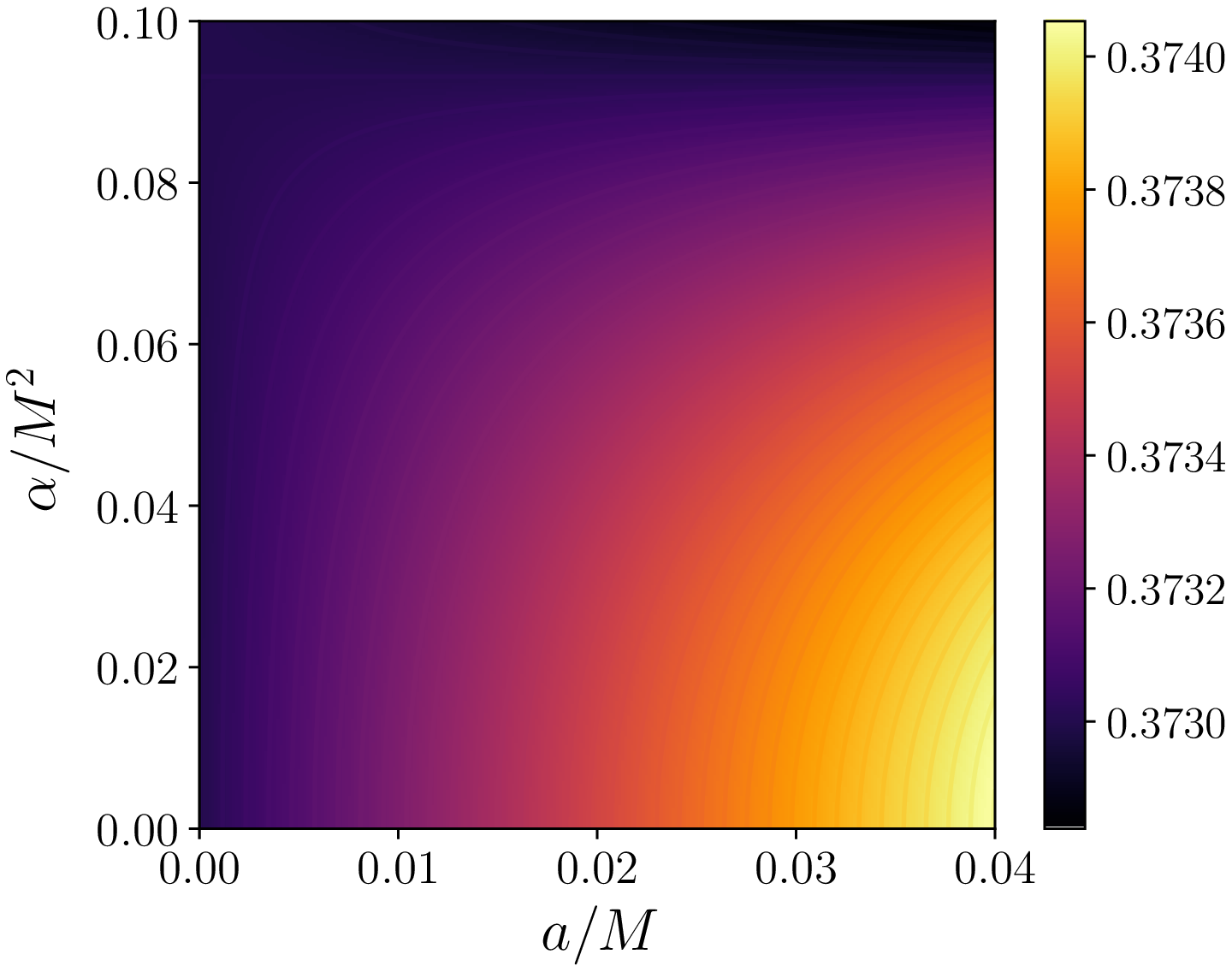}}
     \qquad
     \subfloat[$\textrm{Im}(\omega_g^\textrm{axial})$]{
     \includegraphics[width=0.3\textwidth]{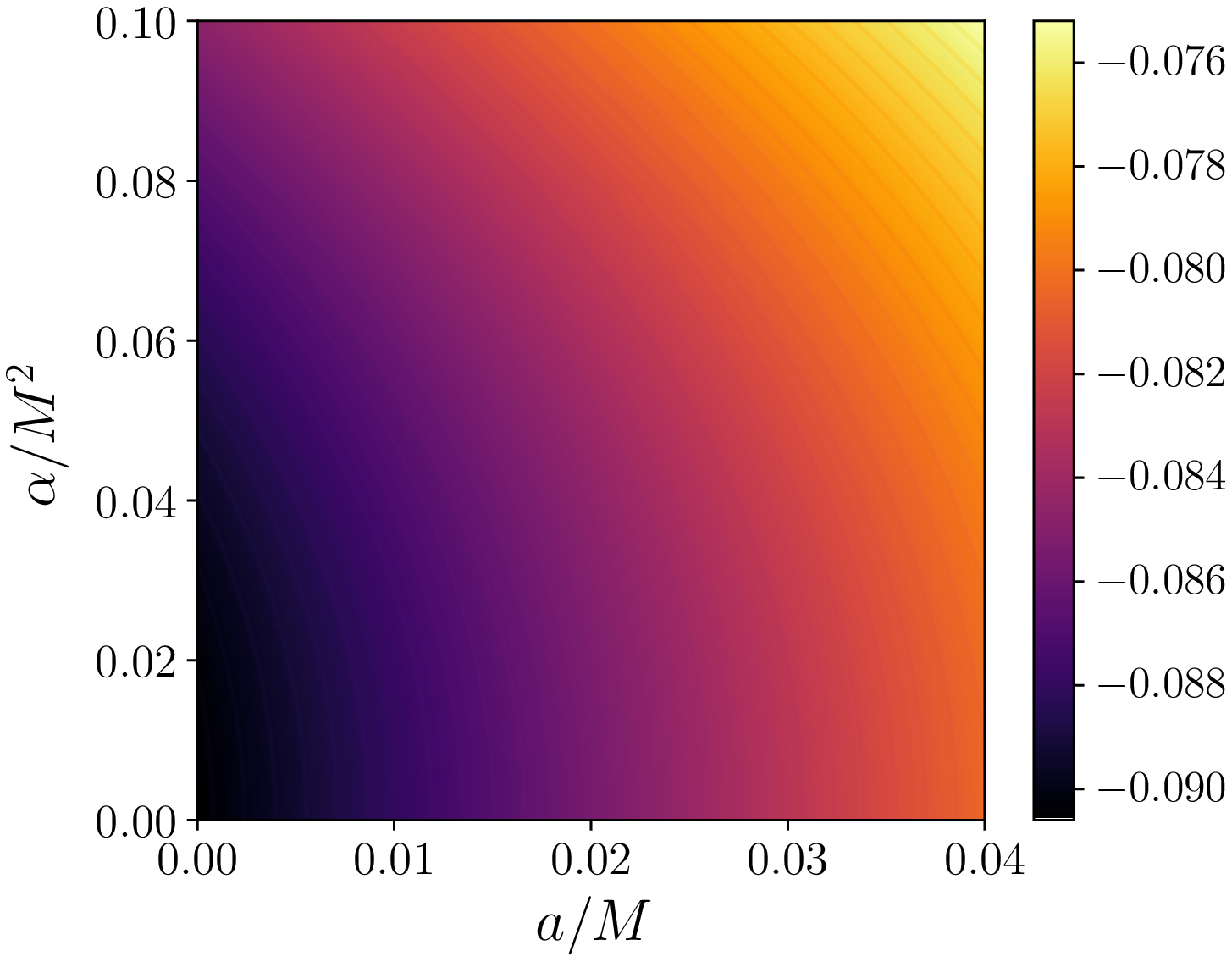}}
     \subfloat[$\textrm{Im}(\omega_s)$]{
     \includegraphics[width=0.3\textwidth]{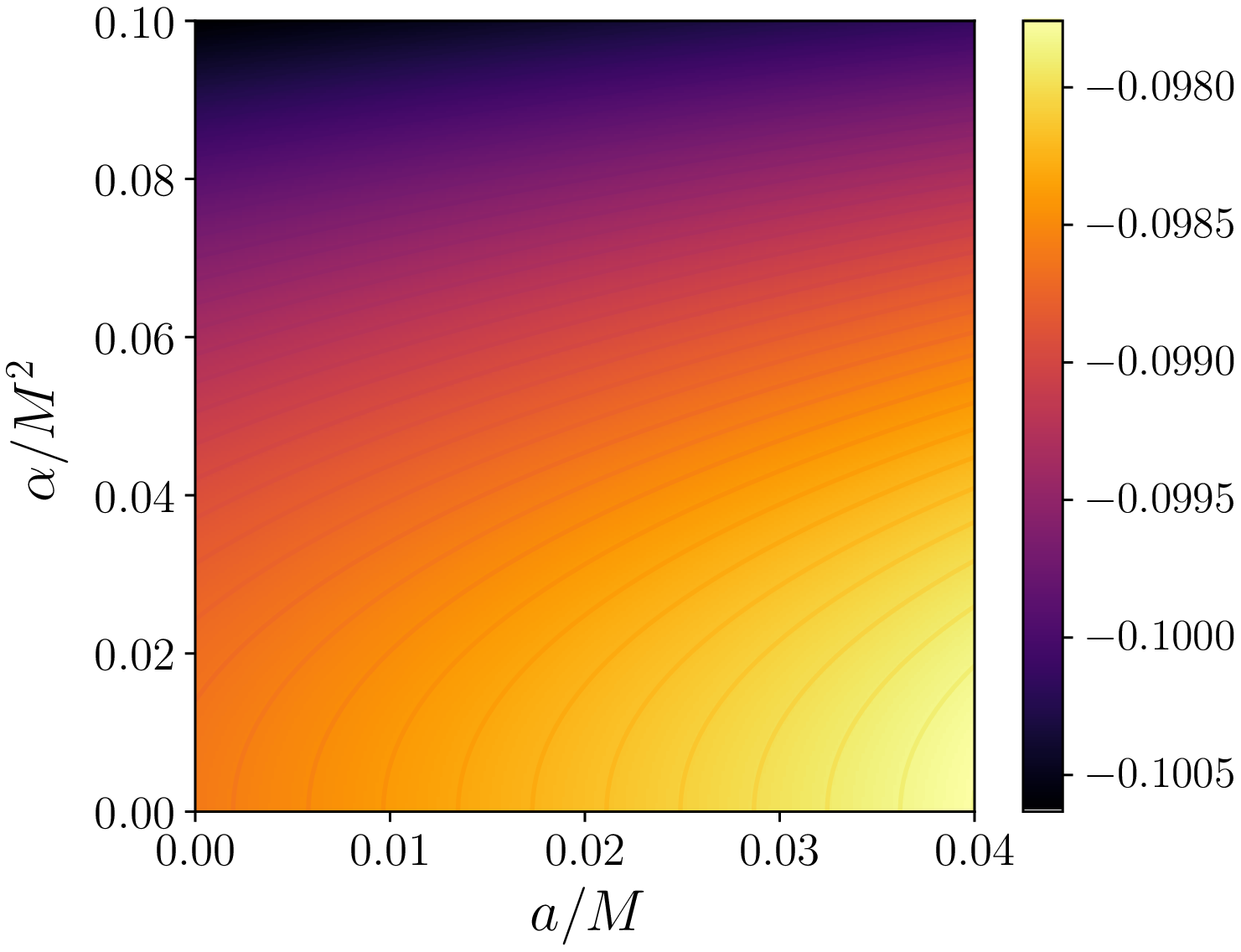}}
     \subfloat[$\textrm{Im}(\omega_g^\textrm{polar})$]{
     \includegraphics[width=0.3\textwidth]{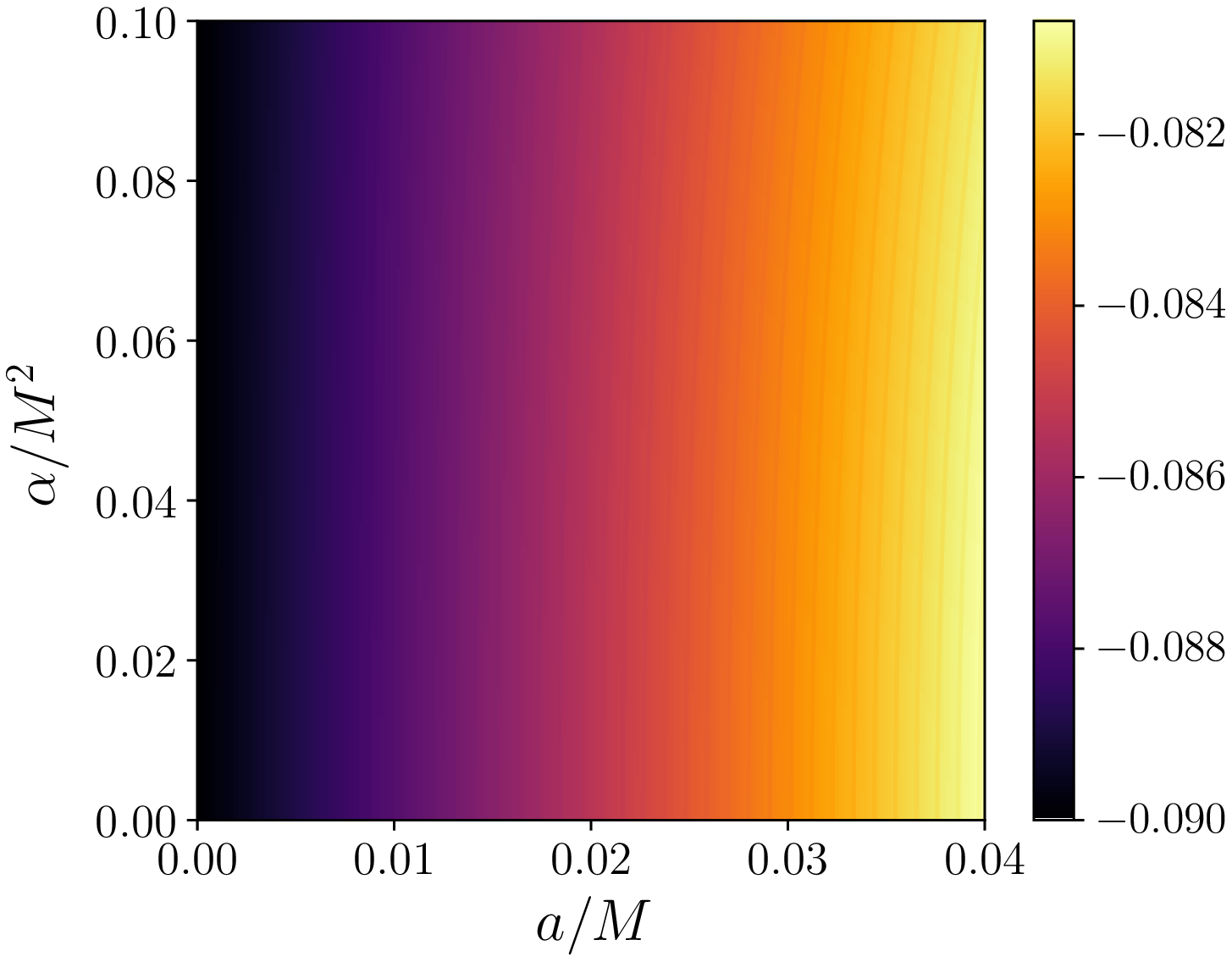}}
     \caption{
    Values of the QNM frequencies as functions of dimensionless 
    spin ($a/M$) and CS coupling ($\alpha / M^2$) relative to their GR values 
    as calculated with Eq.~\eqref{eq:compnorm}.
    Top row: from the left-most to right-most panel we show the values of 
    the real part of the axial gravitational-led, the scalar-led and polar gravitational-led modes.
    Bottom row: the same, but for the imaginary part.
    }
\label{fig:contour}
\end{figure*}

In Fig.~\ref{fig:contour} we show the behavior 
of the real and imaginary parts of the QNM frequencies as a function of 
the dimensionless spin $a/M$ and 
the CS coupling constant $\alpha/M^2$ in the form of a heat map.
This figure clear shows the positive correlation on the dependency of the axial gravitational-led mode with respect to changes in spin and CS coupling, as discussed in Sec.~\ref{sec:results}.
We also see how the imaginary part for the axial gravitational sector is 
almost independent of the strength of the CS coupling. 
Altogether, this figure complements our conclusions from Sec.~\ref{sec:results}.

\begin{figure*}
    \centering
     \subfloat[$\delta \, \textrm{Re}(\omega_g^\textrm{axial})$]{
     \includegraphics[width=0.32\textwidth]{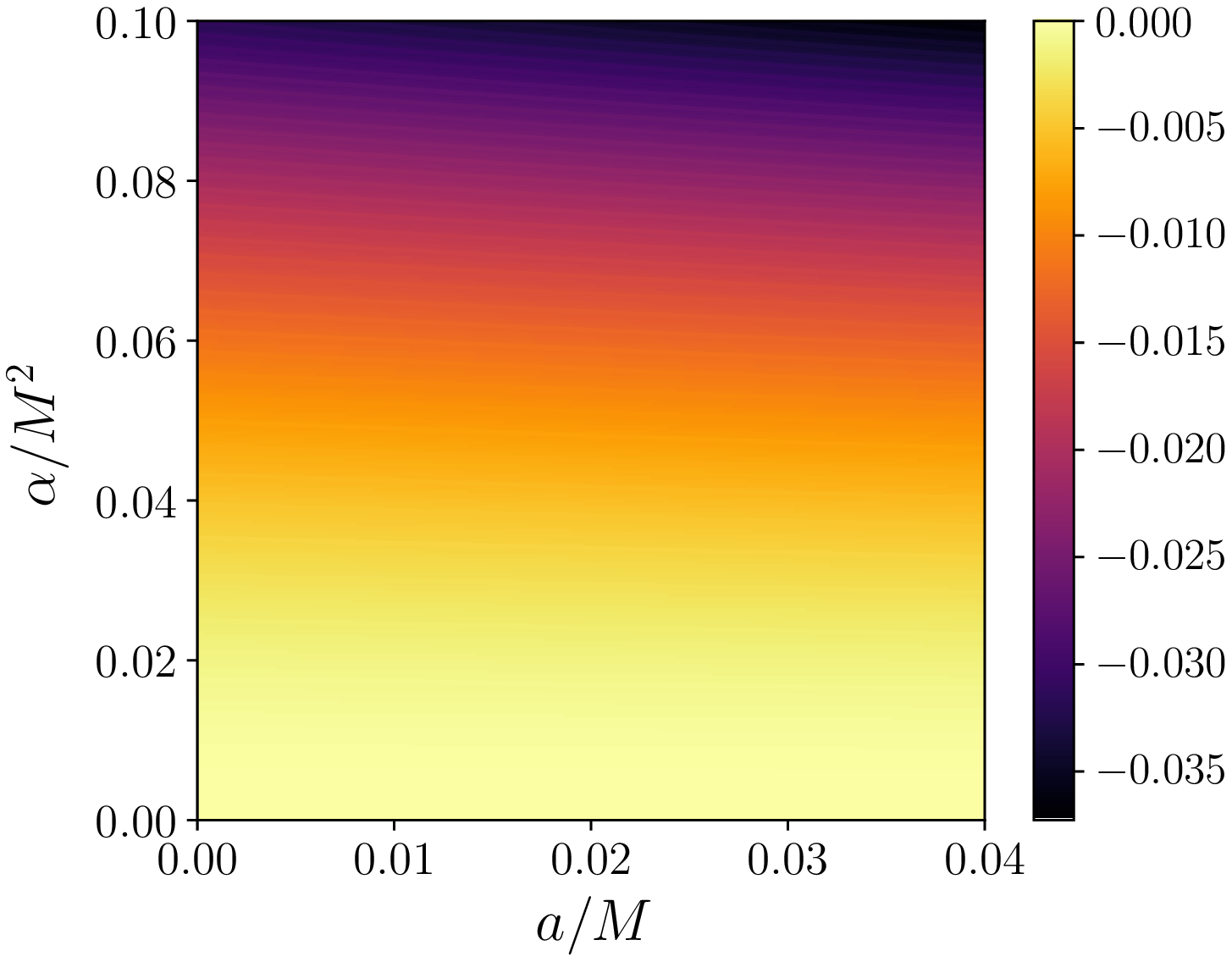}}
     \subfloat[$\delta \, \textrm{Re}(\omega_s)$]{
     \includegraphics[width=0.32\textwidth]{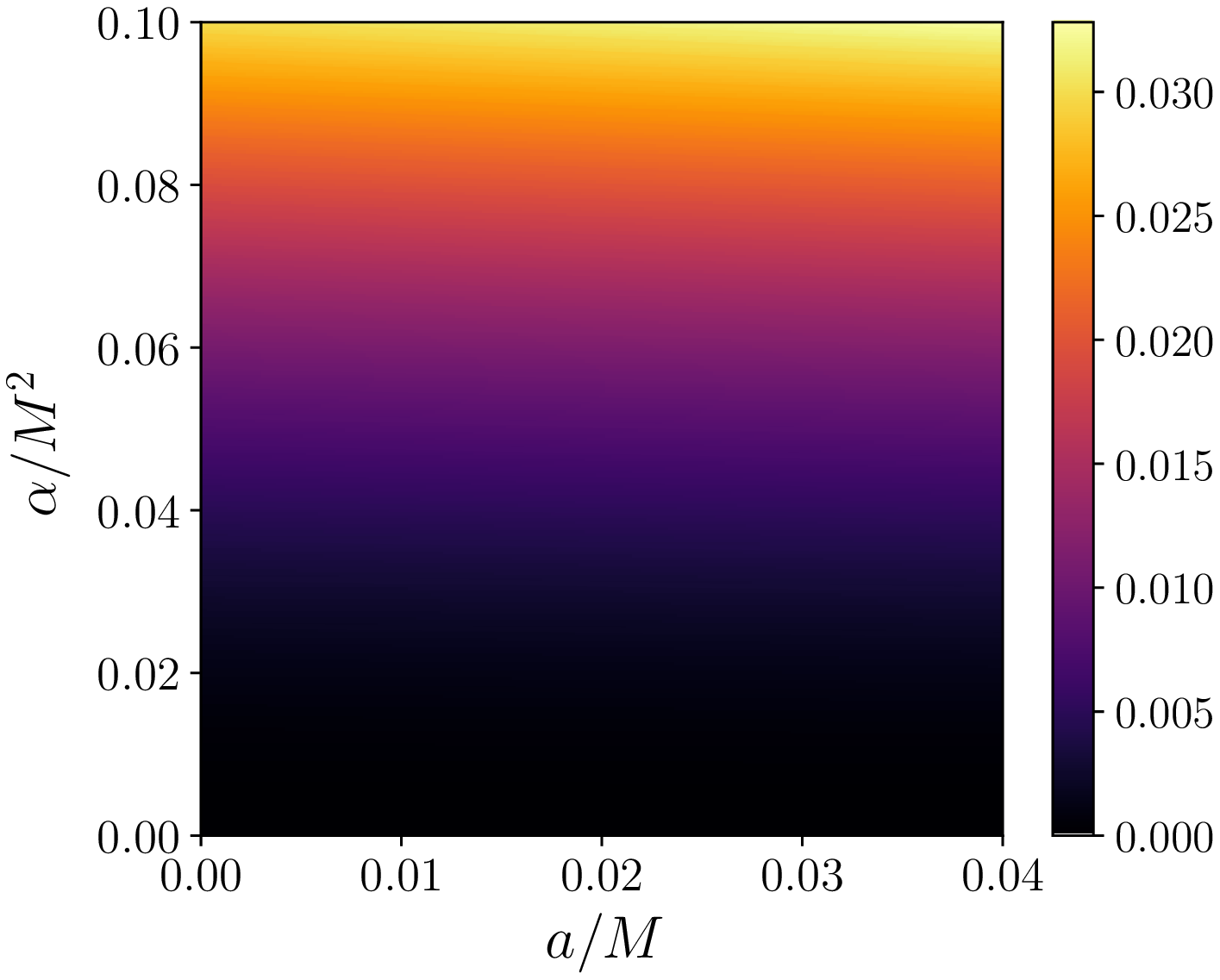}}
     \subfloat[$\delta \, \textrm{Re}(\omega_g^\textrm{polar})$]{
     \includegraphics[width=0.32\textwidth]{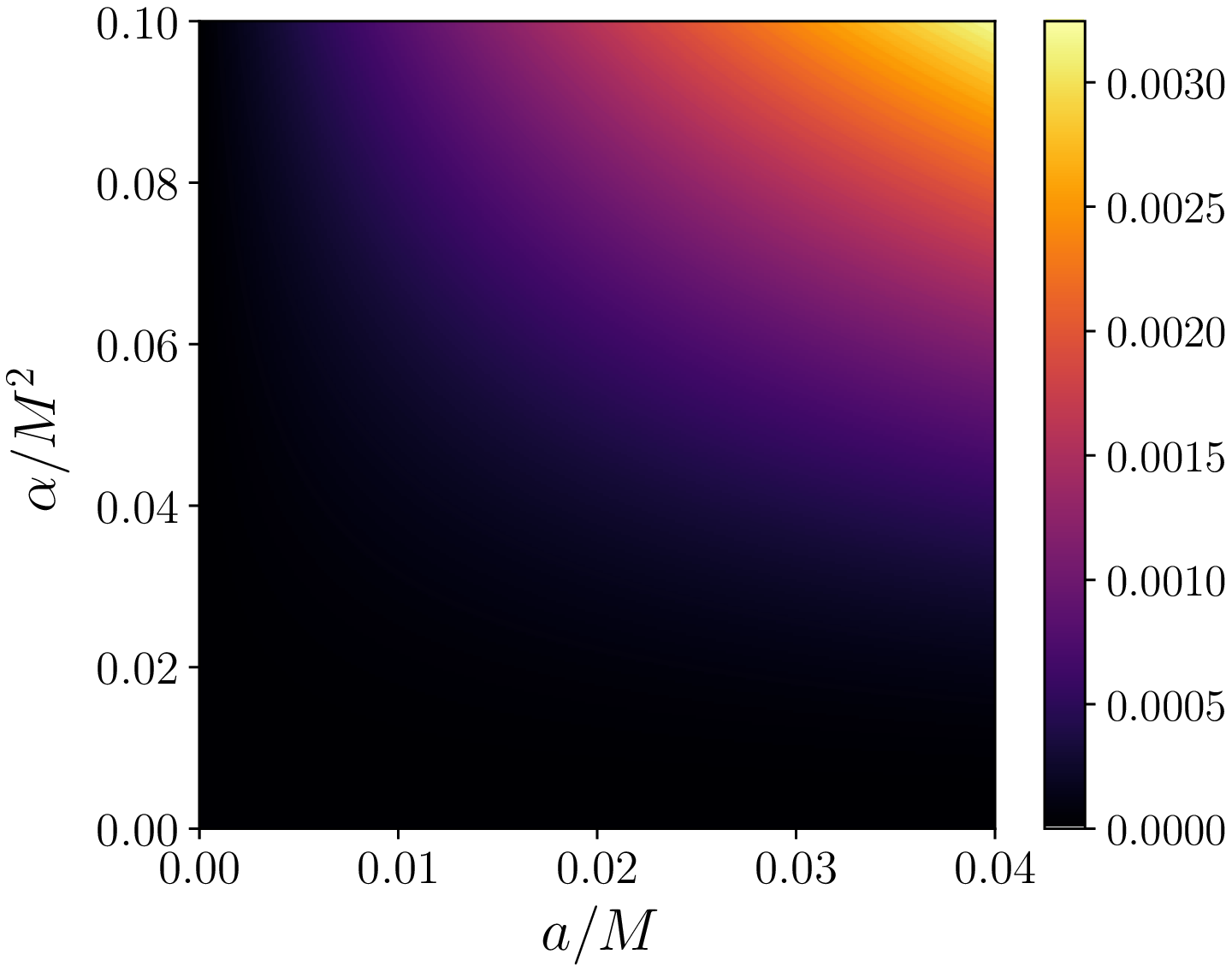}}
     \qquad
     \subfloat[$\delta \, \textrm{Im}(\omega_g^\textrm{axial})$]{
     \includegraphics[width=0.32\textwidth]{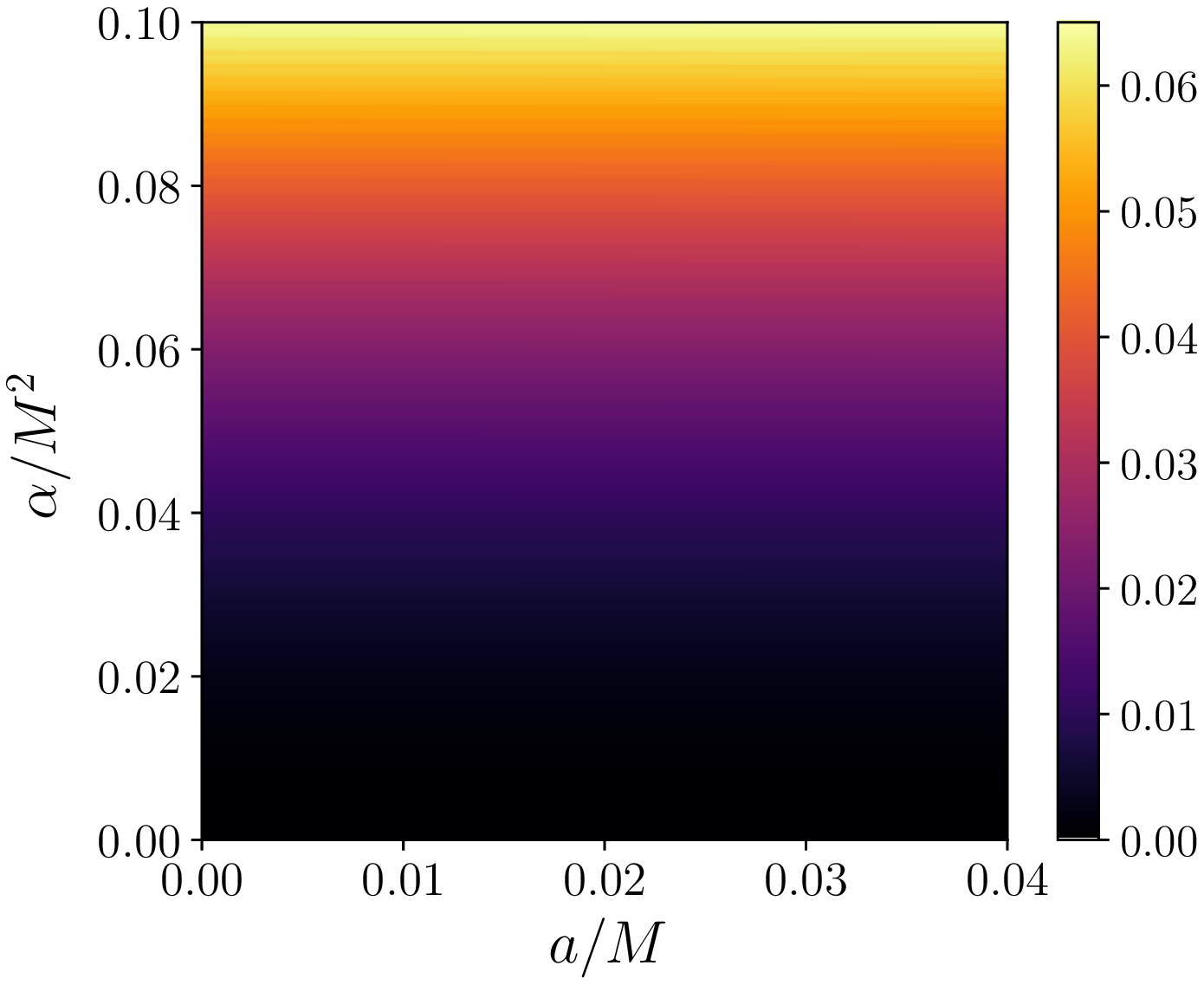}}
     \subfloat[$\delta \, \textrm{Im}(\omega_s)$]{
     \includegraphics[width=0.32\textwidth]{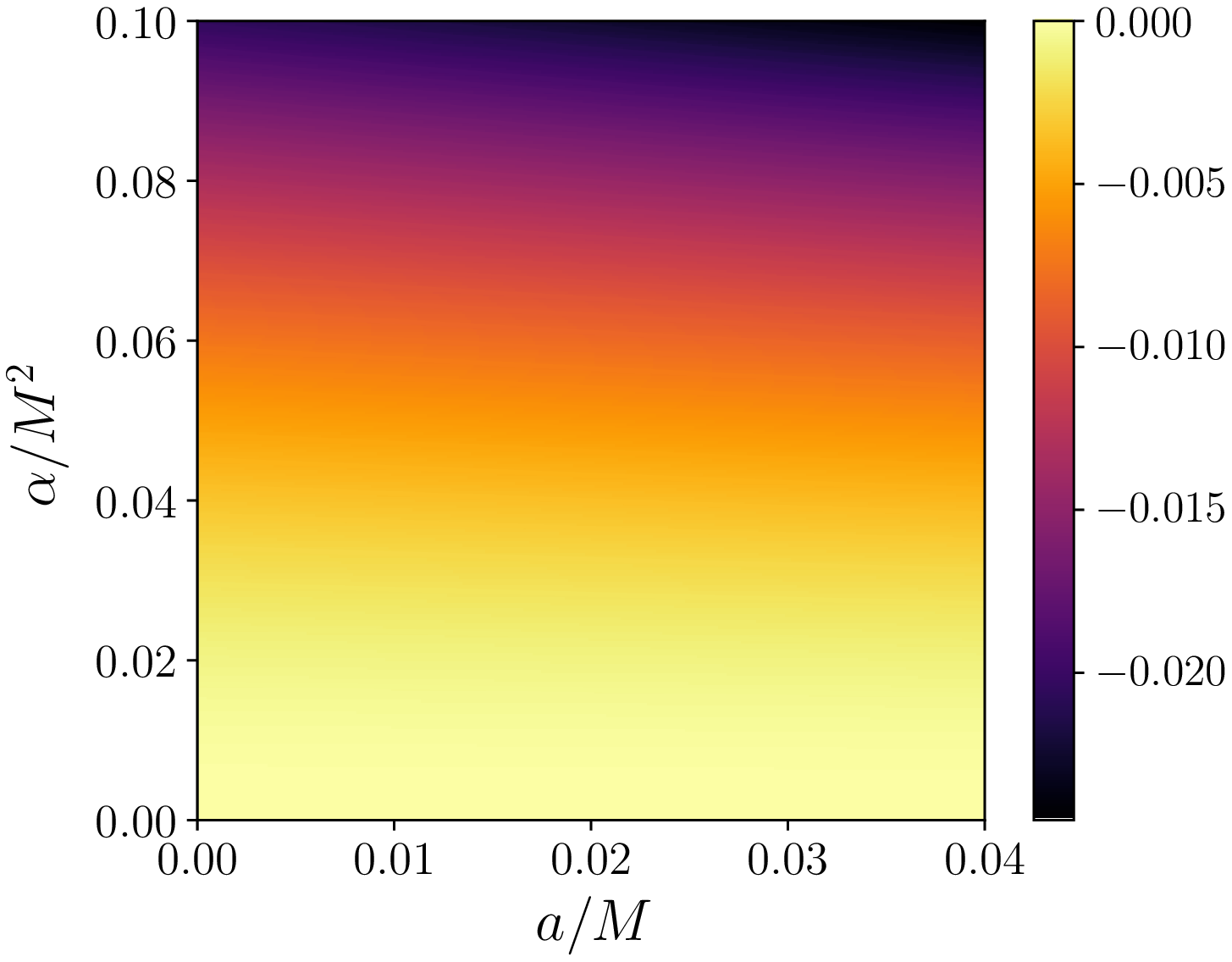}}
     \subfloat[$\delta \, \textrm{Im}(\omega_g^\textrm{polar})$]{
     \includegraphics[width=0.32\textwidth]{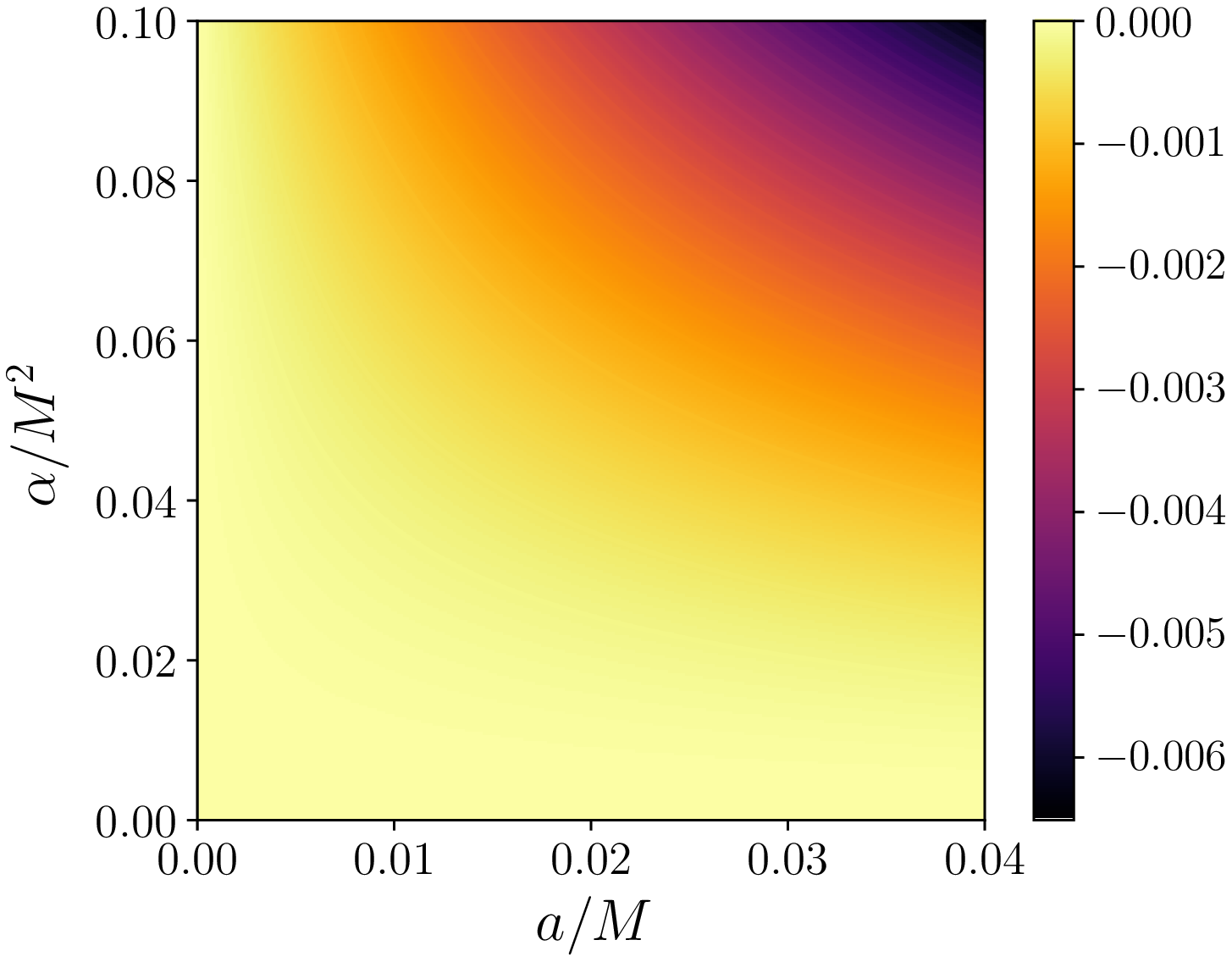}}
     \caption{
    Fractional difference between the dCS QNM frequencies with respect to 
    their GR values 
    [calculated with Eq.~\eqref{eq:compnorm}] as functions of dimensionless 
    spin ($a/M$) and CS coupling ($\alpha / M^2$) parameters.
    Top row: from the left-most to right-most panel we show the fractional changes to 
    the real part of the axial gravitational-led, the scalar-led and polar gravitational-led modes.
    Bottom row: the same, but for the imaginary part.
    In all cases, we see that the deviations become larger as we increase 
    both $a/M$ and $\alpha/M^2$.
    For the range of these two parameters considered here, the largest deviation 
    (of about $6\%$) occurs for the imaginary part of axial gravitational-led mode (bottom-left panel).
    }
\label{fig:compare_contour}
\end{figure*}

In addition, in Fig.~\ref{fig:compare_contour} we present a second heat map
that shows the fractional difference between the dCS QNM frequencies relative to their GR values, as function of both $a/M$ and $\alpha / M^2$.
To do so, we used the fitting formulas~\eqref{eq:fullfit} and~\eqref{eq:polarfit} to calculate
\begin{subequations}
\label{eq:compnorm}
\begin{align}
    \delta(\rm{Re} \, \omega) &= 1 - (\rm{Re} \, \omega)/(\rm{Re} \, \omega_{\rm{GR}}) \,, \\ 
    \delta(\rm{Im} \, \omega) &= 1 - (\rm{Im} \, \omega)/(\rm{Im} \, \omega_{\rm{GR}}) \,,
\end{align}
\end{subequations}
%
%
This complements the results shown in previous section and include information about how strong the deviation gets as the strength of the CS coupling is increased.
We see in particular that the imaginary part of the axial gravitational-led 
mode is the most sensitive, with a maximum percent change of about $6\%$ relative
to its GR value.
%

\section{Discussion} \label{sec:discussions}

We investigated the QNMs of slowly-rotating BHs in dCS gravity. 
We began by finding the perturbation equations that describe the evolution of scalar and tensorial perturbations. 
These triply-coupled set of ODEs generalize the slowly rotating versions of the Regge-Wheeler and the Zerilli-Moncrief master equations to the case of dCS gravity.
Using symmetry arguments, we showed that not all terms in these equations contribute to the QNM frequencies, thus simplifying our set of equations into two sets: two coupled equations for the axial and the scalar
sectors, and an homogeneous equation describing the polar sector; the latter was found to be independent of the coupling with the scalar field but included a CS modification to the effective potential thereby modifying the QNM frequencies.

We then solved these equations to calculate numerically the QNM frequencies. 
We found that the dCS corrections to the QNM frequencies scale with the square of the CS coupling. 
This was verified by means of a Fermi estimate as well as numerically. 
We also found that, in general, slowly-rotating BHs in dCS gravity have a decay time that is mostly independent of the CS coupling, so these BHs return to their stationary configuration on the same time-scale as in GR, but oscillating faster.
Indeed, their (real) frequency of oscillation increases with the CS coupling at fixed spin-value for the axial gravitational-led modes, whereas it decreases for scalar-led modes while remaining effectively remaining almost constant (slightly decreasing) for polar gravitation-led modes. 
Finally, we constructed fitting functions for the real and imaginary parts of the gravitational and scalar QNM frequencies as a function of the BH mass, spin and CS coupling for the fundamental ($n=0$) mode and all $\ell \leqslant 4$ harmonics.
We also found that the modes decay for all values of spin and CS coupling, a strong evidence that BHs described by Eq.~\eqref{eq:ds2} are stable against gravito-scalar perturbations, thus extending the results of~\cite{Molina:2010fb,Kimura:2018nxk} to include spin. 
We have also verified our findings with those of~\cite{Cano:2020cao} who calculated the scalar QNM frequencies for a rotating BH solution in dCS gravity for a scalar field satisfying a homogeneous wave equation instead of Eq.~\eqref{eq:CSfield}. 

What are the observational implications of our findings? 
Since GW detectors are capable of measuring only gravitational effects through GWs, the scalar modes of ringdown are not detectable with current technology. 
As a result, let us now focus only on the gravitational sector. Specifically in the axial gravitational sector, we find a degeneracy through a positive correlation between $a$ and $\alpha$ for the oscillatory frequencies; this is because the QNM frequency increases with increasing $a$ while keeping $\alpha$ constant and vice versa. This correlation is, however, found to break when considering the decay rates, because a change in the CS coupling barely affects the decay rate whereas a change in the spin parameter leads to longer lived modes. This result suggests that it may be possible to separate the effects of spin and the CS coupling for a loud enough ringdown observation.

The work presented here allows for many extensions along different directions. First, as it should be clear from our paper, we have here only considered perturbations to leading-order in the spin parameter. A natural extension of our work then would be to go to higher order in spin, which should produce more accurate results for BHs that are not as slowly spinning. 
Extending this calculation to second order, however, will be very difficult because of mode coupling between the odd- and even-parity sectors, just as in the GR case~\cite{Pani:2012bp}. 

Another interesting extension of our work would be to develop a continued fraction approach for the calculation of the QNM frequencies in dCS gravity. Such a method was introduced by Leaver long ago, but its extension to modified gravity theories is not obvious~\cite{Molina:2010fb}. One could therefore use dCS gravity as a toy problem to extend such methods, and then compare the results from the continued fraction method to the numerical results found in this paper. 
Additionally, this method can then be used in calculating the overtones for QNMs in dCS gravity, which is a limitation of the direct integration method used by us. Knowing the fundamental mode and at least one overtone, one can in principle constrain these modified theories of gravity using BH spectroscopy~\cite{Isi:2021iql,Ota:2019bzl,Ota:2021ypb}.

The results that we presented in this paper can also serve as a way to verify numerical simulation of BH binaries that result in slowly-rotating BH remnants, such as the head-on collisions performed in~\cite{Okounkova:2019dfo}.
A natural and important extension of our work, would be to extend the validity of 
our calculations to larger values of spin.
Currently, the only way to find the QNM frequencies of not-slowly-spinning BHs in dCS 
is through numerical relativity simulations of BH mergers which are computationally 
expensive. 
Another interesting path for future research would then be to find a modified Teukolsky equation for dCS gravity. 
Such a task, however, may not be possible given that the BHs of dCS gravity are Petrov type I and not Petrov type D~\cite{Yagi:2012ya,Owen:2021eez}, as assumed in the work of Teukolsky~\cite{Teukolsky:1973ha,Press:1973zz}.

Another interesting calculation would be to map our theory-specific numerical results to 
the theory-agnostic QNM parametrization introduced in~\cite{Maselli:2019mjd} (see also~\cite{Carullo:2021dui}). 
Our numerical results could also be used to quantify the error 
in theory-agnostic QNMs calculations due to the geometrical optics approximation,
as done in~\cite{Glampedakis:2017dvb,Glampedakis:2019dqh,Silva:2019scu}.
Yet another possible avenue for future work is to apply the tools developed here to study the oscillation spectra of rotating neutron stars in dCS gravity~\cite{Yagi:2013mbt,Gupta:2017vsl}. Such an analysis could have applications to GW asteroseismology~\cite{Andersson:1997rn,Pratten:2019sed}.

Finally, these results technically allow for the construction of ringdown templates that could be used by the LIGO-Virgo-Kagra collaboration to
place constraints on dCS gravity. This would of course only be possible for ringdown signals produced by slowly-rotating BH remnants, which in turn only occurs when the inspiraling binary components have the right spin magnitude and orientation prior to merger.
Whether the remnant is spinning or not, however, cannot currently be determined accurately enough because the signals detected so far do not have sufficiently high signal-to-noise ratio~\cite{Colpi:2016fup}, and thus the posteriors on the spin are very wide~\cite{LIGOScientific:2018mvr,LIGOScientific:2020ibl}. 
Moreover, such ringdown tests would require the unambiguous detection of more than one QNM mode~\cite{Kokkotas:1999bd,Ota:2019bzl,Forteza:2020hbw}
As the signal-to-noise ratio increases, it may be possible to carry out such a test, and in the meantime, it would be highly desirable to extend our results to more rapidly rotating BH backgrounds. 

\section*{Acknowledgements}
We thank Takahiro Tanaka, Helvi Witek, Pablo~A.~Cano and Thomas Hertog for useful discussions.
We also thank Emanuele Berti, Leo~C.~Stein and Leonardo Gualtieri for providing comments on the initial draft of this paper.
Some of our algebraic work used the package {\sc xAct}~\cite{xact} for Mathematica.
P.K.W. and N.Y. are supported by NSF grants No. PHY-1759615,~PHY-1949838  and  NASA  ATP  Grant  No.~17-ATP17-0225,   No.~NNX16AB98G  and  No.~80NSSC17M0041.

\appendix

\section{Spherical Harmonics -- orthogonality relations} 
\label{appendix:sphharm}

In this appendix, we are providing some useful orthogonality relations for
scalar, vector and tensor spherical harmonics. The scalar spherical harmonics
satisfy a fundamental identity,
\be
Y^{\ell m}_{,\theta \theta} + \cot \theta \, Y^{\ell m}_{,\theta} + \frac{1}{\sin^2 \theta} Y^{\ell m}_{,\phi \phi} = -\ell (\ell +1)Y^{\ell m} \,.
\ee
They also satisfy the orthogonality relation given by
\be
\langle Y^{\ell m}, Y^{\ell 'm'} \rangle = \delta^{\ell \ell '} \delta^{mm'} \,.
\ee
Vector spherical harmonics hold the following orthogonality relations,
\be
\langle Y^{\ell m}_{a}, Y^{\ell m}_{a} \rangle = \langle S^{\ell m}_{a}, S^{\ell m}_{a} \rangle = (\ell -1)(\ell +2) + 2 \,,
\ee
where the polar and axial vector harmonics have been defined as
\begin{align}
Y^{\ell m}_{a} &= \left(Y^{\ell m}_{,\theta} , Y^{\ell m}_{,\phi}\right) \,, \nonumber \\
S^{\ell m}_{a} &= \left(- \frac{Y^{\ell m}_{,\theta}}{\sin \theta}, \sin \theta\, Y^{\ell m}_{,\theta}  \right) \,.
\end{align}
Finally, the tensor spherical harmonics are given by
\begin{align}
\langle Z^{\ell m}_{ab}, Z^{\ell 'm'}_{ab} \rangle  &= \langle S^{\ell m}_{ab}, S^{\ell 'm'}_{ab} \rangle \nonumber \\ &= 2\ell  (\ell -1)(\ell +1)(\ell +2) \delta^{\ell \ell'} \delta^{mm'} \,.
\end{align}
There are additional orthogonality relations which can be found
in~\cite{Pani:2013pma}. The additional relations that we have used in this work
are
\begin{align}
	\cos\theta \, Y^{\ell m} = q_{\ell +1,m} Y^{\ell +1,m} +& q_{\ell ,m} Y^{\ell -1,m} \,, \nonumber \\
	\sin \theta \, Y^{\ell m}_{,\theta} = q_{\ell +1,m} \ell  Y^{\ell +1,m} -& q_{\ell ,m} (\ell +1) Y^{\ell -1,m} \nonumber \\
	A_{\ell 'm'}~ \langle Y^{\ell m},\, \sin \theta \, Y^{\ell 'm'}_{,\theta} \rangle & = (\ell -1) q_{\ell m} A_{\ell -1,m} \nonumber \\ -(\ell +2)q_{\ell +1,m}&A_{\ell +1,m} \,,
\end{align}
where $q_{\ell m}$ is defined in Eq.~\eqref{eq:qlm}. $A_{\ell m}$ is the operator
defined to separate the angular dependence of the linearized field equations
within the slow-rotation approximation.

\onecolumngrid

\section{Coefficients of the perturbation equations} \label{appendix:coefficientsofPE}

In this Appendix, we list the explicit forms of the coefficients appearing in
the perturbation equations presented in Sec.~\ref{sec:slowrotpert}. These
coefficients have prefactors of a combination of $\alpha, a$ and $m$ which have
already been shown in the expression for the perturbation equations. However,
it is worth noting that there is no direct correspondence between these
coefficients.
The potentials are given by:
\begin{align}
    V^S_{\rm eff} &= \left( 1 - \frac{2M}{r} \right)
\left[
    \frac{\ell(\ell+1)}{r^2} + \frac{2M}{r^3}
\right] + 2 a m \omega \left[ \frac{4M}{r^3} - \alpha^2 \frac{189 M^2+120 M r+70 r^2}{112 \kappa 
   r^8} \right]\,,
\\
%
V^A_{\rm eff} &=  \left( 1 - \frac{2M}{r} \right) \left[ \frac{\ell(\ell+1)}{r^2} - \frac{6M}{r^3} +\frac{a m}{\omega} \frac{24 M \left(3 r - 7 M\right) }{\ell(\ell+1) r^6} \right]
+ 2 a m \omega \left[ \frac{4M}{r^3} \right.
\nonumber \\
&\quad - \left. \alpha^2 \frac{1}{112 \kappa  \lambda_{\ell} \ell (\ell+1) M
   r^{13} \omega ^2}\sum_{i=0}^{6}\Theta_i(r, \ell, \omega) M^i r^{6-i}\right]
 \,,
\\
%
V^P_{\rm eff} &= \left(1 - \frac{2M}{r}\right) \left[ \frac{2M}{r^3} + \frac{1}{3}\lambda_{\ell} \left(\frac{1}{r^2} + \frac{2 \lambda_{\ell} (\ell^2 + \ell +1)}{(6M + \lambda_{\ell} r)^2}\right) \right] 
\nonumber \\
&\quad +  \frac{4amM}{r^8 \ell (\ell + 1)\left(\lambda_{\ell} r+6 M\right)^4 \omega}  \left[ \sum_{i=0}^{7}\xi_i(r, \ell, \omega) M^i r^{7-i} - \frac{\alpha^2}{448 M r^5 \kappa} \sum_{i=0}^{10}\upsilon_i(r, \ell, \omega) M^i r^{10-i} \right]\,.
\end{align}
where $\lambda_{\ell} = (\ell+2)(\ell-1)$ and the functions $\xi_i$ are the same as those found for slowly rotating Kerr BHs in GR~\cite{Pani:2013pma}. We also provide all other functions in a Mathematica notebook that can be made available upon request.
The other functions in the perturbed field equations that contribute to the QNM frequencies [See Eqs.~\eqref{eqs:perturbed_final}] are given by
\begin{subequations}
\begin{align}
    g(r) &= \frac{6 i   \lambda_{\ell} \ell  (\ell +1) M}{r^5 \omega }\,, \\
	h(r) &= -\frac{i    \left(r^4 \omega ^2 \left(12 (2 \ell
		(\ell +1)-1) M^2+15 M r+5 r^2\right)+144 M^3 (2
		M-r)\right)}{2 M r^9 \omega ^2}\,, \\
	j(r) &= \frac{72 i    M^2  (r-2 M)}{r^8 \omega ^2}\,,
\end{align}
\end{subequations}
and
\begin{subequations}
\begin{align}
    v(r) &= -\frac{6 i   M \omega }{\kappa  r^5}\,, \\
    n(r) &= \frac{i    \left(-4224 M^4+3306 M^3 r+48 M^2 r^2
	\left(r^2 \omega ^2-15\right)+5 M r^3+15
	r^4\right)}{4 \kappa  l (l+1) M r^9}\,, \\
    p(r) &= \frac{12 i    M  (12 M-5 r) (2 M-r)}{\kappa  \ell
	(\ell +1) r^8}\,. 
\end{align}
\end{subequations}
All the other functions such as $k_i,p_i,s_i$ and $r_i$ ($i=1,\dots,4)$ are rather lengthy and non-illuminating and hence not provided here. Instead these can be found in a Mathematica notebook which can be made available upon request.

\section{QNM frequencies of different multipoles and fitting coefficients}
\label{appendix:qnm}

In this appendix, we have tabulated the values of QNM frequencies for all multipoles of $\ell=2$ and $\ell=3$. $\ell=4$ values have been compiled into a data file available upon request.
In the following tables, we have shown the values for the real and imaginary parts of the QNM frequencies for both gravitational- and scalar-led modes.

We then proceed to present the numerical values for the fitting functions shown in Eqs.~\eqref{eq:fullfit} for both the gravitational and scalar QNM frequencies. We have also calculated the average percent error in these fits and presented these in our tables below.

\subsection*{Axial gravitational and scalar sectors}
\begin{table}[h]
	\begin{tabular}{c @{\hskip 0.12in} c @{\hskip 0.12in} c @{\hskip 0.12in} c @{\hskip 0.12in} c @{\hskip 0.12in} c @{\hskip 0.12in} c }
		\hline
		\hline
		$a/M$&$\alpha/M^2$& $m = -2$ & $m=-1$ & $m=0$ & $m=1$ & $m=2$ \\ \hline
		
		&$0.0$	& $ .3737,.0889 $ & $ .3737,.0889 $ & $ .3737,.0889 $ & $ .3737,.0889 $ & $ .3737,.0889 $ \\[0.1cm]
		
		$0.0$&$0.05$	& $ .3757,.0879 $ & $ .3757,.0879 $ & $ .3757,.0879 $ & $ .3757,.0879 $ & $ .3756,.0879 $ \\[0.1cm]
		
		&$0.1$	& $ .3828,.0846 $ & $ .3829,.0846 $ & $ .3829,.0846 $ & $ .3829,.0846 $ & $ .3828,.0846 $ \\[0.1cm] \hline
		
		&$0.0$	& $ .3723,.0881 $ & $ .3729,.0887 $ & $ .3737,.0889 $ & $ .3743,.0887 $ & $ .3749,.0881 $ \\[0.1cm]
		
		$0.01$&$0.05$	& $ .3744,.0872 $ & $ .375,.0877 $ & $ .3757,.0879 $ & $ .3763,.0877 $ & $ .3770,.0871 $ \\[0.1cm]
		
		&$0.1$	& $ .3816,.0839 $ & $ .3822,.0844 $ & $ .3829,.0846 $ & $ .3836,.0844 $ & $ .3845,.0839 $ \\[0.1cm] \hline
		
		&$0.0$	& $ .3709,.0859 $ & $ .3723,.0881 $ & $ .3737,.0889 $ & $ .3749,.0881 $ & $ .3764,.0857 $ \\[0.1cm]
		
		$0.02$&$0.05$	& $ .373,.085 $ & $ .3744,.0872 $ & $ .3757,.0879 $ & $ .377,.0871 $ & $ .3786,.0849 $ \\[0.1cm]
		
		&$0.1$	& $ .3806,.0821 $ & $ .3816,.0839 $ & $ .3829,.0846 $ & $ .3845,.0839 $ & $ .3865,.0823 $ \\[0.1cm] \hline
		
		&$0.0$	& $ .3693,.0829 $ & $ .3716,.0871 $ & $ .3737,.0889 $ & $ .3756,.0870 $ & $ .3783,.0825 $ \\[0.1cm]
		
		$0.03$&$0.05$	& $ .3716,.0822 $ & $ .3737,.0862 $ & $ .3829,.0846 $ & $ .3778,.0862 $ & $ .3806,.0819 $ \\[0.1cm]
		
		&$0.1$	& $ .3796,.0796 $ & $ .3811,.0831 $ & $ .3829,.0846 $ & $ .3854,.0832 $ & $ .3889,.0799 $ \\[0.1cm] \hline
		
		&$0.0$	& $ .3678,.0796 $ & $ .3709,.0859 $ & $ .3737,.0889 $ & $ .3764,.0857 $ & $ .3805,.0792 $ \\[0.1cm]
		
		$0.04$&$0.05$	& $ .3702,.0791 $ & $ .373,.085 $ & $ .3671,.0926 $ & $ .3786,.0849 $ & $ .3829,.0788 $ \\[0.1cm]
		
		&$0.1$	& $ .3786,.0769 $ & $ .3806,.0821 $ & $ .3829,.0846 $ & $ .3865,.0823 $ & $ .3915,.0773 $ \\ \hline
		\hline
	\end{tabular}
    \caption{QNM frequencies for axial
    gravitational-led sector with $n=0, \,\ell =2$ for slowly rotating BHs in dCS gravity. The format
    used is $M$($\textrm{Re}(\omega),-\textrm{Im}(\omega)$). To save space, the leading zeros have been omitted.}
    \label{tab:grv_l2_all_m}
    \end{table}
    %
    %
    \begin{table}[h]
	\begin{tabular}{c @{\hskip 0.12in} c @{\hskip 0.12in} c @{\hskip 0.12in} c @{\hskip 0.12in} c @{\hskip 0.12in} c @{\hskip 0.12in} c }
		\hline
		\hline
		$a/M$&$\alpha/M^2$& $m = -2$ & $m=-1$ & $m=0$ & $m=1$ & $m=2$ \\ \hline
		
		&$0.0$	& $ .4836,.0967 $ & $ .4836,.0967 $ & $ .4836,.0967 $ & $ .4836,.0967 $ & $ .4836,.0967 $ \\[0.1cm]
		
		$0.0$&$0.05$	& $ .4810,.0967 $ & $ .4810,.0967 $ & $ .4810,.0967 $ & $ .4810,.0967 $ & $ .4810,.0967 $ \\[0.1cm]
		
		&$0.1$	& $ .4720,.0970 $ & $ .4720,.0970 $ & $ .4720,.0970 $ & $ .4720,.0970 $ & $ .4720,.0970 $ \\[0.1cm] \hline
		
		&$0.0$	& $ .4817,.0976 $ & $ .4828,.0969 $ & $ .4836,.0967 $ & $ .4843,.0969 $ & $ .4846,.0967 $ \\[0.1cm]
		
		$0.01$&$0.05$	& $ .4790,.0976 $ & $ .4801,.0970 $ & $ .4810,.0967 $ & $ .4816,.097 $ & $ .4819,.0976 $ \\[0.1cm]
		
		&$0.1$	& $ .470,.0979 $ & $ .4711,.0973 $ & $ .4720,.0970 $ & $ .4726,.0972 $ & $ .4729,.0979 $ \\[0.1cm] \hline
		
		&$0.0$	& $ .4787,.1010 $ & $ .4817,.0976 $ & $ .4836,.0967 $ & $ .4846,.0976 $ & $ .4839,.1004 $ \\[0.1cm]
		
		$0.02$&$0.05$	& $ .4759,.1010 $ & $ .4790,.0976 $ & $ .4810,.0967 $ & $ .4819,.0976 $ & $ .4812,.1004 $ \\[0.1cm]
		
		&$0.1$	& $ .4667,.1012 $ & $ .4701,.0979 $ & $ .4720,.0970 $ & $ .4729,.0979 $ & $ .4718,.1008 $ \\[0.1cm] \hline
		
		&$0.0$	& $ .4719,.1085 $ & $ .4803,.0988 $ & $ .4836,.0967 $ & $ .4846,.0987 $ & $ .4782,.1044 $ \\[0.1cm]
		
		$0.03$&$0.05$	& $ .4689,.1082 $ & $ .4776,.0988 $ & $ .4810,.0967 $ & $ .4819,.0987 $ & $ .4754,.1040 $ \\[0.1cm]
		
		&$0.1$	& $ .4577,.1082 $ & $ .4686,.0992 $ & $ .4720,.0970 $ & $ .4727,.0991 $ & $ .4652,.1038 $ \\[0.1cm] \hline
		
		&$0.0$	& $ .4537,.107 $ & $ .4787,.1007 $ & $ .4836,.0967 $ & $ .4840,.1004 $ & $ .4692,.1005 $ \\[0.1cm]
		
		$0.04$&$0.05$	& $ .4512,.107 $ & $ .4759,.1010 $ & $.4810,.0967 $ & $ .4812,.1006 $ & $ .4666,.1002 $ \\[0.1cm]
		
		&$0.1$	& $ .4424,.1049 $ & $ .4667,.1012 $ & $ .4720,.0970 $ & $ .4718,.1008 $ & $ .4573,.0994 $ \\ \hline
		\hline
	\end{tabular}
	\caption{Same as Table~\ref{tab:grv_l2_all_m} but for the scalar-led sector with $n=0, \, \ell =2$.
    }
    \label{tab:sca_l2_all_m}
\end{table}

\begin{table*}[h]
	\begin{tabular}{c @{\hskip 0.12in} c @{\hskip 0.12in} c @{\hskip 0.12in} c @{\hskip 0.12in} c @{\hskip 0.12in} c @{\hskip 0.12in} c @{\hskip 0.12in} c @{\hskip 0.12in} c }
		\hline \hline
		$a/M$&$\alpha/M^2$& $m = -3$ & $m = -2$ & $m=-1$ & $m=0$ & $m=1$ & $m=2$ & $m=3$ \\ \hline
		
		&$0.0$	& $ .5994,.0927 $ &$ .5994,.0927 $ & $ .5994,.0927 $ & $ .5994,.0927 $ & $ .5994,.0927 $ & $ .5994,.0927 $& $ .5994,.0927 $ \\[0.1cm]
		
		$0.0$&$0.05$	& $ .6078,.0911 $ &$ .6078,.0911 $ & $ .6078,.0911 $ & $ .6078,.0911 $ & $ .6078,.0911 $ & $ .6078,.0911 $& $ .6078,.0911 $ \\[0.1cm]
		
		&$0.1$	& $ .6381,.1251 $ &$ .6381,.1251 $ & $ .6381,.1251 $ & $ .6381,.1251 $ & $ .6381,.1251 $ & $ .6381,.1251 $& $ .6380,.1251 $ \\[0.1cm]\hline
		
		&$0.0$	& $ .5970,.0911 $ &$ .5979,.0919 $ & $ .5987,.0925 $ & $ .5994,.0927 $ & $ .6001,.0925 $ & $ .6007,.0919 $ & $ .6013,.0911 $\\[0.1cm]
		
		$0.01$&$0.05$& $ .6055,.0896 $ & $ .6063,.0904 $ & $ .6069,.0909 $ & $ .6078,.0911 $ & $ .6085,.0909 $& $ .6092,.0904 $ & $ .6100,.0895 $ \\[0.1cm]
		
		&$0.1$	& $ .6347,.1263 $&$ .6363,.1256 $ & $ .6373,.1251 $ & $ .6381,.1251 $ & $ .6387,.1254 $ & $ .6389,.1260 $& $ .6386,.1267 $ \\[0.1cm] \hline
		
		&$0.0$	& $ .5941,.0874 $ & $ .5961,.0900 $ & $ .5979,.0919 $ & $ .5994,.0927 $ & $ .6007,.0919 $ & $ .6019,.0899 $& $ .6035,.0871 $ \\[0.1cm]
		
		$0.02$&$0.05$& $ .6032,.0859 $ & $ .6048,.0885 $ & $ .6063,.0904 $ & $ .6092,.0904 $ & $ .5799,.0955 $ & $ .6108,.0884 $& $ .6128,.0858 $\\[0.1cm]
		
		&$0.1$	& $ .6265,.1254 $ & $ .6322,.1267 $ & $ .6363,.1256 $ & $ .6381,.1251 $ & $ .6390,.1260 $ & $ .6375,.1269 $& $ .6354,.1250 $\\[0.1cm] \hline
		
		&$0.0$	& $ .5912,.0829 $ & $ .5942,.0874 $ & $ .5971,.0911 $ & $ .5994,.0927 $ & $ .6013,.0911 $ & $ .6034,.0871 $& $ .6064,.0825 $\\[0.1cm]
		
		$0.03$&$0.05$& $ .6008,.0816 $ & $ .6032,.0859 $ & $ .6055,.0896 $ & $ .6078,.0911 $ & $ .6100,.0895 $ & $ .6128,.0858 $& $ .6162,.0816 $\\[0.1cm]
		
		&$0.1$	& $ .6198,.1209 $ & $ .6265,.1254 $ & $ .6347,.1263 $ & $ .6381,.1251 $ & $ .6386,.1267 $ & $ .6354,.1250 $& $ .6348,.1200 $\\[0.1cm] \hline
		
		&$0.0$	& $ .5883,.0787 $ & $ .5922,.0844 $ & $ .5961,.0900 $ & $.5994,.0927  $ & $ .6019,.0899 $ & $ .6053,.0841 $& $ .6099,.0782 $\\[0.1cm]
		
		$0.04$&$0.05$	& $ .5984,.0775 $ & $ .6016,.0829 $ & $ .6048,.0885 $ & $ .6078,.0911 $ & $ .6108,.0884 $ & $ .6150,.0830 $& $ .6202,.0777 $\\[0.1cm]
		
		&$0.1$	& $ .6146,.1161 $ & $ .6218,.1225 $ & $ .6322,.1267 $ & $ .6381,.1251 $ & $ .6375,.1269 $ & $ .6347,.1217 $& $ .6363,.1150 $\\ \hline
		\hline
	\end{tabular}
    \caption{Same as Table~\ref{tab:grv_l2_all_m} but for the axial gravitational-led sector with $n=0, \, \ell =3$. 
    }
    \label{tab:grv_l3_all_m}
\end{table*}

\clearpage

\begin{table*}[h]
	\begin{tabular}{c @{\hskip 0.12in} c @{\hskip 0.12in} c @{\hskip 0.12in} c @{\hskip 0.12in} c @{\hskip 0.12in} c @{\hskip 0.12in} c @{\hskip 0.12in} c @{\hskip 0.12in} c }
		\hline \hline
		$a/M$&$\alpha/M^2$& $m = -3$ & $m = -2$ & $m=-1$ & $m=0$ & $m=1$ & $m=2$ & $m=3$ \\ \hline
		
		&$0.0$	& $ .6753,.0965 $ &$ .6753,.0965 $ & $ .6753,.0965 $ & $ .6753,.0965 $ & $ .6753,.0965 $ & $ .6753,.0965 $& $ .6753,.0965 $ \\[0.1cm]
		
		$0.0$&$0.05$	& $ .6656,.0969 $ &$ .6656,.0969 $ & $ .6656,.0969 $ & $ .6656,.0969 $ & $ .6656,.0969 $ & $ .6656,.0969 $& $ .6656,.0969 $ \\[0.1cm]
		
		&$0.1$	& $ .6381,.1251 $ &$ .6381,.1251 $ & $ .6381,.1251 $ & $ .6381,.1251 $ & $ .6381,.1251 $ & $ .6381,.1251 $& $ .6381,.1251 $ \\[0.1cm]\hline
		
		&$0.0$	& $ .6714,.0974 $ &$ .6731,.0969 $ & $ .6744,.0966 $ & $ .6753,.0965 $ & $ .6759,.0966 $ & $ .6761,.0968 $ & $ .6758,.0971 $\\[0.1cm]
		
		$0.01$&$0.05$& $ .6616,.0979 $ & $ .6634,.0973 $ & $ .6647,.0970 $ & $ .6656,.0969 $ & $ .6661,.0973 $& $ .6662,.0973 $ & $ .6658,.0976 $ \\[0.1cm]
		
		&$0.1$	& $ .6347,.1263 $&$ .6363,.1256 $ & $ .6373,.1251 $ & $ .6381,.1251 $ & $ .6387,.1254 $ & $ .6390,.1260 $ & $ .6386,.1267 $ \\[0.1cm] \hline
		
		&$0.0$	& $ .6630,.0989 $ & $ .6693,.0979 $ & $ .6731,.0969 $ & $ .6753,.0965 $ & $ .6761,.0968 $ & $ .6750,.0974 $& $ .6720,.0969 $ \\[0.1cm]
		
		$0.02$&$0.05$& $ .6525,.0994 $ & $ .6593,.0985 $ & $ .6634,.0973 $ & $ .6656,.0969 $ & $ .6662,.0973 $ & $ .6648,.0979 $& $ .6614,.0971 $\\[0.1cm]
		
		&$0.1$	& $ .6265,.1254 $ & $ .6322,.1267 $ & $ .6363,.1256 $ & $ .6381,.1251 $ & $ .6390,.1260 $ & $ .6375,.1269 $& $ .6354,.1250 $\\[0.1cm] \hline
		
		&$0.0$	& $ .6515,.0961 $ & $ .6630,.0989 $ & $ .6715,.0973 $ & $ .6753,.0965 $ & $ .6758,.0972 $ & $ .6720,.0969 $& $ .6679,.0919 $\\[0.1cm]
		
		$0.03$&$0.05$& $ .6411,.0959 $ & $ .6526,.0994 $ & $ .6616,.0979 $ & $ .6656,.0969 $ & $ .6658,.0976 $ & $ .6614,.0971 $& $ .6572,.0917 $\\[0.1cm]
		
		&$0.1$	& $ .6198,.1209 $ & $ .6265,.1254 $ & $ .6347,.1263 $ & $ .6381,.1251 $ & $ .6386,.1267 $ & $ .6354,.1250 $& $ .6348,.1200 $\\[0.1cm] \hline
		
		&$0.0$	& $ .6428,.0903 $ & $ .6551,.0976 $ & $ .6693,.0979 $ & $.6753,.0965  $ & $ .6750,.0974 $ & $ .6689,.0939 $& $ .6671,.0855 $\\[0.1cm]
		
		$0.04$&$0.05$& $ .6326,.0902 $ & $ .6446,.0976 $ & $ .6593,.0985 $ & $ .6656,.0969 $ & $ .6648,.0979 $ & $ .6582,.0938 $& $ .6563,.0852 $\\[0.1cm]
		
		&$0.1$	& $ .6146,.1161 $ & $ .6218,.1225 $ & $ .6322,.1267 $ & $ .6381,.1251 $ & $ .6375,.1269 $ & $ .6347,.1218 $& $ .6363,.1150 $\\ \hline
		\hline
	\end{tabular}
	\caption{Same as Table~\ref{tab:grv_l2_all_m} but for the scalar-led sector with $n=0, \, \ell =3$. 
    }
    \label{tab:sca_l3_all_m}
\end{table*}

\begin{table*}[h]
	\begin{center}
		\begin{tabular}{c @{\hskip 0.1in} c @{\hskip 0.1in} c @{\hskip 0.1in} c @{\hskip 0.1in} c @{\hskip 0.1in} c @{\hskip 0.1in} c @{\hskip 0.1in} c}
			\hline \hline
			$m$& $f_1$ & $f_2$ & $f_3$& $f_4$ & $f_5$ & $f_6$ & \% error\\[0.1cm] \hline
			$2$	 & $ 0.7814 $&$ 9.4099$ & $-0.4092$ & $-8.2154$ & $0.4546$ & $3.5368$ & $ 0.4 $ \\[0.1cm]
			$1$	 & $ 0.7156$ & $4.0846$ & $-0.3431$ & $-2.8854$ & $0.2028$ & $5.2968 $ & $ 0.2 $ \\ [0.1cm]
			$0$	 & $ 0.6864$ & $0.60529$ & $-0.3137$ & $0.60431 $ & $ \approx 0 $ & $ \approx 0 $ & $ 0.02 $ \\ [0.1cm]
			$-1$ & $ 0.8109 $& $-21.339 $& $-0.4382$ & $22.535$ & $-0.2552$ & $24.736 $ & $ 0.3 $ \\ [0.1cm]
			$-2$ & $  0.8669$ & $-24.001$ & $-0.4942$ & $25.188$ & $-0.4277$ & $34.662 $ & $ 0.3 $ \\ \hline
            \hline
		\end{tabular}
		
		{\vskip 0.15in}

	    \begin{tabular}{c @{\hskip 0.1in} c @{\hskip 0.1in} c @{\hskip 0.1in} c @{\hskip 0.1in} c @{\hskip 0.1in} c @{\hskip 0.1in} c @{\hskip 0.1in} c}
			\hline \hline
			$m$& $f_1$ & $f_2$ & $f_3$& $f_4$ & $f_5$ & $f_6$ & \% error\\[0.1cm] \hline
			$2$	 & $ 0.5454 $&$ -1.4771 $& $-0.6360 $& $2.0641 $& $0.395 $& $-1.2499 $ & $ 0.7 $ \\[0.1cm]
			$1$	 & $ 0.4953 $& $-1.5226 $& $-0.5854 $& $2.1208 $& $0.1549 $& $-1.6087 $ & $ 0.5 $ \\ [0.1cm]
			$0$	 & $ 0.4327 $& $0.44203 $& $-0.5222 $& $0.15251 $ & $ \approx 0 $ & $ \approx 0 $ & $ 0.04 $ \\ [0.1cm]
			$-1$ & $ 0.4951 $& $0.13356 $& $-0.5851 $& $0.46337 $& $0.1453 $& $-0.18332 $ & $ 0.6 $ \\ [0.1cm]
			$-2$ & $ 0.505 $& $3.826 $& $-0.5956 $& $-3.2381 $& $0.404 $&$ -1.4359 $ & $ 0.6 $ \\ \hline
            \hline
		\end{tabular}
	\end{center}
	\caption{Same as Table~\ref{table:fitsl2m2} but for the axial gravitational-led sector and the $n=0,\,\ell=2$ mode for real (top) and imaginary (bottom) parts.}
	\label{table:gravfitl2}
\end{table*}

\clearpage

\begin{table*}[h]
	\begin{center}
		\begin{tabular}{c @{\hskip 0.1in} c @{\hskip 0.1in} c @{\hskip 0.1in} c @{\hskip 0.1in} c @{\hskip 0.1in} c @{\hskip 0.1in} c @{\hskip 0.1in} c}
			\hline \hline
			$m$& $g_1$ & $g_2$ & $g_3$& $g_4$ & $g_5$ & $g_6$ & \% error\\[0.1cm] \hline
			$2$	 & $0.0358 $&$ -1.309 $& $0.4527 $& $-0.15518 $& $0.875 $&$ 5.1813$ & $ 0.4 $ \\[0.1cm]
			$1$	 & $ 0.4108 $&$ -8.4164 $& $0.07448 $& $7.0422 $& $0.267 $&$ 8.7201 $ & $ 0.6 $ \\ [0.1cm]
			$0$	 & $ 0.755 $&$ -1.3916 $& $-0.2705 $& $-0.014878 $ & $ \approx 0 $ & $ \approx 0 $ & $ 0.1 $ \\ [0.1cm]
			$-1$ & $ 0.2332 $&$ -5.1311 $& $0.2517 $& $3.7652 $& $0.5398 $&$ 4.8022 $ & $ 0.7 $ \\ [0.1cm]
			$-2$ & $ 0.09331 $&$ 32.873 $& $0.3645 $& $-33.706 $& $0.6074 $&$ 7.751 $ & $ 0.4 $ \\ \hline
            \hline
		\end{tabular}
		
		{\vskip 0.15in}

	    \begin{tabular}{c @{\hskip 0.1in} c @{\hskip 0.1in} c @{\hskip 0.1in} c @{\hskip 0.1in} c @{\hskip 0.1in} c @{\hskip 0.1in} c @{\hskip 0.1in} c}
			\hline \hline
			$m$& $g_1$ & $g_2$ & $g_3$& $g_4$ & $g_5$ & $g_6$ & \% error\\[0.1cm] \hline
			$2$	 & $ 0.4943 $& $-0.41208$ & $-0.5929 $& $0.20877 $& $0.03482$ & $-1.4985 $ & $ 1.6 $ \\[0.1cm]
			$1$	 & $ 0.4565 $& $0.34382 $& $-0.5519 $& $-0.53517 $& $-0.1934 $& $-2.8791 $ & $ 1.5 $ \\ [0.1cm]
			$0$	 & $ 0.5237 $& $-0.66988 $& $-0.6199 $& $0.47395 $ & $ \approx 0 $ & $ \approx 0 $ & $ 0.2 $ \\ [0.1cm]
			$-1$ & $ 0.4395 $& $-1.6133 $& $-0.5347 $& $1.4324 $& $-0.2306 $& $-6.1669 $ & $ 1.2 $ \\ [0.1cm]
			$-2$ & $ 0.6585 $&$ -13.538 $& $-0.79 $& $11.594 $& $-0.5339 $&$ 38.21 $ & $ 1.9 $ \\ \hline
            \hline
		\end{tabular}
	\end{center}
	\caption{Same as Table~\ref{table:fitsl2m2} but for the scalar-led sector and the $n=0,\,\ell=2$ mode for real (top) and imaginary (bottom) parts.}
	\label{table:gravfitl2}
\end{table*}

\begin{table*}[h]
	\begin{center}
		\begin{tabular}{c @{\hskip 0.1in} c @{\hskip 0.1in} c @{\hskip 0.1in} c @{\hskip 0.1in} c @{\hskip 0.1in} c @{\hskip 0.1in} c @{\hskip 0.1in} c}
			\hline \hline
			$m$& $f_1$ & $f_2$ & $f_3$& $f_4$ & $f_5$ & $f_6$ & \% error\\[0.1cm] \hline
			$3$	 & $ 1.233 $&$ -14.389 $& $-0.6342 $& $17.589 $& $0.4971 $& $-3.4131 $ & $ 0.2 $ \\[0.1cm]
			$2$	 & $ 1.058 $& $-23.33  $& $-0.4581 $& $25.753 $& $0.4438 $& $-41.871 $ & $ 0.3 $ \\[0.1cm]
			$1$	 & $ 0.9376 $& $-40.147 $& $-0.339 $& $44.144 $& $0.2295 $& $6.0011 $ & $ 0.1 $ \\ [0.1cm]
			$0$	 & $ 0.8008 $& $1.8508 $& $-0.202 $& $2.0534 $ & $ \approx 0 $ & $ \approx 0 $ & $ 0.1 $ \\ [0.1cm]
			$-1$ & $ 0.4219 $& $-11.592 $& $0.1769 $& $15.583 $& $0.4033 $& $9.1724 $ & $ 0.05 $ \\ [0.1cm]
			$-2$ & $ 0.3243 $& $-20.236 $& $0.2745 $& $24.276 $& $0.5383 $& $24.935 $ & $ 1.2 $ \\ [0.1cm]
			$-3$ & $ 0.2394 $& $-19.856 $& $0.3595 $& $23.803 $& $0.6471 $& $29.555 $ & $ 1.6 $ \\ \hline
            \hline
		\end{tabular}
		
		{\vskip 0.15in}

	    \begin{tabular}{c @{\hskip 0.1in} c @{\hskip 0.1in} c @{\hskip 0.1in} c @{\hskip 0.1in} c @{\hskip 0.1in} c @{\hskip 0.1in} c @{\hskip 0.1in} c}
			\hline \hline
			$m$& $f_1$ & $f_2$ & $f_3$& $f_4$ & $f_5$ & $f_6$ & \% error\\[0.1cm] \hline
			$3$	 & $ 0.5702 $& $3.2919 $& $-0.6672 $& $0.25096 $& $0.5459 $& $0.94723 $ & $ 3.9 $ \\[0.1cm]
			$2$	 & $ 0.5219 $& $-8.7958 $& $-0.6164 $& $10.14 $& $0.4255 $& $-41.399 $ & $ 0.5 $ \\[0.1cm]
			$1$	 & $ 0.5218 $& $-0.1413 $& $-0.6106 $& $-3.3617 $& $0.1519 $& $-19.551 $ & $ 3.5 $ \\ [0.1cm]
			$0$	 & $ 0.4929 $& $0.6282 $& $-0.5811 $& $-4.164 $ & $ \approx 0 $ & $ \approx 0 $ & $ 2.5 $ \\ [0.1cm]
			$-1$ & $ 0.5013 $& $0.74239 $& $-0.5901 $& $-4.2191 $& $0.1513 $& $-19.53 $ & $ 4.5 $ \\ [0.1cm]
			$-2$ & $ 0.5151 $& $-4.5501 $& $-0.6043 $& $0.9326 $& $0.3959 $& $-20.313 $ & $ 4.6 $ \\[0.1cm]
			$-3$ & $ 0.5943 $& $4.3365 $& $-0.6836 $& $-8.0688 $& $0.5481 $& $-17.372 $ & $ 4.2 $ \\ \hline
            \hline
		\end{tabular}
	\end{center}
	\caption{Same as Table~\ref{table:fitsl2m2} but for the axial gravitational-led sector and the $n=0,\,\ell=3$ mode.}
	\label{table:l3fitgrav}
\end{table*}

\clearpage

\begin{table*}[t]
	\begin{center}
		\begin{tabular}{c @{\hskip 0.1in} c @{\hskip 0.1in} c @{\hskip 0.1in} c @{\hskip 0.1in} c @{\hskip 0.1in} c @{\hskip 0.1in} c @{\hskip 0.1in} c @{\hskip 0.1in} c @{\hskip 0.1in} c}
			\hline \hline
			$m$& $g_1$ & $g_2$ & $g_3$& $g_4$ & $g_5$ & $g_6$ & \% error\\[0.1cm] \hline
			$3$	 & $ 0.3327 $& $47.358 $& $0.3429 $& $-51.198 $& $0.7484 $& $-75.948 $& $ 0.3$ \\[0.1cm]
			$2$	 & $ 0.2805 $& $15.019 $& $0.3965 $& $-18.848 $& $0.5196 $& $-9.7528 $& $ 0.2$ \\[0.1cm]
			$1$	 & $ 0.5433 $& $-3.161 $& $0.1326 $& $-0.54931 $& $0.2796 $& $1.0592 $& $ 0.1$ \\ [0.1cm]
			$0$	 & $ 0.7772 $& $5.2469 $ & $  -0.102 $& $-8.9636 $& $-0.002274 $& $0.2666 $& $ 0.0 $ \\ [0.1cm]
			$-1$ & $ 0.3250 $& $-2.8674 $& $0.3509 $& $-0.83273 $& $0.508 $& $-0.5176 $ & $ 0.1 $ \\ [0.1cm]
			$-2$ & $ 0.0359 $& $6.9322 $& $0.6421 $& $-10.778 $& $0.9033 $& $-9.3872 $& $ 0.2 $ \\[0.1cm]
			$-3$ & $ 0.1106 $& $-4.7023 $& $0.5367 $& $2.9615 $& $0.842 $& $6.626 $& $ 3.0$ \\\hline
            \hline
		\end{tabular}
		
		{\vskip 0.15in}

	    \begin{tabular}{c @{\hskip 0.1in} c @{\hskip 0.1in} c @{\hskip 0.1in} c @{\hskip 0.1in} c @{\hskip 0.1in} c @{\hskip 0.1in} c @{\hskip 0.1in} c @{\hskip 0.1in} c @{\hskip 0.1in} c}
			\hline \hline
			$m$& $g_1$ & $g_2$ & $g_3$& $g_4$ & $g_5$ & $g_6$ & \% error\\[0.1cm] \hline
			$3$	 & $ 0.7327 $& $-7.5698 $& $-0.8297 $& $4.5338 $& $0.4339 $& $-2.4478 $& $ 3.5 $ \\[0.1cm]
			$2$	 & $ 0.4974 $& $0.56638 $& $-0.5926 $& $-3.6759 $& $0.2108 $& $0.9019 $& $ 3.1 $ \\[0.1cm]
			$1$	 & $ 0.4444 $& $7.1089 $& $-0.5377 $& $-10.199 $& $-0.03791 $& $-1.074 $& $ 2.8 $ \\ [0.1cm]
			$0$	 & $ 0.5231 $& $-0.92939 $& $-0.6165 $& $-2.1366 $ & $ \approx 0 $ & $ \approx 0 $ & $ 2.5 $ \\ [0.1cm]
			$-1$ & $ 0.4454 $& $7.1024 $& $-0.5385 $& $-10.188 $& $-0.08306 $& $2.8145 $& $ 2.7 $ \\ [0.1cm]
			$-2$ & $ 0.4852 $& $-2.369 $& $-0.5797 $& $-0.76727 $& $0.03384 $& $15.112 $& $ 2.8 $ \\[0.1cm]
			$-3$ & $ 0.6379 $& $-0.79127 $& $-0.731 $& $-2.4876 $& $0.3901 $& $1.2141 $& $ 5.1 $ \\ \hline
            \hline
		\end{tabular}
	\end{center}
	\caption{Same as Table~\ref{table:fitsl2m2} but for the scalar-led sector and the $n=0,\,\ell=3$ mode.}
	\label{table:l3fitsca}
\end{table*}

\subsection*{Polar gravitational sector}

\begin{table*}[h]
	\begin{tabular}{c @{\hskip 0.12in} c @{\hskip 0.12in} c @{\hskip 0.12in} c @{\hskip 0.12in} c @{\hskip 0.12in} c @{\hskip 0.12in} c }
		\hline
		\hline
		$a/M$&$\alpha/M^2$& $m = -2$ & $m=-1$ & $m=0$ & $m=1$ & $m=2$ \\ \hline
		
		&$0.0$	& $ .3737,.0889 $ & $ .3737,.0889 $ & $ .3737,.0889 $ & $ .3737,.0889 $ & $ .3737,.0889 $ \\[0.1cm]
		
		$0.0$&$0.05$	& $ .3737,.0889 $ & $ .3737,.0889 $ & $ .3737,.0889 $ & $ .3737,.0889 $ & $ .3737,.0889 $ \\[0.1cm]
		
		&$0.1$	& $ .3737,.0889 $ & $ .3737,.0889 $ & $ .3737,.0889 $ & $ .3737,.0889 $ & $ .3737,.0889 $ \\[0.1cm] \hline
		
		&$0.0$	& $ .3726,.0881 $ & $ .3730,.0888 $ & $ .3737,.0889 $ & $ .3744,.0888 $ & $ .3751,.0882 $ \\[0.1cm]
		
		$0.01$&$0.05$	& $ .3726,.0881 $ & $ .3730,.0888 $ & $ .3737,.0889 $ & $ .3744,.0888 $ & $ .3751,.0882 $ \\[0.1cm]
		
		&$0.1$	& $ .3726,.0881 $ & $ .3731,.0887 $ & $ .3737,.0889 $ & $ .3744,.0887 $ & $ .3752,.0883 $ \\[0.1cm] \hline
		
		&$0.0$	& $ .3712,.0861 $ & $ .3725,.0882 $ & $ .3737,.0889 $ & $ .3751,.0882 $ & $ .3768,.0859 $ \\[0.1cm]
		
		$0.02$&$0.05$	& $ .3712,.0861 $ & $ .3725,.0882 $ & $ .3737,.0889 $ & $ .3751,.0882 $ & $ .3768,.0859 $ \\[0.1cm]
		
		&$0.1$	& $ .3712,.0861 $ & $ .3726,.0883 $ & $ .3737,.0889 $ & $ .3752,.0883 $ & $ .3768,.0861 $ \\[0.1cm] \hline
		
		&$0.0$	& $ .3699,.0832 $ & $ .3719,.0872 $ & $ .3737,.0889 $ & $ .3759,.0873 $ & $ .3789,.0829 $ \\[0.1cm]
		
		$0.03$&$0.05$	& $ .3699,.0832 $ & $ .3719,.0874 $ & $ .3737,.0889 $ & $ .3759,.0873 $ & $ .3789,.0829 $ \\[0.1cm]
		
		&$0.1$	& $ .3699,.0832 $ & $ .3719,.0875 $ & $ .3737,.0889 $ & $ .3759,.0874 $ & $ .3789,.0832 $ \\[0.1cm] \hline
		
		&$0.0$	& $ .3682,.0801 $ & $ .3711,.0862 $ & $ .3737,.0889 $ & $ .3767,.0859 $ & $ .3811,.0799 $ \\[0.1cm]
		
		$0.04$&$0.05$	& $ .3682,.0805 $ & $ .3713,.0861 $ & $ .3737,.0889 $ & $ .3810,.0801 $ & $ .3829,.0788 $ \\[0.1cm]
		
		&$0.1$	& $ .3682,.0809 $ & $ .3713,.0863 $ & $ .3737,.0889 $ & $ .3766,.0859 $ & $ .3805,.0809 $ \\ \hline
		\hline
	\end{tabular}
    \caption{QNM frequencies for polar
    gravitational-led sector with $n=0, \,\ell =2$ for slowly rotating BHs in dCS gravity. The format
    used is $M$($\textrm{Re}(\omega),-\textrm{Im}(\omega)$). To save space, the leading zeros have been omitted.}
    \label{tab:pol_grv_l2_all_m}
    \end{table*}
    
\clearpage
    %
    %
\begin{table*}[h]
	\begin{tabular}{c @{\hskip 0.12in} c @{\hskip 0.12in} c @{\hskip 0.12in} c @{\hskip 0.12in} c @{\hskip 0.12in} c @{\hskip 0.12in} c @{\hskip 0.12in} c @{\hskip 0.12in} c }
		\hline \hline
		$a/M$&$\alpha/M^2$& $m = -3$ & $m = -2$ & $m=-1$ & $m=0$ & $m=1$ & $m=2$ & $m=3$ \\ \hline
		
		&$0.0$	& $ .5994,.0927 $ &$ .5994,.0927 $ & $ .5994,.0927 $ & $ .5994,.0927 $ & $ .5994,.0927 $ & $ .5994,.0927 $& $ .5994,.0927 $ \\[0.1cm]
		
		$0.0$&$0.05$	& $ .5994,.0927 $ &$ .5994,.0927 $ & $ .5994,.0927 $ & $ .5994,.0927 $ & $ .5994,.0927 $ & $ .5994,.0927 $& $ .5994,.0927 $ \\[0.1cm]
		
		&$0.1$	& $ .5994,.0927 $ &$ .5994,.0927 $ & $ .5994,.0927 $ & $ .5994,.0927 $ & $ .5994,.0927 $ & $ .5994,.0927 $& $ .5994,.0927 $ \\[0.1cm]\hline
		
		&$0.0$	& $ .5972,.0911 $ &$ .5980,.0920 $ & $ .5987,.0925 $ & $ .5994,.0927 $ & $ .6001,.0925 $ & $ .6008,.0919 $ & $ .6014,.0911 $\\[0.1cm]
		
		$0.01$&$0.05$& $ .5972,.0912 $ & $ .5980,.0920 $ & $ .5988,.0925 $ & $ .5994,.0927 $ & $ .6001,.0925 $& $ .6008,.0920 $ & $ .6014,.0911 $ \\[0.1cm]
		
		&$0.1$	& $ .5971,.0913 $&$ .5980,.0921 $ & $ .5988,.0925 $ & $ .5994,.0927 $ & $ .6001,.0926 $ & $ .6008,.0919 $& $ .6015,.0912 $ \\[0.1cm] \hline
		
		&$0.0$	& $ .5947,.0874 $ & $ .5963,.0901 $ & $ .5980,.0919 $ & $ .5994,.0927 $ & $ .6008,.0919 $ & $ .6022,.0899 $& $ .6038,.0872 $ \\[0.1cm]
		
		$0.02$&$0.05$& $ .5947,.0875 $ & $ .5963,.0901 $ & $ .5980,.0919 $ & $ .5994,.0927 $ & $ .6008,.0920 $ & $ .6022,.0900 $& $ .6038,.0872 $\\[0.1cm]
		
	&$0.1$	& $ .5946,.0877 $ & $ .5961,.0901 $ & $ .5980,.0920 $ & $ .5994,.0927 $ & $ .6008,.0920 $ & $ .6021,.0902 $& $ .6037,.0876 $\\[0.1cm] \hline
		
		&$0.0$	& $ .5917,.0831 $ & $ .5945,.0874 $ & $ .5972,.0911 $ & $ .5994,.0927 $ & $ .6015,.0911 $ & $ .6038,.0872 $& $ .6069,.0827 $\\[0.1cm]
		
		$0.03$&$0.05$& $ .5915,.0832 $ & $ .5945,.0875 $ & $ .5972,.0912 $ & $ .5994,.0927 $ & $ .6015,.0911 $ & $ .6038,.0873 $& $ .6068,.0828 $\\[0.1cm]
		
		&$0.1$	& $ .5912,.0835 $ & $ .5942,.0877 $ & $ .5973,.0913 $ & $ .5994,.0927 $ & $ .6015,.0912 $ & $ .6037,.0876 $& $ .6066,.0832 $\\[0.1cm] \hline
		
		&$0.0$	& $ .5888,.0787 $ & $ .5927,.0845 $ & $ .5964,.0901 $ & $.5994,.0927  $ & $ .6022,.0899 $ & $ .6047,.0869 $& $ .6103,.0785 $\\[0.1cm]
		
		$0.04$&$0.05$	& $ .5987,.0790 $ & $ .5925,.0846 $ & $ .5964,.0901 $ & $ .5994,.0927 $ & $ .6022,.0900 $ & $ .6045,.0869 $& $ .6102,.0787 $\\[0.1cm]
		
		&$0.1$	& $ .5882,.0794 $ & $ .5922,.0849 $ & $ .5965,.0903 $ & $ .5994,.0927 $ & $ .6022,.0902 $ & $ .6041,.0872 $& $ .6099,.0792 $\\ \hline
		\hline
	\end{tabular}
    \caption{Same as Table~\ref{tab:grv_l2_all_m} but for the polar gravitational-led sector with $n=0, \, \ell =3$. 
    }
    \label{tab:pol_grv_l3_all_m}
\end{table*}

\begin{table*}[h]
	\begin{center}
		\begin{tabular}{c @{\hskip 0.1in} c @{\hskip 0.1in} c @{\hskip 0.1in} c @{\hskip 0.1in} c @{\hskip 0.1in} c}
			\hline \hline
			$m$& $f_1$ & $f_3$ & $f_5$ & $f_6$ & \% error\\[0.1cm] \hline
			$2$	 & $ 0.9072 $ & $-0.5342 $& $0.4833 $& $-5.5714 $  & $ 0.1 $ \\[0.1cm]
			$1$	 & $ 0.7535 $& $-0.3800 $& $0.2582 $& $-0.77801 $  & $ 0.1 $ \\ [0.1cm]
			$0$	 & $ 0.6868 $& $-0.3132 $ & $\approx 0 $ & $ \approx 0 $ & $ 0.02 $ \\ [0.1cm]
			$-1$ & $ 0.6921 $& $-0.3185 $& $-0.1322 $& $-0.3766 $  & $ 0.01 $ \\ [0.1cm]
			$-2$ & $ 0.7417 $& $-0.3681 $& $-0.2339 $& $0.2931 $  & $ 0.01 $ \\ \hline
            \hline
		\end{tabular}
		
		{\vskip 0.15in}

	    \begin{tabular}{c @{\hskip 0.1in} c @{\hskip 0.1in} c @{\hskip 0.1in} c @{\hskip 0.1in} c @{\hskip 0.1in} c}
			\hline \hline
			$m$& $f_1$ & $f_3$ & $f_5$ & $f_6$ & \% error\\[0.1cm] \hline
			$2$	 & $ 0.5117 $& $-0.6017 $& $0.3824 $& $-2.171 $  & $ 0.7 $ \\[0.1cm]
			$1$	 & $ 0.5005 $& $-0.5899 $& $0.1328 $& $0.1733 $  & $ 0.4 $ \\ [0.1cm]
			$0$	 & $ 0.4351 $& $-0.5239 $ & $\approx 0 $ & $ \approx 0 $ & $ 0.02 $ \\ [0.1cm]
			$-1$ & $ 0.5001 $& $-0.5895 $& $0.1335 $& $0.2283 $  & $ 0.4 $ \\ [0.1cm]
			$-2$ & $ 0.5158 $& $-0.6058 $& $0.3924 $& $-2.0636 $  & $ 0.7 $ \\ \hline
            \hline
		\end{tabular}
	\end{center}
	\caption{Same as Table~\ref{table:fitsl2m2} but for the polar gravitational-led sector and the $n=0,\,\ell=2$ mode for real (top) and imaginary (bottom) parts.}
	\label{table:pol_gravfitl2}
\end{table*}

\clearpage 

\begin{table*}[htb]
	\begin{center}
		\begin{tabular}{c @{\hskip 0.1in} c @{\hskip 0.1in} c @{\hskip 0.1in} c @{\hskip 0.1in} c @{\hskip 0.1in} c @{\hskip 0.1in} c @{\hskip 0.1in} c}
			\hline \hline
			$m$& $f_1$ & $f_3$ & $f_5$ & $f_6$ & \% error\\[0.1cm] \hline
			$3$	 & $ 1.039 $& $-0.4406 $& $0.7068 $& $-1.2427 $ & $ 0.1 $ \\[0.1cm]
			$2$	 & $ 1.048 $& $-0.449 $& $0.3923 $& $-0.5770 $ & $ 0.1 $ \\[0.1cm]
			$1$	 & $ 0.8921 $& $-0.2927 $& $0.2417 $& $-0.1956 $ & $ 0.1 $ \\ [0.1cm]
			$0$	 & $ 0.7998 $& $-0.2004 $&  $ \approx 0 $ & $ \approx 0 $ & $ 0.1 $ \\ [0.1cm]
			$-1$ & $ 0.413 $& $0.1865 $& $0.4049 $& $-0.2653 $ & $ 0.1 $ \\ [0.1cm]
			$-2$ & $ 0.3065 $& $0.2931 $& $0.5563 $& $-0.5452 $ & $ 0.1 $ \\ [0.1cm]
			$-3$ & $ 0.2187 $& $0.381 $& $0.6709 $& $-0.7868 $ & $ 0.1 $ \\ \hline
            \hline
		\end{tabular}
		
		{\vskip 0.15in}

	    \begin{tabular}{c @{\hskip 0.1in} c @{\hskip 0.1in} c @{\hskip 0.1in} c @{\hskip 0.1in} c @{\hskip 0.1in} c @{\hskip 0.1in} c @{\hskip 0.1in} c}
			\hline \hline
			$m$& $f_1$ & $f_3$ & $f_5$ & $f_6$ & \% error\\[0.1cm] \hline
			$3$	 & $ 0.6027 $& $-0.6965 $& $0.5076 $& $1.8342 $ & $ 0.9 $ \\[0.1cm]
			$2$	 & $ 0.5094 $& $-0.6032 $& $0.3823 $& $-2.1546 $ & $ 0.7 $ \\[0.1cm]
			$1$	 & $ 0.5022 $& $-0.5953 $& $0.1296 $& $0.2921 $ & $ 0.3 $ \\ [0.1cm]
			$0$	 & $ 0.5364 $& $-0.6291 $&  $ \approx 0 $ & $ \approx 0 $ & $ 0.1 $ \\ [0.1cm]
			$-1$ & $ 0.4608 $& $-0.5539 $& $0.1397 $& $-1.0851 $ & $ 0.3 $ \\ [0.1cm]
			$-2$ & $ 0.5074 $& $-0.6012 $& $0.375 $& $-2.0335 $ & $ 0.7 $ \\[0.1cm]
			$-3$ & $ 0.5912 $& $-0.6851 $& $0.5286 $& $-0.7794 $ & $ 0.8 $ \\ \hline
            \hline
		\end{tabular}
	\end{center}
	\caption{Same as Table~\ref{table:fitsl2m2} but for the polar gravitational-led sector and the $n=0,\,\ell=3$ mode.}
	\label{table:pol_l3fitgrav}
\end{table*}

\twocolumngrid


\bibliography{ppr_v}

\end{document}